\documentclass[11pt,a4paper]{article}
\pdfoutput=1
\usepackage{jcappub}
\usepackage{aasmacros}

\ifx\pdfoutput\undefined
\usepackage[dvips,bookmarks=false]{hyperref}	% This is for arXiv.org
\else
\usepackage{hyperref}	% This is for pdftex
\fi
\hypersetup{colorlinks,bookmarksopen,bookmarksnumbered,citecolor=blue,
linkcolor=black,pdfstartview=FitH,urlcolor=blue}

\usepackage{graphicx}
\usepackage{amsmath}
\usepackage{amssymb}
\usepackage{rotating} 
\usepackage{xspace}
\usepackage{color}

\newcommand{\eagle}{{\sc eagle}\xspace}
\newcommand{\apostle}{{\sc apostle}\xspace}

\newcommand{\be}{\begin{equation}}
\newcommand{\ee}{\end{equation}}
\newcommand{\beq}{\begin{equation}}
\newcommand{\eeq}{\end{equation}}

\newcommand{\vect}[1]{\boldsymbol{\rm #1}}
\newcommand{\Msun}{{\rm M}_\odot}

\makeatletter
\renewcommand{\fnum@table}{\textbf{\tablename~\thetable}}
\renewcommand{\fnum@figure}{\textbf{\figurename~\thefigure}}
\makeatother

\title{Simulated Milky Way analogues: implications for dark matter direct searches}

\author[a]{Nassim Bozorgnia,}
\author[a]{Francesca Calore,}
\author[b]{Matthieu Schaller,}
\author[a,c]{Mark Lovell,}
\author[a]{Gianfranco Bertone,}
\author[b]{Carlos S. Frenk,}
\author[d]{Robert A. Crain,}
\author[e, f]{Julio F. Navarro,}
\author[g]{Joop Schaye}
\author[b]{and Tom Theuns}

\affiliation[a]{GRAPPA, University of Amsterdam,\\
Science Park 904,
  1090 GL Amsterdam, Netherlands} 
\affiliation[b]{Institute for Computational Cosmology,\\ 
Durham University, South Road, Durham DH1 3LE, UK}
\affiliation[c]{Instituut-Lorentz for Theoretical Physics,\\ 
Niels Bohrweg 2, NL-2333 CA Leiden,
  Netherlands}
\affiliation[d]{Astrophysics Research Institute, Liverpool John Moores University,\\ 
146 Brownlow Hill, Liverpool L3 5RF, UK}
\affiliation[e]{Department of Physics \& Astronomy, University of Victoria,\\
 Victoria, BC, V8P 5C2, Canada}
 \affiliation[f]{Senior CIfAR Fellow}
\affiliation[g]{Leiden Observatory, Leiden University,\\ 
PO Box 9513, NL-2300 RA Leiden, Netherlands}

\emailAdd{n.bozorgnia@uva.nl}

\abstract{We study the implications of galaxy formation on dark matter direct
  detection using high resolution hydrodynamic simulations of Milky Way-like galaxies simulated 
  within the  \eagle and \apostle projects. We identify
  Milky Way analogues that satisfy observational constraints on the
  Milky Way rotation curve and total stellar mass. We then extract the
  dark matter density and velocity distribution in the Solar neighbourhood for this set of
  Milky Way analogues, and use them to analyse the results of current
  direct detection experiments. For most Milky Way analogues, 
  the  event rates in direct detection experiments obtained from 
  the best fit Maxwellian distribution (with peak speed of 223 -- 289 km$/$s) 
  are similar to those obtained directly from the simulations. As a consequence, the allowed regions and exclusion limits set by direct  detection experiments in the dark matter mass and spin-independent
  cross section plane shift by a few GeV compared to the Standard Halo Model, at low dark matter
  masses. For each dark matter mass, the halo-to-halo variation of the local dark 
matter density results in an overall shift of the allowed regions and 
exclusion limits for the cross section. However, the
  compatibility of the possible hints for a dark matter signal from DAMA and CDMS-Si and
  null results from LUX and SuperCDMS is not improved.}

\arxivnumber{}

\begin{document}
\maketitle

%%%%%%%%%%%%%%%%%%%%%%%%%%%%%%%%%%%%%%%%%%%%%%%%%%%%%%%%%%%%%%%%%%%%%%%%%%%%%%%%%%%%%
\section{Introduction}
\label{sec:introduction}

Dark matter (DM) direct detection searches aim to measure the recoil energy of a
nucleus in a low background detector induced by the scattering of Weakly Interacting
Massive Particles (WIMPs), one of the most well-motivated classes of particle DM
candidates, from the dark halo of the Milky Way (MW). In the last few years, a number
of direct detection experiments have reported hints for signals that could be
interpreted as signatures of DM interactions. Currently, only hints from the
DAMA/LIBRA~\cite{Bernabei:2013xsa} and CDMS-II~\cite{Agnese:2013rvf} experiments
remain. These hints, however, are in tension with the null results from many other
experiments, including LUX (Large Underground Xenon)~\cite{Akerib:2013tjd} and
SuperCDMS~\cite{Agnese:2014aze}, which set the strongest exclusion limits in the
spin-independent DM--nucleon cross section and WIMP mass plane at large ($> 6$~GeV)
and small (3.5--6~GeV) WIMP masses, respectively.

There are large systematic uncertainties in the interpretation of the data from DM direct
detection experiments due to the uncertainties in the DM density and velocity distribution at the position
of the Sun. Indeed, in order to show the preferred regions and exclusion limits in
the DM mass and scattering cross section plane, one needs to make assumptions about
the DM halo properties, namely the local DM density and DM velocity distribution. The
simplest and most commonly adopted DM halo model is the Standard Halo Model (SHM): it
assumes an isothermal, spherical DM halo with an isotropic Maxwell-Botzmann DM
velocity distribution. In this model, the local DM density is usually assumed to be
0.3 GeV$/{\rm cm}^3$; the local velocity distribution in the Galactic rest frame is
assumed to be a Maxwellian with a most probable (peak) speed equal to the local circular speed $v_c$ (usually assumed to be
220 km$/$s), a three-dimensional velocity dispersion of $\sqrt{3/2}~v_c$, and truncated at the
escape speed of 544 km/s from the Galaxy. Throughout this paper, when referring to
the SHM Maxwellian, we consider the Maxwellian distribution with a most probable speed of 230
km$/$s in order to be consistent with the local circular speed assumed in deriving
the observed MW rotation curves~\cite{Iocco:2015xga} (see section~\ref{sec:selection}
for more details).

The local DM density enters as a normalisation factor in the event rate in direct
detection experiments, whilst the DM velocity distribution enters in the event rate
in a more complicated way (see eq.~\eqref{rate}). In particular, different direct
detection experiments with different energy thresholds and target nuclei
probe different DM speed ranges, and thus their dependence on the DM velocity
distribution varies. To overcome the astrophysical uncertainties in the
interpretation of direct detection results, methods to compare the results of
different experiments in a halo-independent way have been developed and applied to
direct detection data~\cite{Fox:2010bz, Fox:2010bu, McCabe:2011sr, Frandsen:2011gi,
  Gondolo:2012rs, HerreroGarcia:2012fu, Frandsen:2013cna, DelNobile:2013cta,
  BHSZ:2013, DelNobile:2013cva, DelNobile:2013gba, Feldstein:2014gza, Fox:2014kua,
  Cherry:2014wia, Feldstein:2014ufa, Bozorgnia:2014gsa, Anderson:2015xaa,
  Blennow:2015gta, Ferrer:2015bta, Gelmini:2015voa}. Such
methods are based on comparing different experiments in the space of the minimum DM
speed, $v_{\rm min}$, required for a DM particle to deposit a recoil energy $E_R$ in
the detector, and thus can only be applied to compare experiments which are sensitive
to the same range of $v_{\rm min}$.

High resolution numerical simulations of galaxy formation can provide information on
the properties of the DM halo. Velocity distributions extracted from high resolution
DM-only (DMO) simulations deviate substantially from a Maxwellian distribution, with
fewer particles at the peak and an excess of particles at large
speeds~\cite{Vogelsberger:2008qb, Kuhlen:2009vh}. DMO simulations can achieve high
resolution, but have significant systematic uncertainties since they neglect the effect of the baryonic
components during the galaxy formation process. Additionally, velocity distributions drawn from isolated simulations of the MW and the accretion of the Sagittarius dwarf galaxy show significant deviations from the best fit Maxwellian distribution, and this has important effects on direct detection rates~\cite{Purcell:2012sh}. Nevertheless, the local DM distribution is 
expected to be very smooth: high resolution N-body simulations predict 
that the density at the Solar radius can differ from the mean over the 
best-fit ellipsoidal equidensity contour by at most 15\% at the 99.9\% 
confidence level~\cite{Vogelsberger:2008qb}.

Recently, realistic cosmological
simulations including baryons have become possible thanks to advances in physical
modeling, as well as better understanding and treatment of numerical issues. 
The hydrodynamic simulation performed by Ling {\it et al.}~\cite{Ling:2009eh}
predicts a DM speed distribution which deviates significantly from the Maxwellian,
and instead gives an excellent fit to a Tsallis~\cite{Tsallis:1987eu} distribution\footnote{See also ref.~\cite{Beraldo:2013ioa} for a variation of the Tsallis distribution function which is anisotropic.}. In
this simulation, the local DM density is found to be 0.37 -- 0.39 GeV$/$cm$^3$.  The
Eris hydrodynamic simulation~\cite{Kuhlen:2013tra} finds a speed distribution that
is not a perfect Maxwellian, with an excess of particles at speeds less than 350 km$/$s compared to the SHM Maxwellian 
and fewer particles at higher speeds. The speed distribution is instead well fit by the fitting function
proposed by Mao {\it et al.}~\cite{Mao:2012hf}. The NIHAO (Numerical Investigation of
a Hundred Astrophysical Objects) simulations~\cite{Butsky:2015pya} find a DM speed
distribution which is very different from the Maxwellian, being more symmetric around
the peak velocity and falling rapidly at large velocities, thus agreeing with the
results of the Eris simulation. All these hydrodynamic simulations help build an understanding of the processes that can shape and modify the local DM density and velocity distribution. However, to make more precise quantitative predictions it is necessary that all these effects are modeled in such a way that the main galaxy population properties are broadly reproduced. This ensures, for instance, that appropriate amounts of feedback are applied and that the corresponding impact of feedback on DM is reasonably realistic.

In this work, we study the implications of simulated MW-like galaxies within the
\eagle (``Evolution and Assembly of GaLaxies and their Environments")~\cite{Schaye:2015,Crain:2015} and \apostle~\cite{Sawala:2015,Fattahi:2015} projects for DM direct detection. The \eagle
project is a suite of hydrodynamic simulations that are calibrated to reproduce the
observed distribution of stellar masses and sizes of low-redshift galaxies as well as
the relation between the stellar mass and black hole mass of galaxies; in this respect they are unique amongst simulations of galaxy formation. In a companion
paper, we have studied the implications of the \eagle and the related \apostle simulations for DM indirect
detection~\cite{Calore:2015oya}. We first consider a selection of MW-like galaxies
within haloes with virial mass in the range $\mathcal{O}(10^{12} - 10^{13}) \,
\Msun$. We then impose a set of selection criteria crucial for accurately predicting
the local DM density and velocity distribution. Namely, we require that the haloes
satisfy observational constraints on the MW rotation curve~\cite{Iocco:2015xga}, and
have stellar masses within the observed MW stellar mass range. We then extract the
local DM density and velocity distribution for the selected haloes, and use them to
analyse current data from direct detection experiments.

The paper is outlined as follows. In section \ref{sec:simulation} we present the set
of \eagle and \apostle cosmological hydrodynamic simulations used in this work. In section
\ref{sec:selection} we discuss in detail the selection criteria used for choosing MW
analogues, and how each criterion is relevant for determining the local DM
distribution. In sections \ref{sec:veldist} and \ref{sec:DMdensity} we present the
local DM velocity distribution and DM density extracted from the simulations. In
section \ref{sec:signals} we introduce the formalism for computing the predicted
signals in DM direct detection experiments. In section \ref{sec:results} we perform
an analysis of current direct detection data and discuss how the allowed regions and
exclusion limits set by different experiments in the DM mass and scattering cross
section plane changes when the DM distribution of simulated haloes is used. Finally,
in section \ref{sec:summary} we summarize the conclusions of this work. In
Appendices~\ref{app:criteria}, \ref{app:f(v)}, and \ref{app:fits}, we provide
additional material relevant for this work.

\section{Simulations}
\label{sec:simulation}

The simulations used in this study comprise parts of the \eagle~\cite{Schaye:2015,Crain:2015}
and \apostle~\cite{Sawala:2015,Fattahi:2015} projects, both of which we describe below.

The \eagle project combines a state-of-the-art hydrodynamical SPH implementation
nicknamed {\sc anarchy} \cite{DallaVecchia:2015,Schaller:2015b} -- built on top of an
optimized version of the {\sc gadget} SPH code, last described by \cite{Springel:2005} -- and a detailed
subgrid model of galaxy formation that features metal-line cooling,
photo-reionization, star formation, and feedback from star formation and active
galactic nuclei~\cite{Wiersma:2008cs, Wiersma:2009wf, Schaye:2007ss, Vecchia:2012wu, Rosas-Guevara:2013yma}. The subgrid model is regulated by a number of parameters, which are
calibrated to give a good fit to the $z=0.1$ galaxy stellar mass function and to the observed 
relation between galaxy stellar mass and size for disc galaxies. From the suite of \eagle simulations we
select candidate MW analogues in the high resolution (25~Mpc)$^3$ box (initial gas
particle mass $2.26\times10^{5}~\Msun$), which adopts the Planck cosmological
parameters \citep{Planck:2014} ($\Omega_{m}=0.307$, $\Omega_{\Lambda}=0.693$,
$\Omega_{b}=0.0482$, $h=0.678$, $\sigma_{8}=0.83$, and $n_{s}=0.961$) and the \eagle
model parameters re-calibrated for high resolution (see ref.~\cite{Schaye:2015} for a
discussion of model parameter calibration). Throughout the paper, we refer to this
simulation as \eagle HR. Notice that in other \eagle papers, this simulation is referred to as Recal-L025N0752.

As we will discuss in section~\ref{sec:selection}, the initial set of \eagle HR haloes with the virial mass in the range $\mathcal{O}(10^{12} - 10^{13}) \, \Msun$ consists of 61 objects. To increase our initial number of haloes, we also consider haloes within the \apostle project. The \apostle project applies the \eagle model in a series of zoomed
 simulations of Local Group-analogue systems. Each Local Group volume contains two
 haloes of mass $\sim10^{12}~\Msun$, each of which hosts a potential MW galaxy
 analogue. We picked our candidate galaxies from the 12 intermediate resolution, comparable in resolution to \eagle HR, 
 volumes as presented in ref.~\cite{Sawala:2015}. We refer to these simulations
 collectively as \apostle IR. These simulations were run with the same code as
 \eagle, used the \eagle reference parameters as calibrated for the \eagle
 intermediate resolution simulations, and were run with the WMAP7 cosmological
 parameters ($\Omega_{m}=0.272$, $\Omega_{\Lambda}=0.728$, $\Omega_{b}=0.0455$,
 $h=0.704$, $\sigma_{8}=0.81$, and $n_{s}=0.967$); the initial gas particle mass is
 $1.2\times10^{5}~\Msun$.

For both \eagle HR and \apostle IR, companion simulations were run with all of the
matter content treated as collisionless, and in this work we compare the results of
these DMO runs with those that use the full hydrodynamical treatment.

 Notice that we have not performed resolution tests for the simulations used in this work. Lower resolution simulations compared to \eagle HR and \apostle IR cannot be used to test for numerical convergence  because there would be insufficient particles to probe the DM distribution in the Solar neighbourhood. Re-calibrated higher resolution simulations are not available either. The caveat is that without such a test, it remains to be demonstrated that the conclusions of this work are robust with respect to the numerical resolution.

\section{Selection of Milky Way-like galaxies}
\label{sec:selection}

In ref.~\cite{Calore:2015oya} we identified simulated galaxies that are ``good" MW
analogues according to the following three criteria: (i) their rotation curves
provide good fits to the recent compilation of observed MW rotation curve from
ref.~\cite{Iocco:2015xga}, (ii) their stellar masses are within the $3\sigma$
observed stellar mass range of the MW, $4.5 \times10^{10}<M_{*}/\Msun<8.3
\times10^{10}$~\cite{McMillan:2011wd}, and (iii) they contain a substantial stellar
disc component. The details of our selection procedure are presented in
ref.~\cite{Calore:2015oya}. Note that in deriving the measured MW rotation curves
from ref.~\cite{Iocco:2015xga}, we assume as fiducial values a local circular
speed of $v_c = 230$ km/s, a local galactocentric distance of $R_0 = 8$ kpc, and
the component of the Solar peculiar velocity in the direction of the Galactic
rotation of $V_{\odot} = 12.24$ km/s~\cite{Schoenrich:2010}.

We first consider all central galaxies in haloes with
$5\times10^{11}<M_{200}/\Msun<2\times10^{13}$, where $M_{200}$ is the mass enclosed
within the sphere centered on the galaxy's potential minimum, that contains a mean density 200 times the critical density. These
initial sets of galaxies comprise 61 and 24 objects for the \eagle HR and \apostle IR
simulations respectively, and the largest $M_{200}$ in these sets is $1.4 \times
10^{13} \Msun$. The final sets of galaxies that satisfy the three mentioned selection criteria consist of 2 objects for each project at our chose resolution. 

Given the limited statistics once the selection criteria have been applied, we discuss
below how each selection criterion affects the implications for direct detection, in
particular the two main astrophysical inputs for computing direct detection rates,
and examine whether it is possible to relax one or more of the criteria.

\begin{figure}
\begin{center}
  \includegraphics[width=0.49\textwidth]{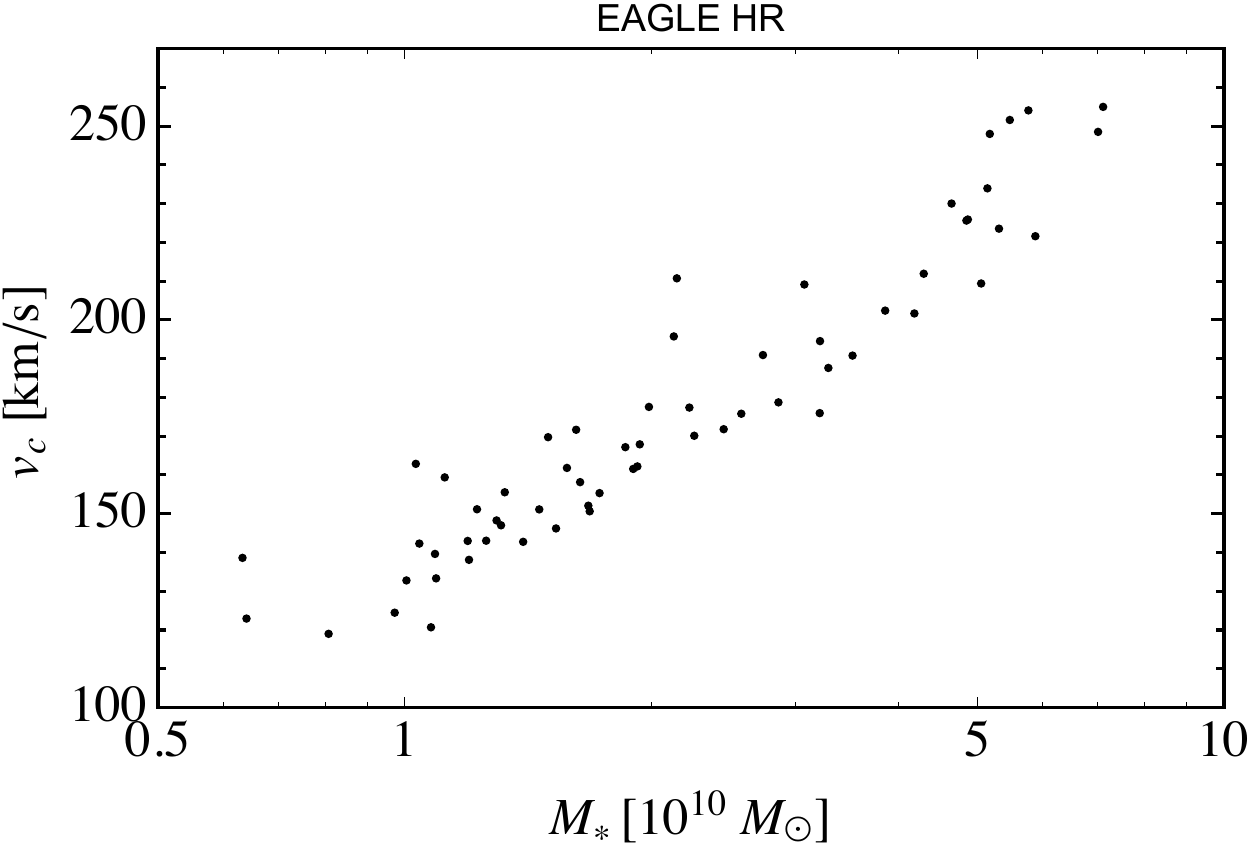}
  \includegraphics[width=0.48\textwidth]{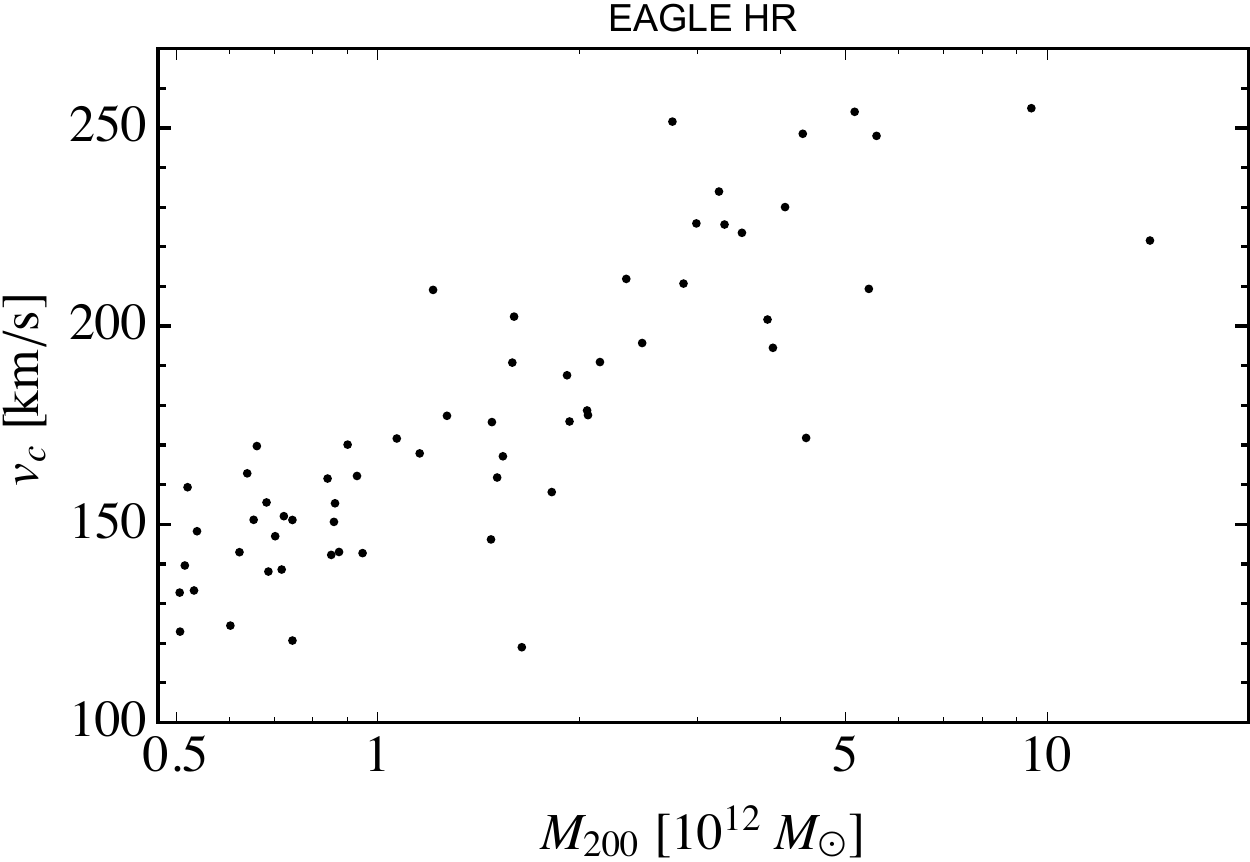}
\caption{\label{fig:vc} The correlation of the local circular speed at $R_0=8$~kpc 
with the total stellar mass, $M_{*}$ (left panel) and the virial mass,
  $M_{200}$ (right panel) for all the haloes with $5\times10^{11}<M_{200}/\Msun \leq
  1.4 \times10^{13}$ in the \eagle HR simulation.}
\end{center}
\end{figure}

The local DM density and velocity distribution of the simulated haloes are important
parameters for direct DM detection. Among the three initial
criteria for identifying MW analogues, criteria (i) and (ii) are important because
they directly affect the local circular speed, $v_c$, at the Solar radius, and
hence the peak of the local DM speed distribution in the Galactic reference frame. In
particular, the fit to the observed MW rotation curve constrains $v_c$. Also, as can
be seen from figure \ref{fig:vc}, the stellar mass, $M_{*}$, is strongly correlated
with $v_c$ at 8 kpc, while the correlation of $M_{200}$ with $v_c$ is weaker. The Pearson's correlation coefficient between $v_c$ and $M_{*}$ is 0.94, while it is 0.68 between $v_c$ and $M_{200}$.

The reason is that $v_c$ at the Solar circle depends on the total mass in stars,
gas and DM enclosed within the sphere of radius 8 kpc.  The enclosed stellar
mass within the Solar circle is a large fraction of the total stellar mass, (0.5
-- 0.9)$M_{*}$ depending on the simulated halo in the \eagle HR and \apostle IR
run, since the enclosed stellar mass does not increase significantly with
galactocentric distance beyond the Solar circle. On the other hand, the enclosed
DM mass increases significantly with galactocentric distance, and the total mass
enclosed within 8 kpc is only (0.01 -- 0.1)$M_{200}$. Over the small halo mass
range probed, we find little correlation between the DM mass enclosed within a
sphere of radius 8 kpc and $M_{200}$. The scatter in galaxy stellar mass at a
fixed halo mass dominates the variation in $v_c$.  A stronger correlation between
$v_c$ and $M_{*}$ compared to the correlation with $M_{200}$ is hence found.

On the other hand, having a substantial stellar disc component, as specified by criterion
(iii), for the simulated galaxies does not have an important effect on the local DM
density and speed distributions. In particular, haloes which satisfy criteria (i) and
(ii) but are spheroidal have DM velocity distributions similar to those with a
disc. The local DM density range for all haloes satisfying criteria (i) and (ii) is
slightly increased compared to the range for haloes satisfying all three criteria
(see appendix~\ref{app:criteria}). This is due to the addition of more haloes to our
sample, but without a significant trend in a specific direction. Thus, we believe
that the predictions for direct detection searches remain unchanged for galaxies that
do not have a substantial disc component.

Analogously to ref.~\cite{Calore:2015oya}, we define a
second set of MW-like galaxies  for each project at our chosen resolution, which satisfy only criteria (i) and (ii). This set
consists of 12 galaxies in the \eagle HR, and 2 galaxies in the \apostle IR
simulations.
Note that only 2 galaxies in the \apostle IR run have a stellar mass compatible with
the observed range of MW stellar mass, and both of these galaxies also have a
dominant stellar disc component. Therefore, it is not possible to increase the number
of MW analogues in the \apostle IR run by relaxing the criterion on the stellar
disc. In table~\ref{tab:criteria} we list the parameters of the MW-like galaxies
satisfying criteria (i) and (ii). In the main sections of this paper we only consider
the set of galaxies which satisfy these two criteria. For completeness and
comparison, we list the results for the two galaxies in the \eagle HR simulation (E9 and E11 in table~\ref{tab:criteria})
which satisfy all three criteria in Appendix \ref{app:criteria}.

\begin{table}[t!]
    \centering
    \begin{tabular}{|c|c|c|c|c|c|c|}
      \hline
        Halo Name  & $M_{*}   [ \times 10^{10} \, \Msun]$ & $M_{200} [ \times 10^{12} \,  \Msun]$  &  D/T & $\chi^2/(N-1)$ \\
       \hline
       E1 & 5.88 & 14.26 & 0.18 & 105.54 \\
       E2 & 7.12 & 9.48 &  0.33 & 324.31 \\
       E3 & 5.77 & 5.16 &  0.35 & 190.97 \\
       E4 & 5.14 & 3.24 &  0.18 & 44.70 \\
       E5 & 5.18 & 5.57 &  0.14 & 160.27 \\
       E6 & 5.05 & 5.42 &  0.10 & 266.83 \\
       E7 & 7.02 & 4.32 &  0.25 & 160.58 \\
       E8 & 4.65 & 4.06 &  0.13 & 55.07 \\
       E9 & 5.31 & 3.50 & 0.45 & 74.55 \\
       E10 & 4.85 & 3.30 &  0.36 & 59.65 \\
       E11 & 5.48 &  2.76 & 0.46 & 220.96 \\
       E12 & 4.87 & 2.99 &  0.25 & 45.45 \\
       \hline
       A1 & 4.88 & 1.64 &  0.70 & 221.27  \\
       A2 & 4.48 & 2.15 &  0.50 & 51.04 \\
      \hline
    \end{tabular}
    \caption{Total stellar mass, halo mass, disc-to-total mass ratio $D/T$ (defined in eq.~(3.4) of ref.~\cite{Calore:2015oya}), and
      reduced $\chi^2$ for the fit to the rotation curve data (for $R > 2.5$~kpc) of
      the MW-like galaxies that satisfy our selection criteria (i) and (ii) in the
      \eagle HR (haloes E1 to E12) and \apostle IR (haloes A1 and A2) runs.}
    \label{tab:criteria}
  \end{table}

In computing the goodness of fit to the observed MW rotation curves from
ref.~\cite{Iocco:2015xga}, we have considered measurements for galactocentric
distances $R>2.5$~kpc, since at smaller radii the gravitational potential of the Galactic bulge can break the assumption of circular orbits for the tracers in ref.~\cite{Iocco:2015xga}. The reduced $\chi^2$ for the fit to the rotation curve data is in the range 44.70 -- 324.31
(51.04 --221.27), for the \eagle HR (\apostle IR) simulation. Since we are interested in the local DM distribution, we would
like to check how the goodness of fit changes if a smaller range of $R$ is
considered. In particular, we consider four additional ranges of $R$ for the fit: $R
> 5$~kpc, $R > 7$~kpc, $5~{\rm kpc} <R < 11$~kpc, and $7~{\rm kpc}< R < 9$~kpc. For
all these cases, the goodness of fit is only marginally better for most haloes with
$M_{*}$ within the observed MW stellar mass range in the \eagle HR and \apostle IR
simulations, and the overall range of the reduced $\chi^2$ is slightly smaller. For
the \eagle HR (\apostle IR) simulation, the reduced $\chi^2$ range is 22.28 -- 290.36
(44.6 --213.24), 17.20 -- 182.91 (41.40 --134.16), 24.29 -- 317.92 (48.92 -- 233.42),
and 23.71 -- 268.92 (59.86 -- 196.92) when taking measurements at $R > 5$~kpc, $R >
7$~kpc, $5~{\rm kpc} <R < 11$~kpc, and $7~{\rm kpc}< R < 9$~kpc, respectively. The
reason for this small reduction in the reduced $\chi^2$ values is that the measured
MW rotation curves have very small error bars at small galactocentric distances, $R
<5$~kpc, and therefore the fit gets worse when including the measurements at
$2.5~{\rm kpc}<R<5$~kpc.

As discussed in ref.~\cite{Calore:2015oya}, the halo masses of our selected simulated
galaxies are higher than the MW halo mass expected from abundance matching arguments
(e.g.~$M_{200,\rm MW} = 1.2^{+0.7}_ {-0.4} \times 10^{12} \,
\Msun$~\cite{Busha:2010sg}) probably due to the slightly too efficient feedback in
the \eagle HR simulated haloes of this mass range \cite{Schaye:2015, Crain:2015}. We have checked
that the large halo masses and the mismatch between stellar mass and halo mass do not
affect the recoil rate and other relevant quantities in direct DM detection
experiments. Note that the higher halo mass results in a higher DM escape speed from
the Galaxy, and hence a higher velocity tail of the DM velocity
distribution. However, the halo integral (see section~\ref{sec:haloint}), which
encodes the astrophysical dependence of the recoil rate, is very similar for haloes
which have the highest and smallest halo masses but with stellar masses within the
$3\sigma$ observed MW stellar mass range.

\section{Dark matter velocity distribution}
\label{sec:veldist}

The event rates in a direct detection experiment depend strongly  on the DM velocity
distribution at the position of the Sun. In this section we discuss the DM velocity
distribution in the Galactic rest frame for the \eagle HR and \apostle IR simulated
haloes which satisfy criteria (i) and (ii) as defined in
section~\ref{sec:selection}.

To find the DM velocity distribution at the Solar circle ($R_0=8$~kpc), we consider a torus with a square cross section 
which is aligned with the stellar disk, and has an inner and outer radius of 7 kpc
and 9 kpc from the Galactic centre respectively, and a height extending from -1 kpc
to 1 kpc with respect the the Galactic plane. This region contains a total of 1821
-- 3201 (2160 -- 3024) particles depending on the halo in the \eagle HR (\apostle IR)
simulation. Notice that due to the limited resolution of the simulation, we are not
sensitive to the local variation of the DM velocity distribution within the
torus. Hence we take the average velocity distribution of DM particles in the full
torus, instead of computing it at the azimuthal position of the Sun at 8 kpc in the torus. 

We define a reference frame in the plane of the galaxy to describe the velocity
vector of the simulation particles. In this reference system, the origin is at the
Galactic centre, the $z$-axis is perpendicular to the stellar disk, $r$ is in the
radial direction, and $\theta$ is in the tangential direction. In the case of the DMO simulations, we orientate the torus in the same direction as in the hydrodynamic case. We can then calculate
the velocity distribution for the vertical ($v_z$), radial ($v_r$), and azimuthal
($v_\theta$) components of the DM in addition to the velocity modulus
distribution. This distribution, as well as the three components of the velocity
distribution, are individually normalised to unity, such that $\int dv f(|\vect v|) =
1$ and $\int dv_i f(v_i) = 1$ for $i = z, r, \theta$. Note that the velocity modulus
distribution, $f(|\vect v|)$, normalised in this way is related to the velocity
distribution, $\tilde f(\vect v)$, by,
\beq
f(|\vect v|) = v^2 \int d \Omega_{\vect v} \tilde f(\vect v),
\eeq
such that $\int d^3 v \tilde f(\vect v) =1$. Here $d \Omega_{\vect v}$ is an infinitesimal solid angle around the direction $\vect v$.

In sections \ref{sec:vel-Mod} and \ref{sec:vel-comp} we present the velocity modulus
distribution, as well as the components of the velocity distribution of the selected
simulated haloes, and discuss how they can be described using fitting
functions. Notice that in the analysis of direct detection data in section \ref{sec:results}, 
we use the velocity distributions extracted  directly from the
simulations instead of the best fit functions.

\subsection{Velocity modulus distribution}
\label{sec:vel-Mod}

\begin{figure}[t!]
\begin{center}
  \includegraphics[width=0.49\textwidth]{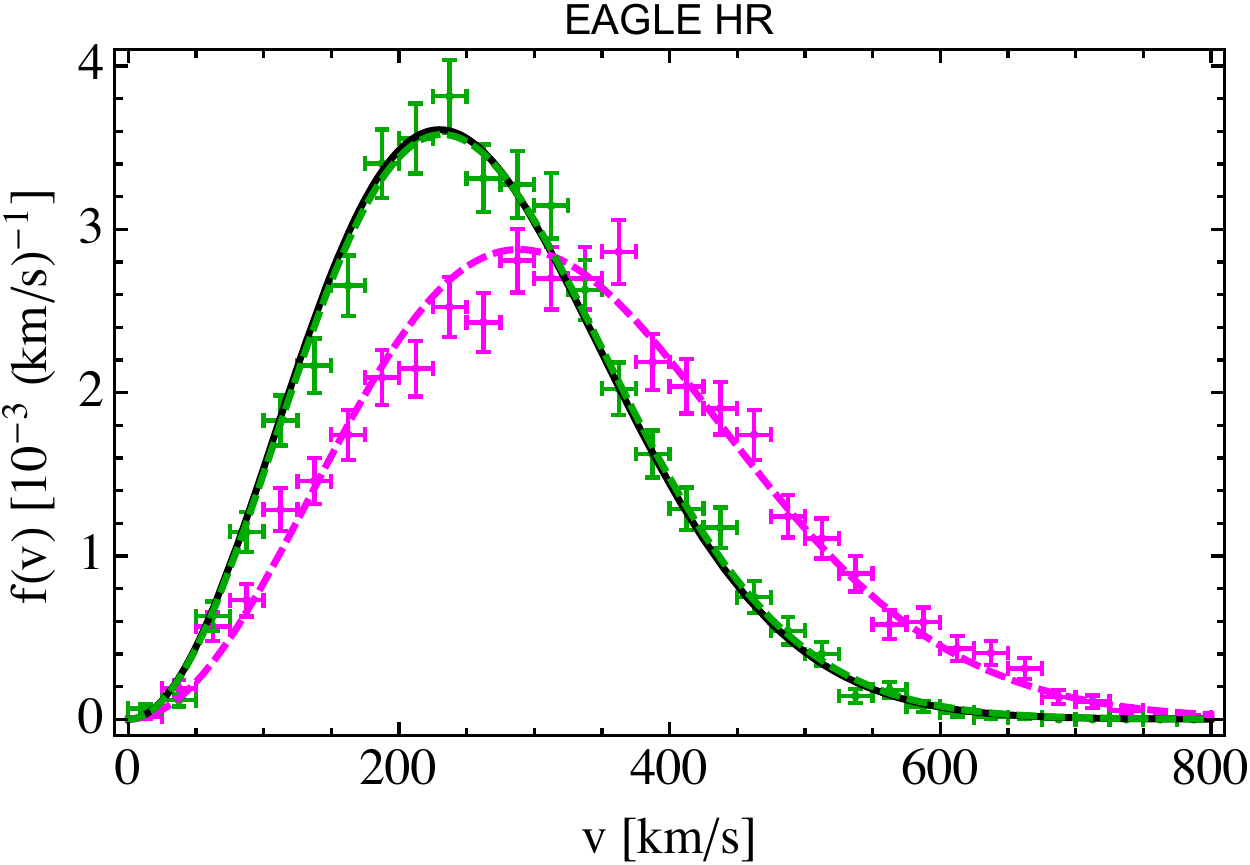}
  \includegraphics[width=0.49\textwidth]{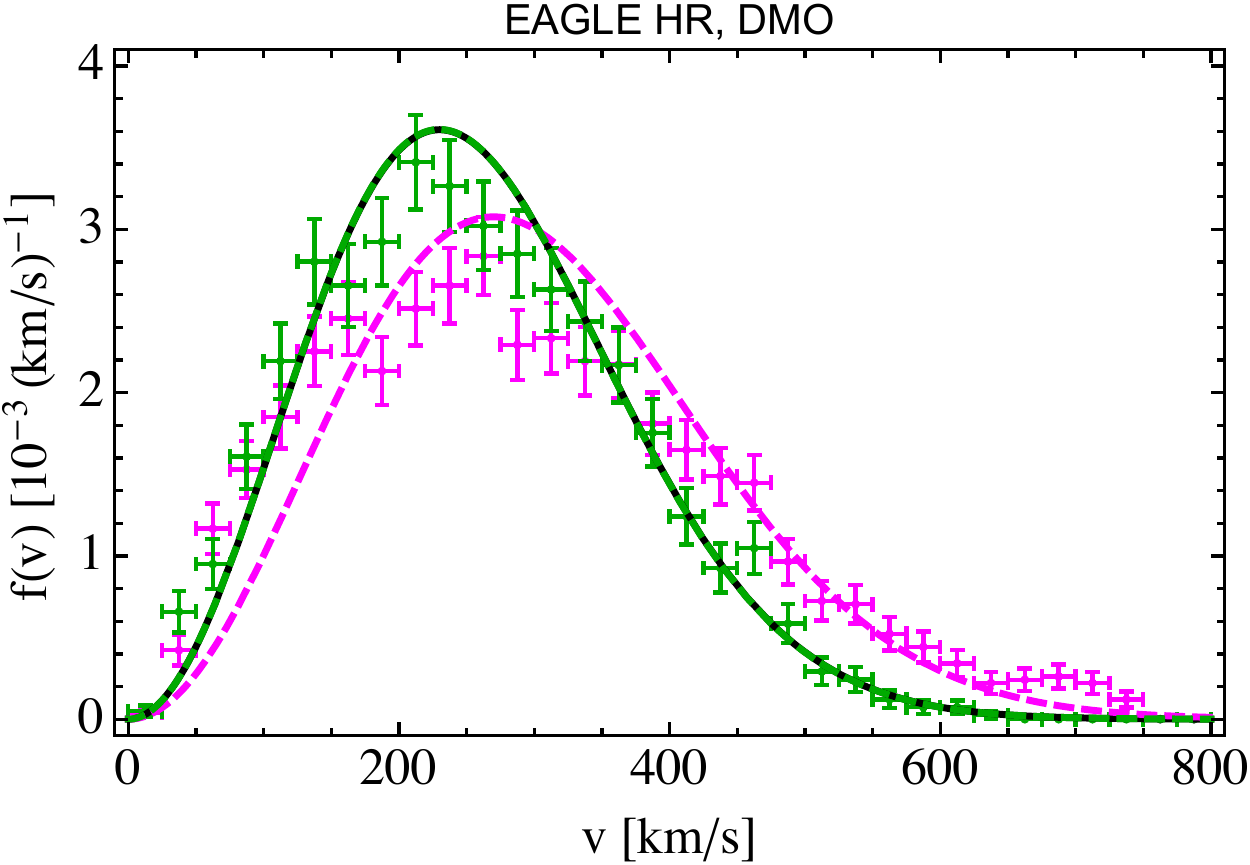}\\
   \vspace{5pt}\includegraphics[width=0.49\textwidth]{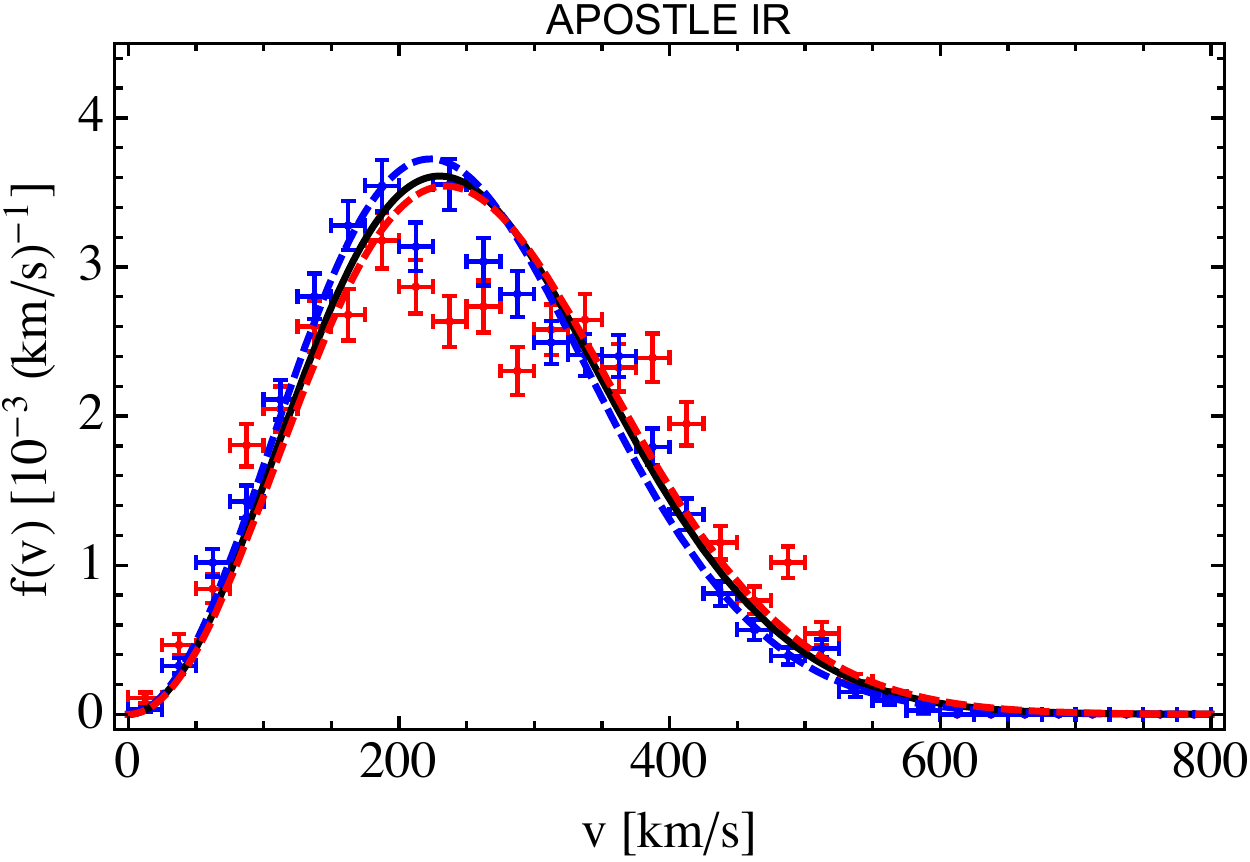}
  \includegraphics[width=0.49\textwidth]{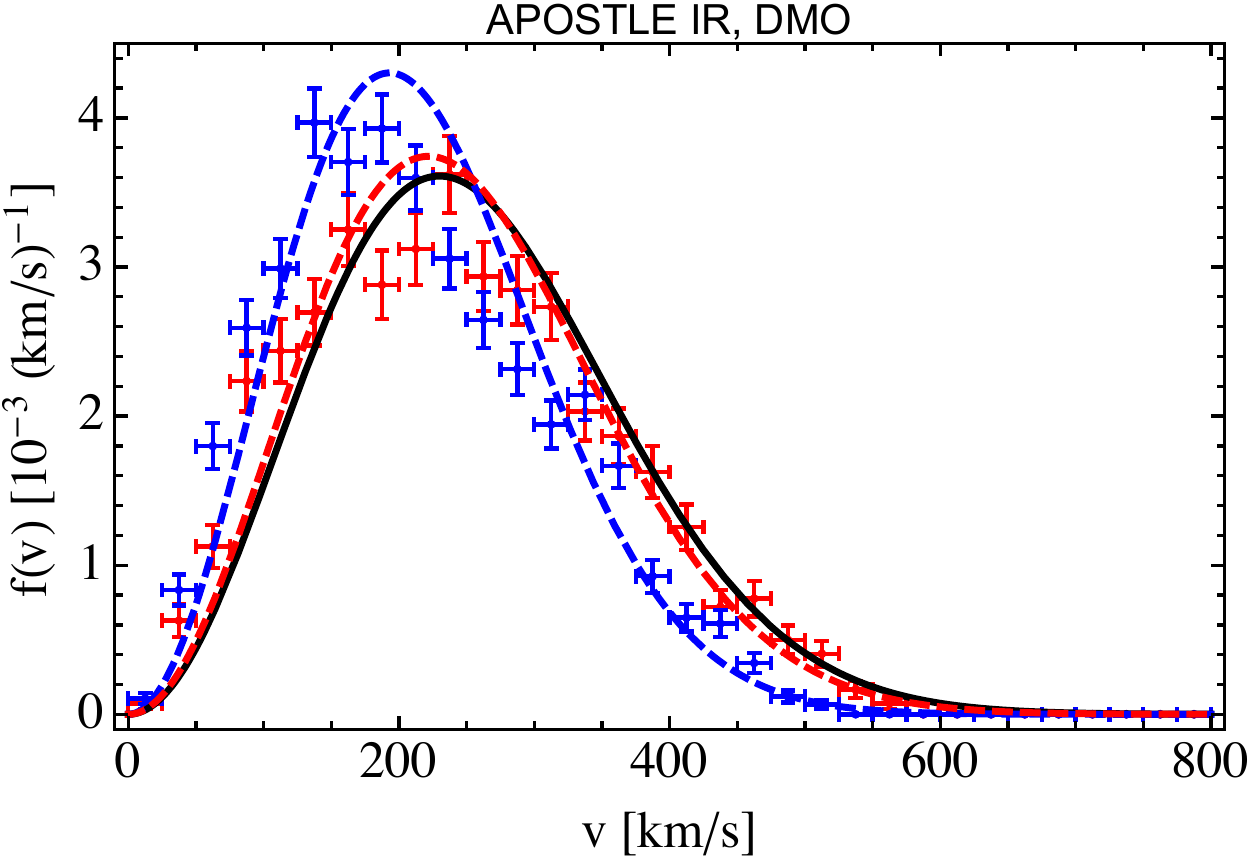}
\caption{\label{fig:fv} DM velocity modulus distributions in the Galactic rest frame
  (coloured data points with 1$\sigma$ error bars) for two haloes in the \eagle HR
  simulation which satisfy our selection criteria and have the speed distributions
  closest to (halo E12, shown in green) and farthest from (halo E3, shown in magenta) the SHM Maxwellian (top left), and two
  haloes in the \apostle IR simulation satisfying our selection criteria (bottom
  left). The right panels show the velocity modulus distributions for the same haloes
  shown in the left panels but in a DMO simulation. The black solid line shows the
  SHM Maxwellian speed distribution (with peak speed of 230 km$/$s), and the coloured
  dashed lines show the best fit Maxwellian distribution for each halo (with matching
  colours).}
\end{center}
\end{figure}

In the left panels of figure~\ref{fig:fv}, we show the local DM velocity modulus
distribution in the Galactic rest frame for two haloes in the \eagle HR (top panel)
and \apostle IR (bottom panel) simulations that satisfy our selection criteria. From
the 12 MW-like \eagle HR haloes, in the top panel we only show the velocity modulus
distribution for the two haloes that have speed distributions closest to (halo E12; reduced
$\chi^2$ of 1.43) and farthest from (halo E3; reduced $\chi^2$ of 23.4) the SHM Maxwellian
distribution (with peak speed of 230 km$/$s). The vertical error bars specify the
$1\sigma$ Poisson error on the data points, and the horizontal error bars specify the
speed bin size. The Maxwellian speed distribution (with peak speed of 230 km$/$s) is
shown as solid black lines, and the best fit Maxwellian speed distributions are also
shown as dashed coloured curves. 

One potential issue with the vertical error bars on the data points is that they might be correlated. Therefore, we have checked that bootstrap-resampling the distributions leads to similar error estimators compared to simple Poisson statistics. Hence, we will use the $1\sigma$ Poisson error on the data points for our analysis in the rest of the paper. Notice also that we have considered different speed bin sizes and found the optimal size to be 25 km/s. A smaller bin size would
increase the noise in the data points due to the small number of particles in each
bin, while a larger bin size would smear out features in the speed distribution.

In the right panels of figure~\ref{fig:fv} we show the velocity modulus distributions
for the same haloes shown in the left panels but in the corresponding DMO simulations.  Comparing the
left and right panels of the figure, it is clear that baryons have an important
effect on the DM speed distribution at the Solar radius. Including baryons in the
simulation leads to a shift of the peak of the local DM speed distribution to higher
speeds. This happens because baryons are dissipative and hence deepen the gravitational potential of the
Galaxy in the inner regions, resulting in more high velocity particles, thus
shifting the peak of the DM speed distribution to larger values.

For comparison, in Appendix~\ref{app:f(v)} we show the velocity modulus distributions
for our selected haloes in the \eagle HR simulation that have the smallest (halo E6) and
largest (halo E4) local DM density. It is clear from figure \ref{fig:fvrho} that the speed distributions for the two haloes are similar, as expected, since the local DM density does not affect the DM speed distributions.

\begin{figure}[h!]
\begin{center}
  \includegraphics[width=0.45\textwidth]{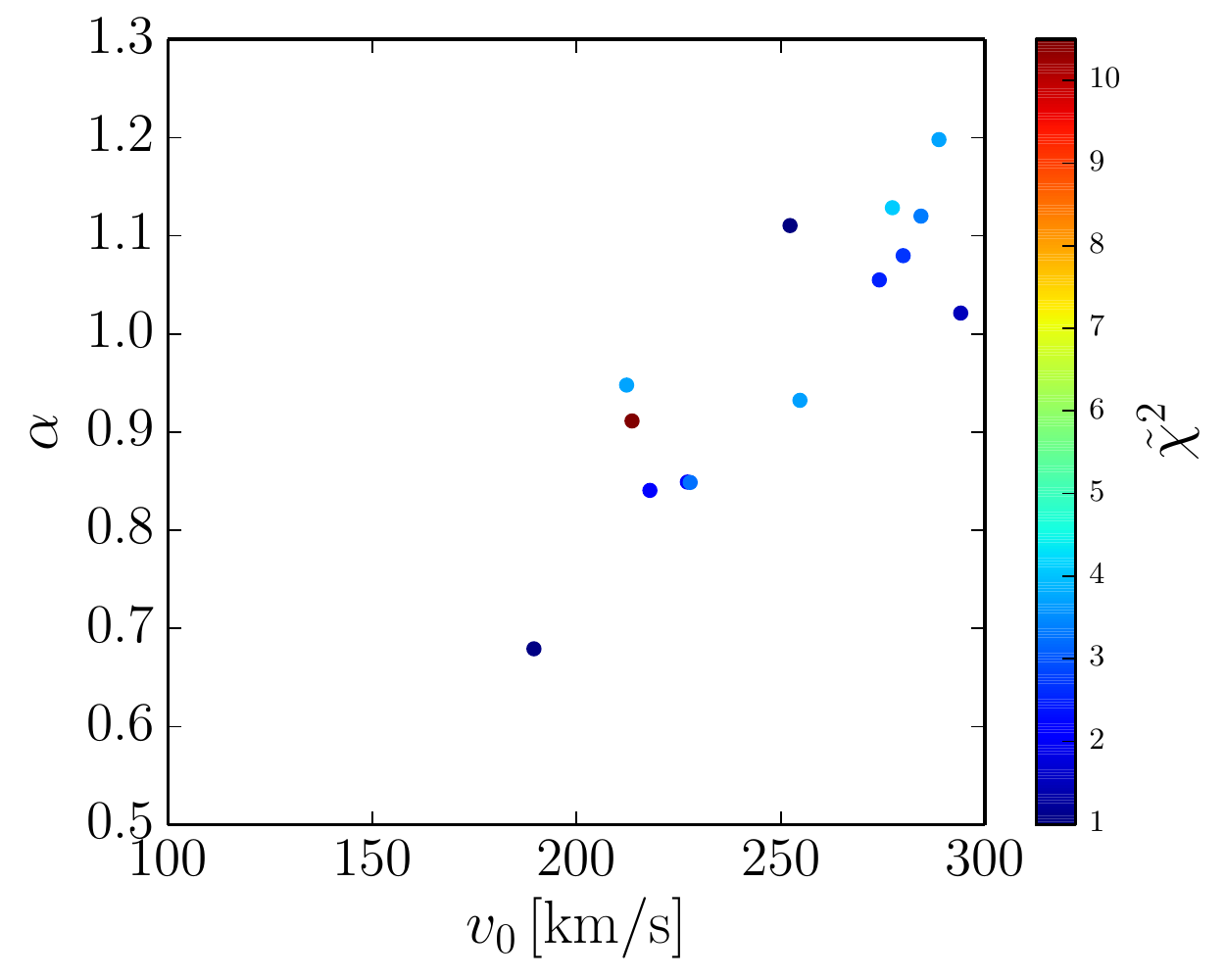}
  \includegraphics[width=0.45\textwidth]{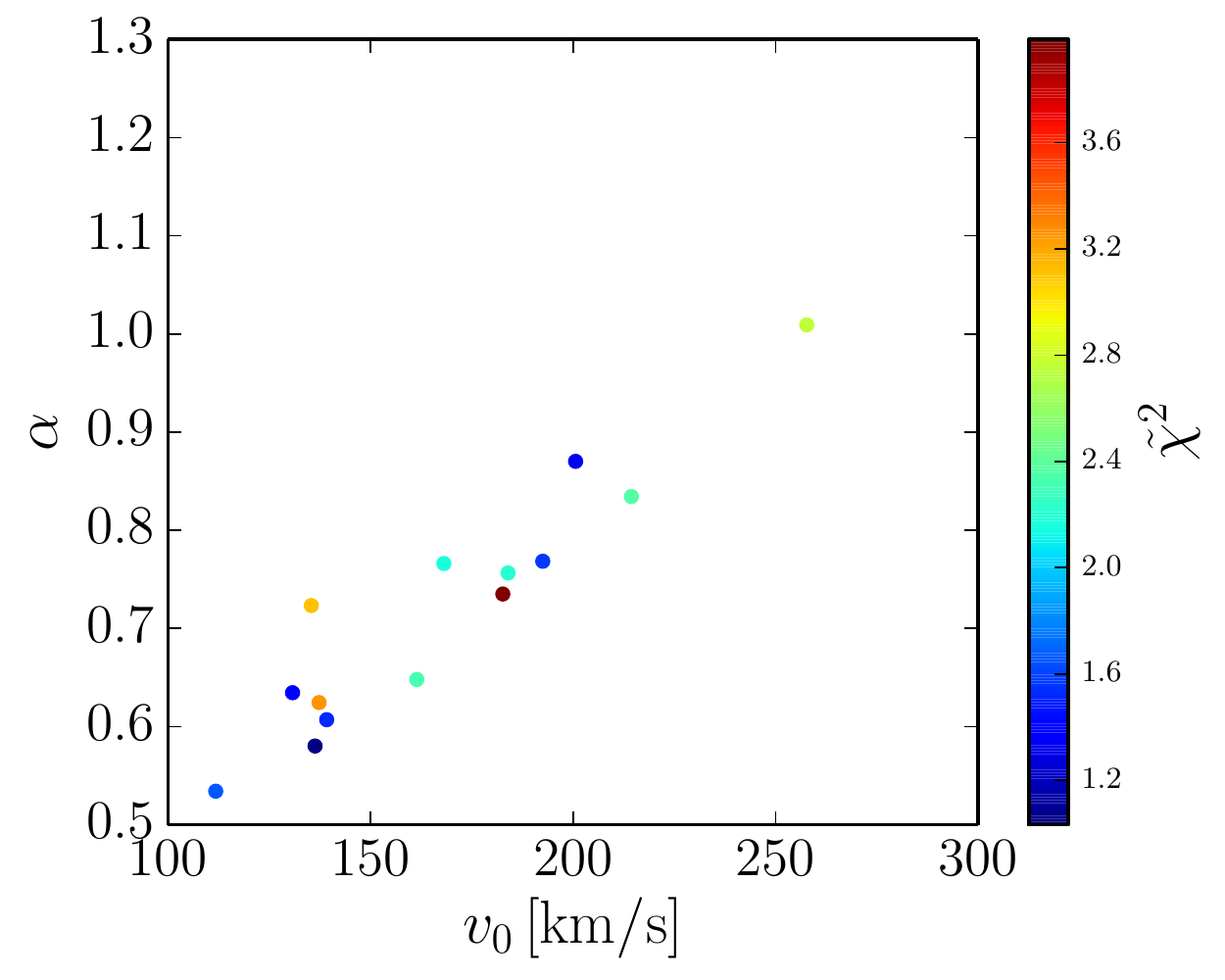}\\
    \vspace{2pt} \includegraphics[width=0.45\textwidth]{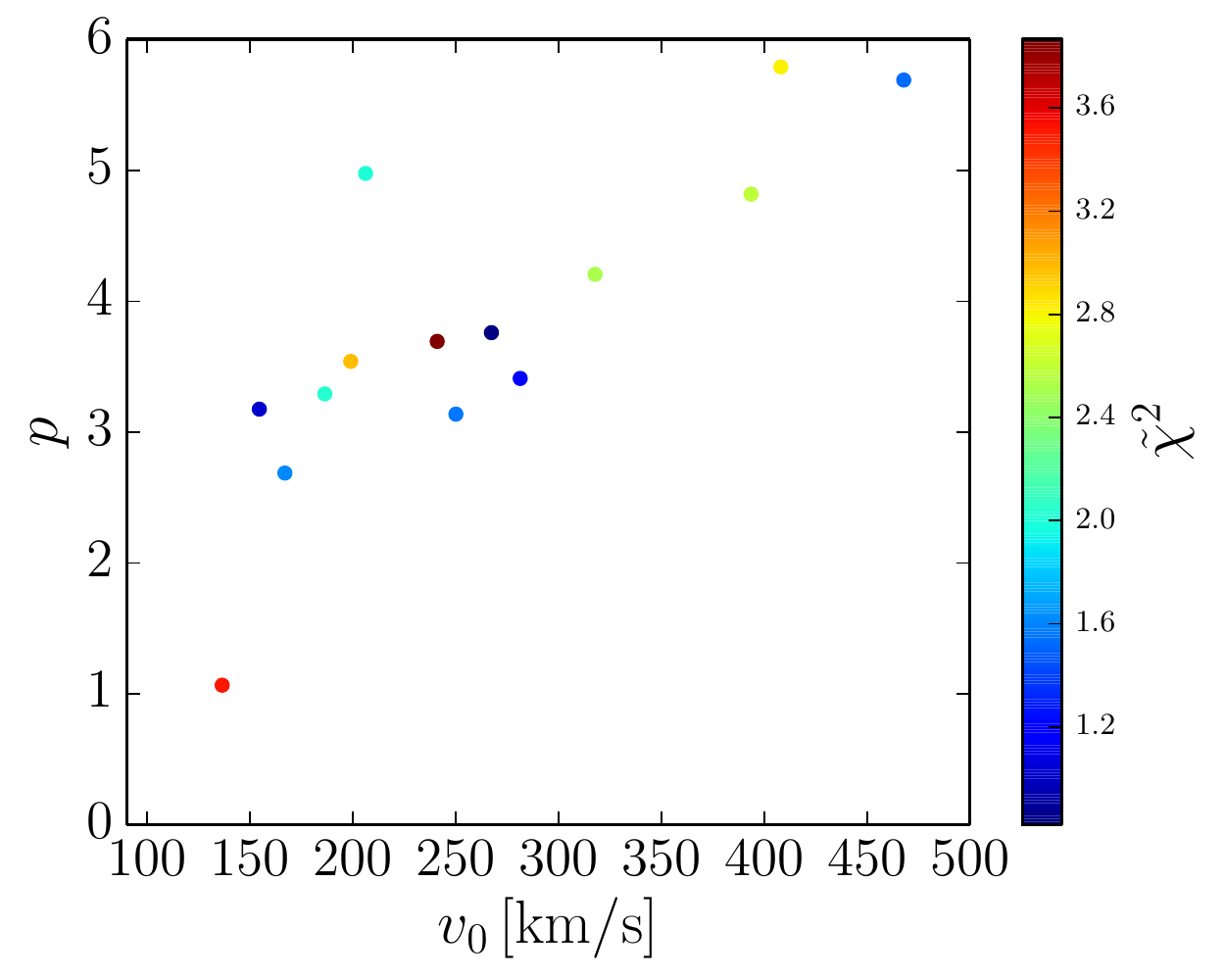}
  \includegraphics[width=0.45\textwidth]{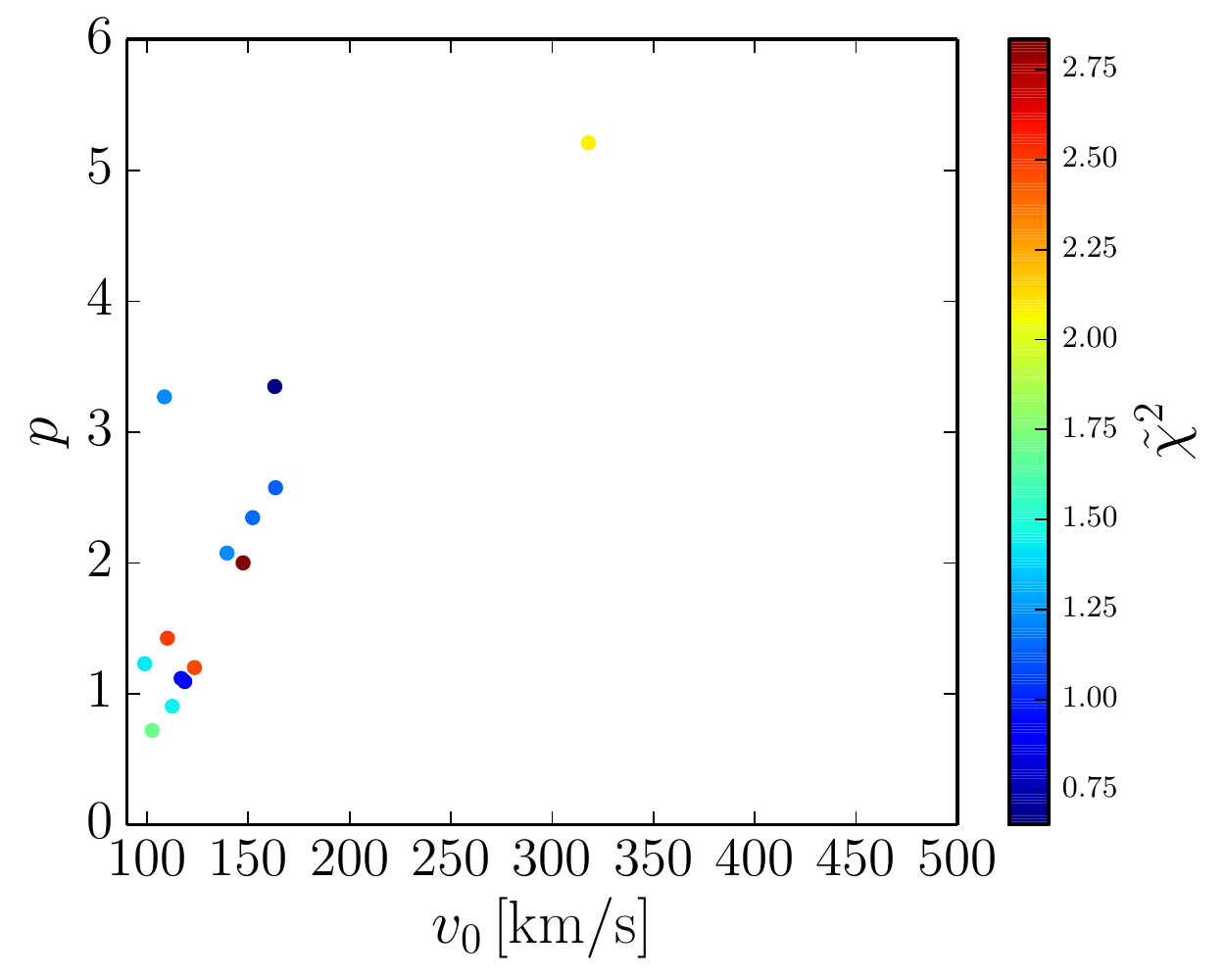}\\
  \vspace{2pt}  \includegraphics[width=0.45\textwidth]{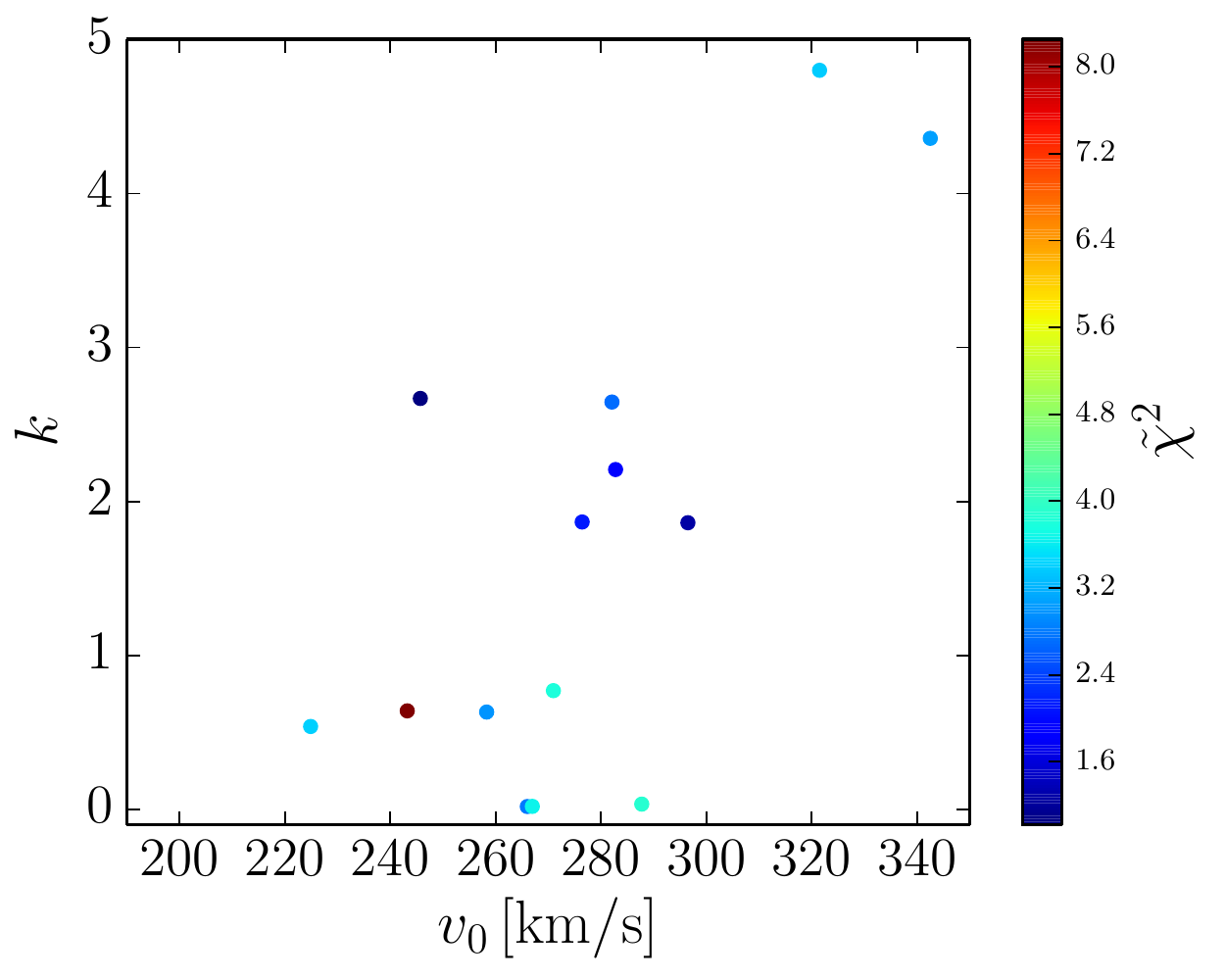}
  \includegraphics[width=0.45\textwidth]{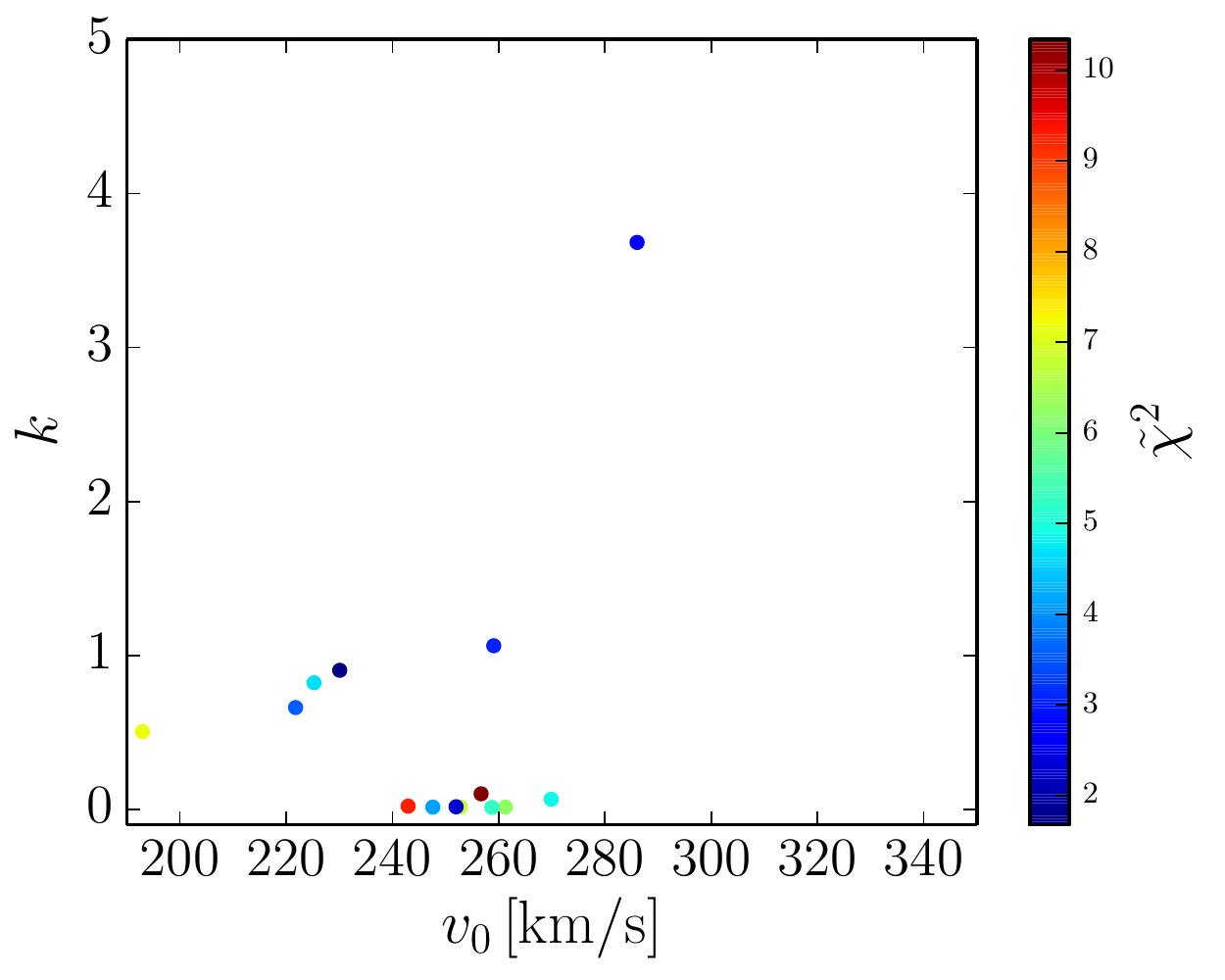}
\caption{\label{fig:scatter} Left panels: Scatter plots showing the results of the
  fits to the simulated DM velocity modulus distribution for the set of 12 (\eagle
  HR) + 2 (\apostle IR) selected MW-like galaxies when adopting, as fitting function,
  a generalized Maxwellian distribution (top panels), and the functions proposed in
  Mao {\it et al.}~\cite{Mao:2012hf} (centre panels) and in Lisanti {\it et
    al.}~\cite{Lisanti:2010qx} (bottom panels). The colour-bar corresponds to the
  value of the reduced $\chi^2$, $\tilde{\chi}^2 \equiv \chi^2/(N - dof)$ (where $dof
  = 2$ for all three cases). Right panels: Same as left panels but for the
  corresponding haloes in the DMO simulations.}
\end{center}
\end{figure}

Note that the local Galactic escape speeds for some haloes in the \eagle HR
simulation are higher than the value of the MW escape speed at the Solar position
found recently by the RAVE survey, $v_{\rm
  esc}=533^{+54}_{-41}~$~km$/$s~\cite{Piffl:2013mla}. Although the RAVE value of the escape speed is actually a lower limit on the true MW escape speed, it is the commonly adopted value. Also, the larger $v_{\rm esc}$ values of the haloes in the \eagle HR simulation are due to the larger
$M_{200}$ of those haloes compared to the MW. However, as discussed in section~\ref{sec:selection}, this
does not affect the predicted signals in direct detection experiments. The local
Galactic escape speeds are in the range of 720 -- 1083 km$/$s (617 -- 646 km$/$s) for
the selected \eagle HR (\apostle IR) haloes. These escape speeds are computed for each simulated halo from the total mass enclosed in a sphere of radius 7 kpc, which is the inner radius of our defined
torus. Therefore, these escape speeds represent an upper limit on those expected at
the Solar circle.

\bigskip

We now discuss how well the DM velocity modulus distributions of the simulated MW
analogues can be fitted with various fitting functions that have been proposed in the
past for the DM velocity distribution. We adopt the following parameterizations of
the DM velocity modulus distribution:

\begin{itemize}

\item Generalized Maxwellian distribution:
\beq \label{eq:genMax}
f(|\vect v|) \propto  |\vect v|^2  \exp[ -( |\vect v| / v_0 )^{2 \alpha} ] \, , 
\eeq
with free parameters $v_0$ and $\alpha$. The case of a standard Maxwellian
distribution is represented by $\alpha = 1$. We also test the goodness of fit for a
standard Maxwellian to the simulated velocity modulus distributions (see
tables~\ref{tab:fvfit} and~\ref{tab:fvfitDMO} in Appendix~\ref{app:fits}).

\item The velocity modulus distribution proposed by Mao {\it et al.}~\cite{Mao:2012hf}:
\beq \label{eq:Mao}
f(|\vect v|) \propto |\vect v|^2  \exp[ - |\vect v| / v_0 ]  ( v_{\rm esc}^2 -  |\vect v|^2)^p ~\Theta (v_{\rm esc} -  |\vect v| )\, , 
\eeq
with free parameters $v_0$ and $p$.

\item The velocity modulus distribution proposed by Lisanti {\it et al.}~\cite{Lisanti:2010qx}:
\beq \label{eq:Lisanti}
f(|\vect v|) \propto |\vect v|^2  \exp[ ( v_{\rm esc}^2 -  |\vect v|^2) / (k \, v_0^2) - 1 ]^k  ~\Theta (v_{\rm esc} -  |\vect v| )\, ,
\eeq
with free parameters $v_0$ and $k$.

\end{itemize}

The normalisation of all the fitting functions is such that the 1D integral in
velocity space of $f(|\vect v|)$ is equal to 1 (integrated from 0 to $v_{\rm esc}$ ).

We perform the fit to the velocity modulus distributions of the simulated haloes with
the above fitting functions, by minimising the $\chi^2$ function:
\beq
\chi^2(\vect p) \equiv \sum_i^N \frac{(y_i - f(v_i, \vect p))^2}{\sigma_i^2} \, , 
\eeq
where $y_i$ is value of the velocity modulus distribution at the velocity $v_i$,
$\sigma_i$ is the corresponding 1$\sigma$ Poisson error, $N$ is the number of data
points in the speed distribution (determined by the number of bins in speed which is 24 -- 32 
depending on the halo in the hydrodynamic simulations), and $\vect p$ are the free parameters determined by the minimisation
procedure.  The local Galactic escape speed, $v_{\rm esc}$, is a fixed parameter and as explained
above it is determined for each simulated halo from the total mass enclosed in a
sphere of radius 7 kpc, thus representing an upper limit on the escape speed at the Solar circle.

As can be seen from tables \ref{tab:fvfit} and \ref{tab:fvfitDMO} in Appendix~\ref{app:fits}, the Maxwellian
distribution with a free peak ($\alpha=1$ in eq.~\eqref{eq:genMax}) provides a better fit to most haloes in the hydrodynamic simulations compared to their DMO counterparts.
However,  the other fitting functions which have one extra free parameter in general provide a slightly better fit to haloes in both the hydrodynamic and DMO simulations.  In particular, the Mao {\it et al.} function~\cite{Mao:2012hf}
provides the best fit for almost all haloes (except a few) for both the hydrodynamic and DMO cases.

In figure~\ref{fig:scatter}, we show the distribution of the best-fit parameters and
compare the goodness of fit of the three fitting functions presented in
eqs.~\eqref{eq:genMax}, \eqref{eq:Mao}, and \eqref{eq:Lisanti} for both
hydrodynamic (left panels) and DMO simulations (right panels). In each panel the
results are shown for the 12 \eagle HR and 2 \apostle IR MW-like haloes. The function
proposed by Mao {\it et al.}~\cite{Mao:2012hf} (centre panels) fits both the
hydrodynamic and DMO simulated haloes slightly better than the other fitting
functions. There is no clear distinction in the goodness
of fit values for DMO and hydrodynamic haloes. However, lower values for both the
parameters $p$ and $v_0$ are preferred by the DMO haloes, while the fit to the
hydrodynamic counterparts spans a wider range in the parameter space.

The generalized Maxwellian distribution (top panels) provides a better fit to the DMO
simulated haloes compared to the hydrodynamic haloes.
This is due to the fact that the local DM speed distributions of the DMO haloes are in general more symmetric around the peak than those obtained from the hydrodynamic haloes. Hence a generalized Maxwellian distribution can parametrize the  speed distribution of the DMO haloes rather well. However, the DM speed distributions of the hydrodynamic haloes have more asymmetry in the high speed tail, and this is better described by the Mao {\it et al.}~\cite{Mao:2012hf} fitting function. While DMO haloes prefer
low $\alpha$ values (0.5 -- 1) and low $v_0$ (100 -- 220 km/s), most of the
hydrodynamic haloes instead have larger best-fit parameters, $\alpha$ in the range
0.8 -- 1.2 and $v_0 >$ 190 km/s.
Analogous results are obtained when using the function proposed by Lisanti {\it et
  al.}~\cite{Lisanti:2010qx} (bottom panels).  For completeness, we list the best
fit parameters and reduced $\chi^2$ values for all fitting functions  in
Appendix~\ref{app:fits}.

As we will discuss in section \ref{sec:haloint}, the relevant quantity for predicting the event rates 
in direct detection experiments is the so-called ``halo integral", which is the DM velocity distribution in the detector 
rest frame divided by the velocity and integrated over the DM velocity range relevant for direct detection. 
Since the halo integral is an integrated quantity, the local features and small fluctuations in the DM velocity distribution 
do not affect the halo integral strongly. However, large fluctuations and the position of the peak of the velocity modulus distribution are especially relevant 
for the halo integral. In particular, a larger peak speed of the velocity modulus distribution translates into a higher velocity tail of the halo integral.
In section \ref{sec:haloint}, we will discuss how the halo integral for the best fit Maxwellian velocity distribution for each halo compares to the halo integral 
computed directly from the simulation for our selected haloes.

\subsection{Velocity distribution components}
\label{sec:vel-comp}

\begin{figure}[h]
\begin{center}
  \includegraphics[width=0.49\textwidth]{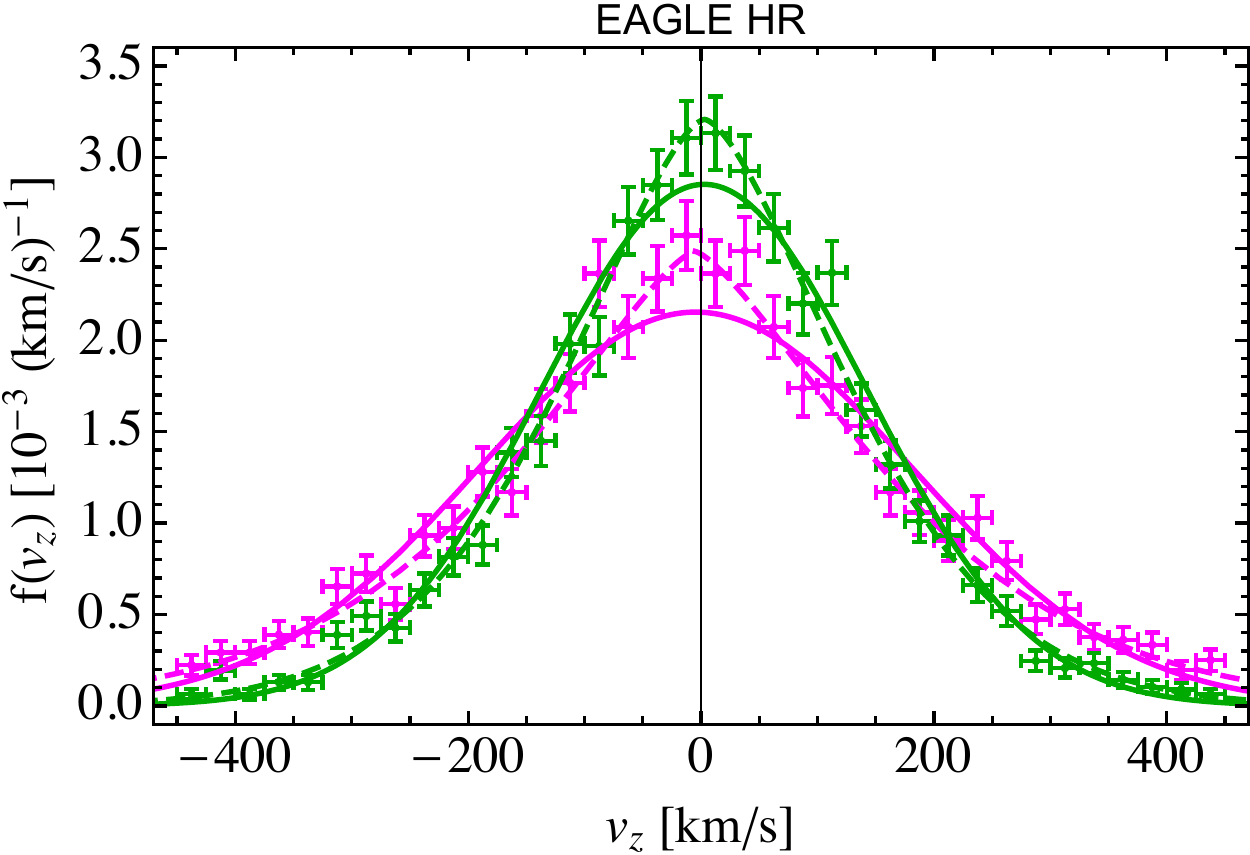}
  \includegraphics[width=0.49\textwidth]{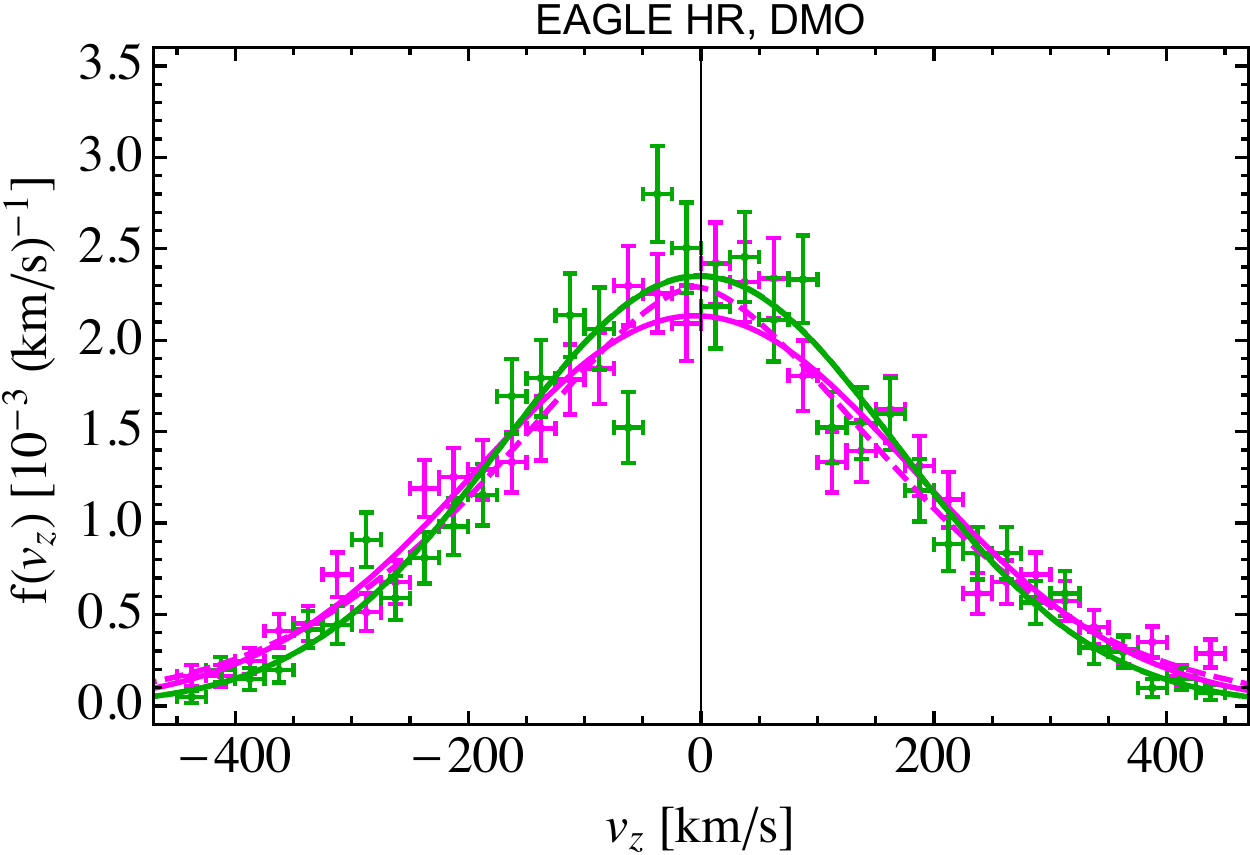}\\
    \vspace{5pt} \includegraphics[width=0.49\textwidth]{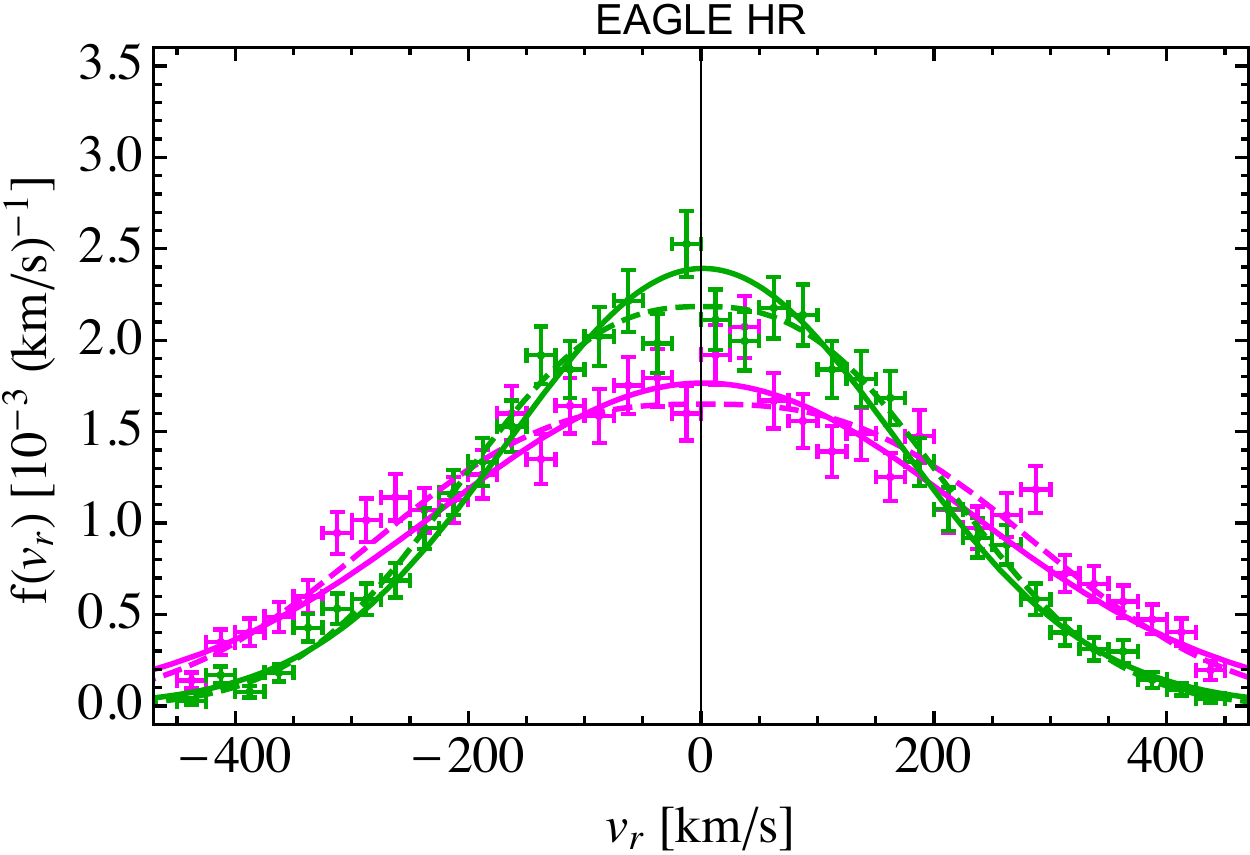}
  \includegraphics[width=0.49\textwidth]{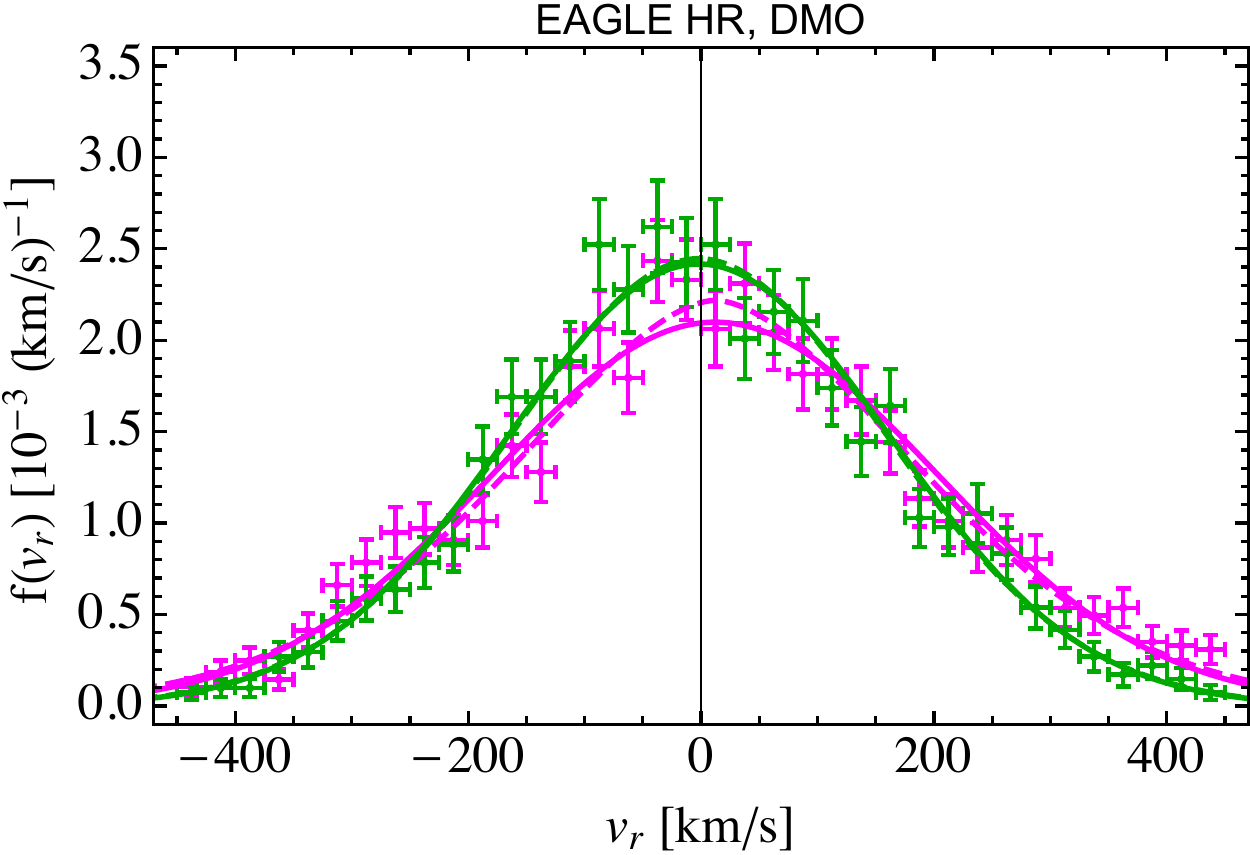}\\
  \vspace{5pt}  \includegraphics[width=0.49\textwidth]{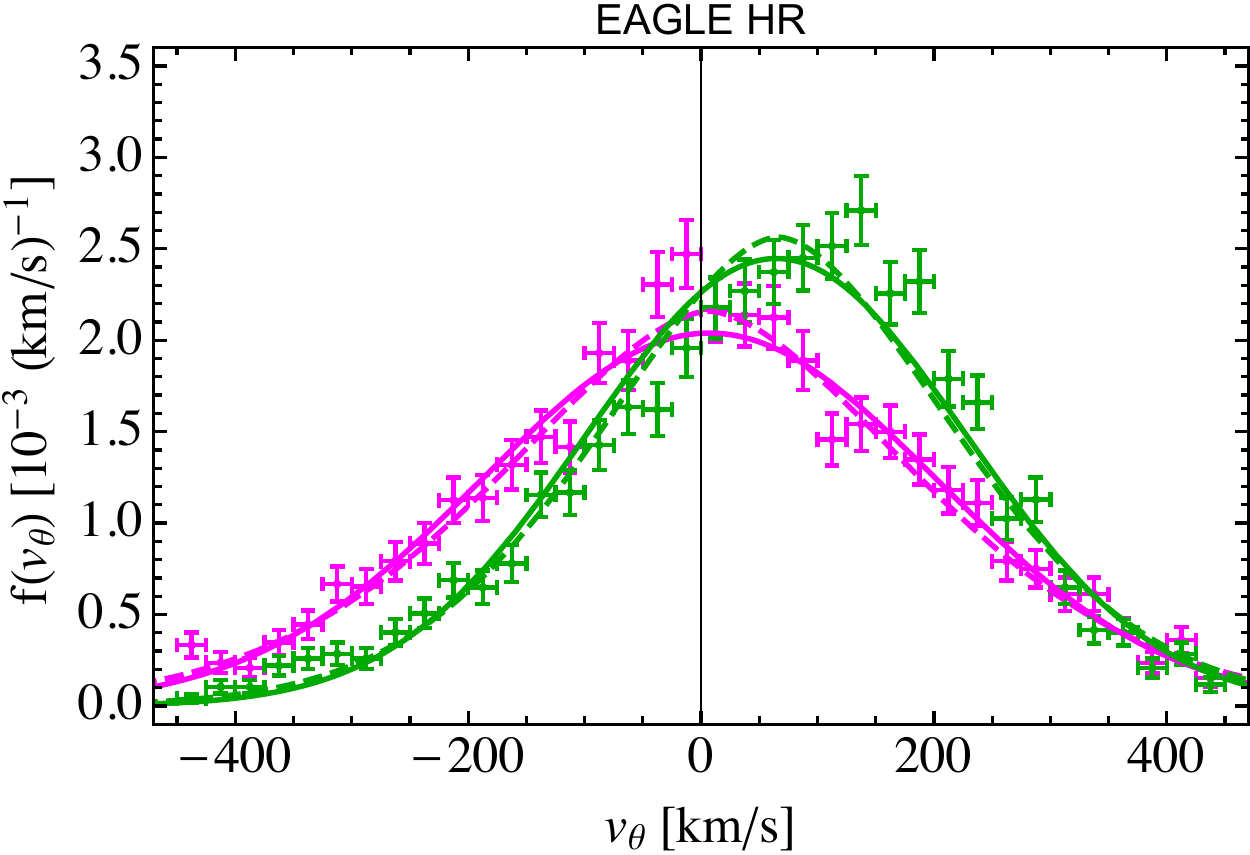}
  \includegraphics[width=0.49\textwidth]{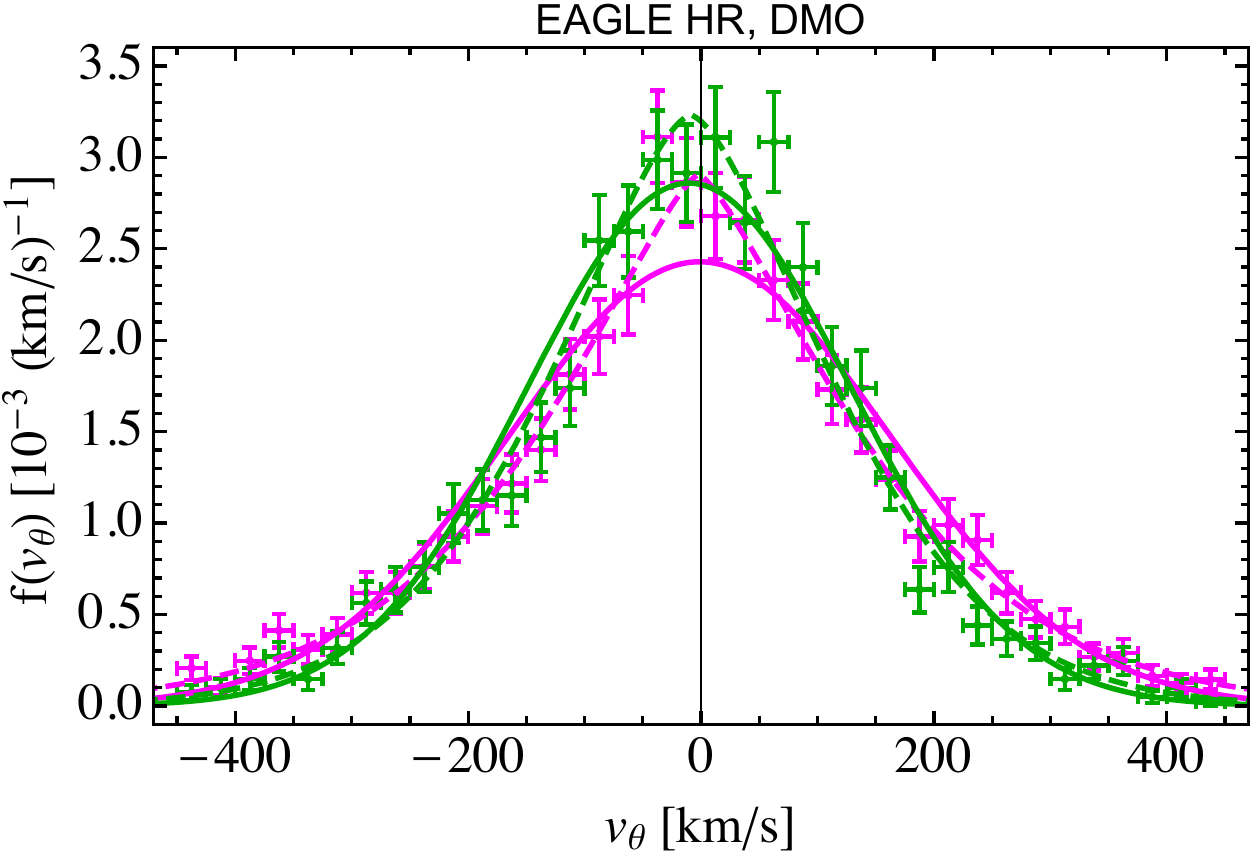}
\caption{\label{fig:fv-comp-EAGLE-HR} Left panels: The vertical (top panel), radial
  (middle panel), and azimuthal (bottom panel) components of the local DM velocity
  distribution in the Galactic rest frame (data points with 1$\sigma$ error bars) for
  the two MW-like haloes in the \eagle HR simulation which have the closest (halo E12, shown in green)
  and farthest (halo E3, shown in magenta) velocity modulus distribution compared to the SHM
  Maxwellian. The best fit Gaussian and generalized Gaussian distributions are also
  plotted as solid and dashed coloured lines, respectively. Right panels: Same as
  left panels but for the corresponding DMO simulation.}
\end{center}
\end{figure}

\begin{figure}
\begin{center}
  \includegraphics[width=0.49\textwidth]{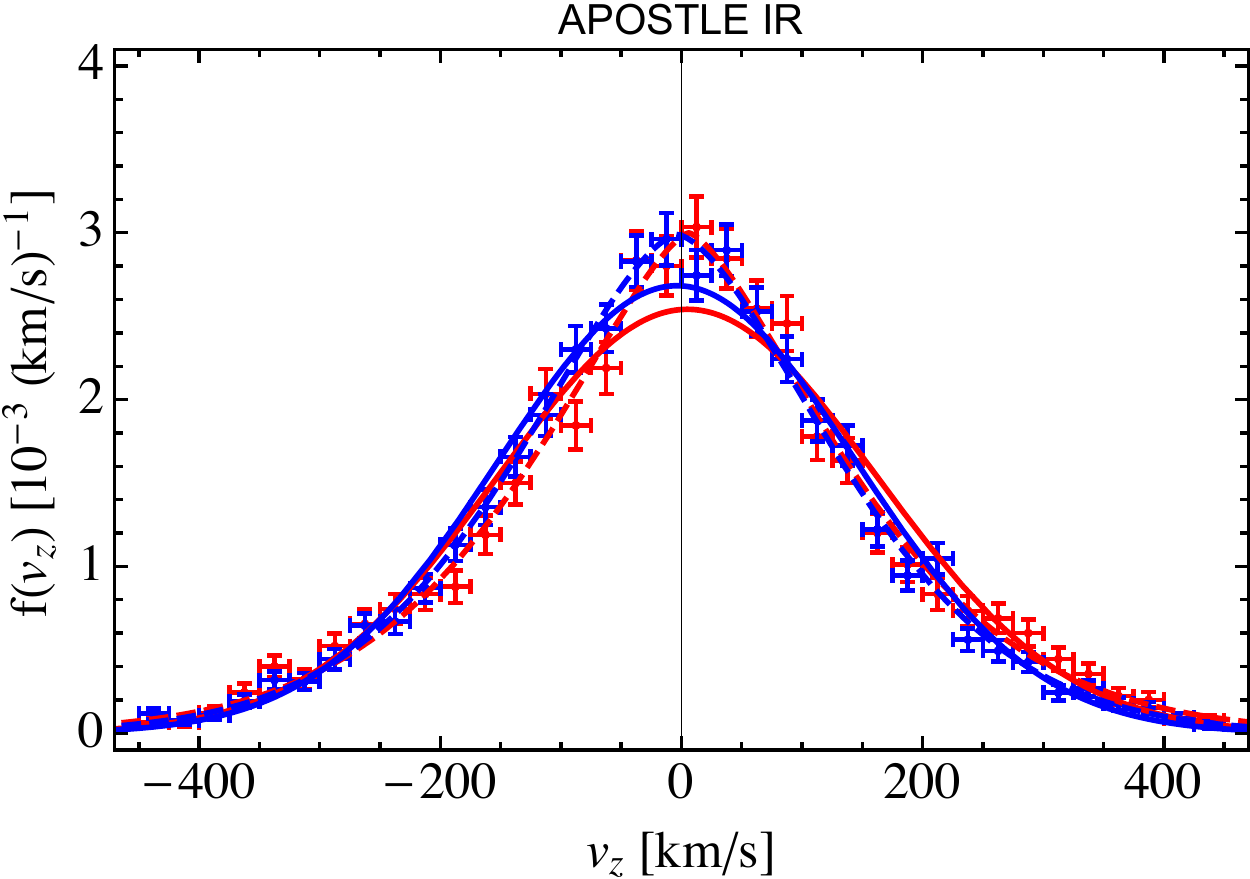}
  \includegraphics[width=0.49\textwidth]{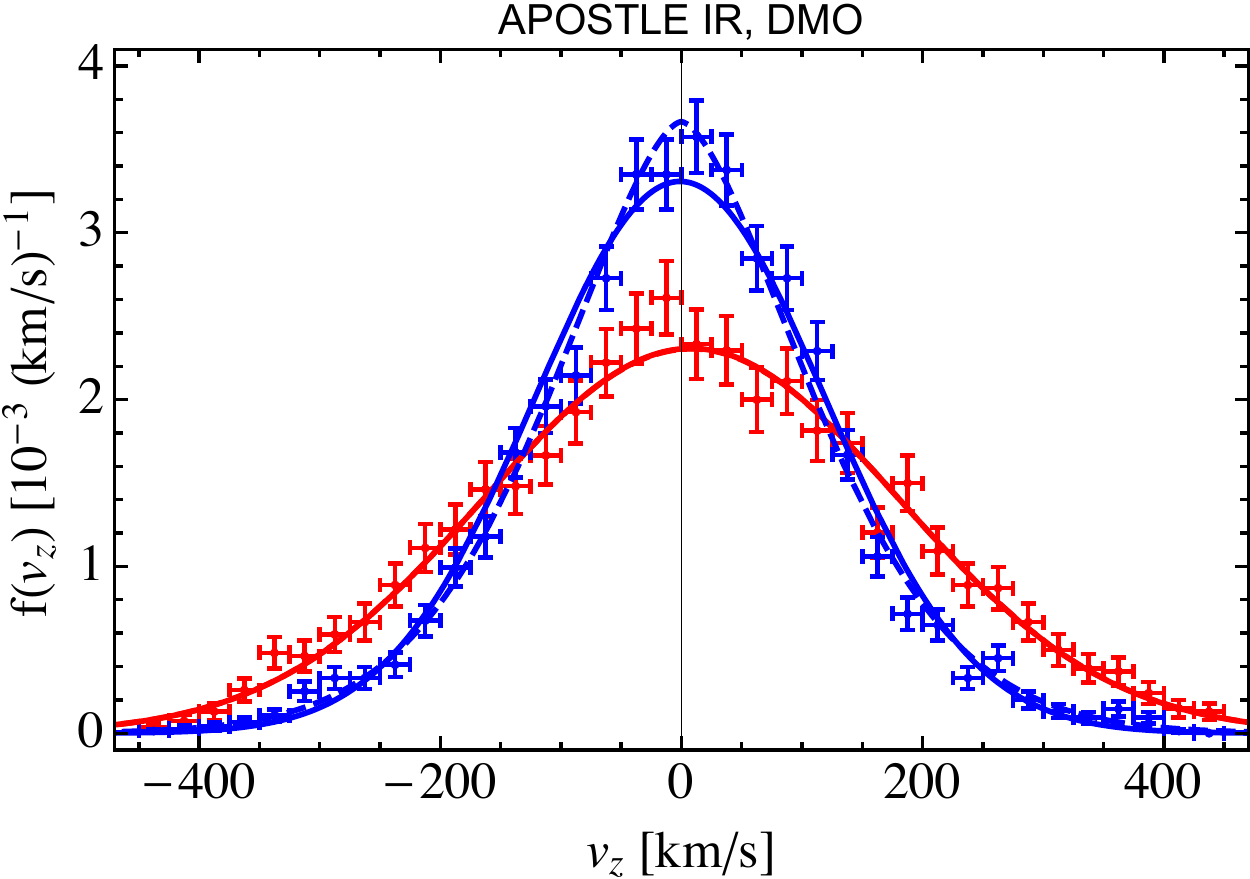}\\
    \vspace{5pt} \includegraphics[width=0.49\textwidth]{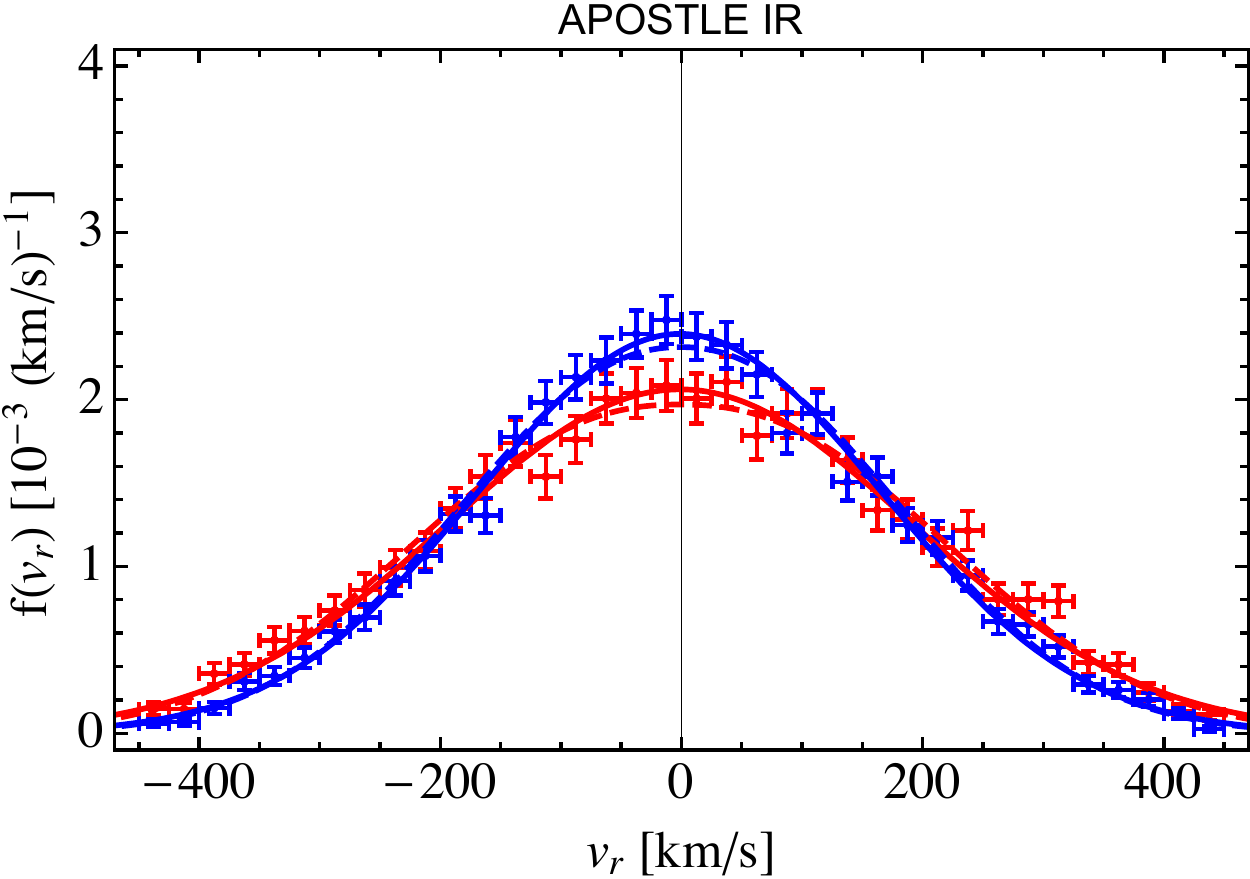}
  \includegraphics[width=0.49\textwidth]{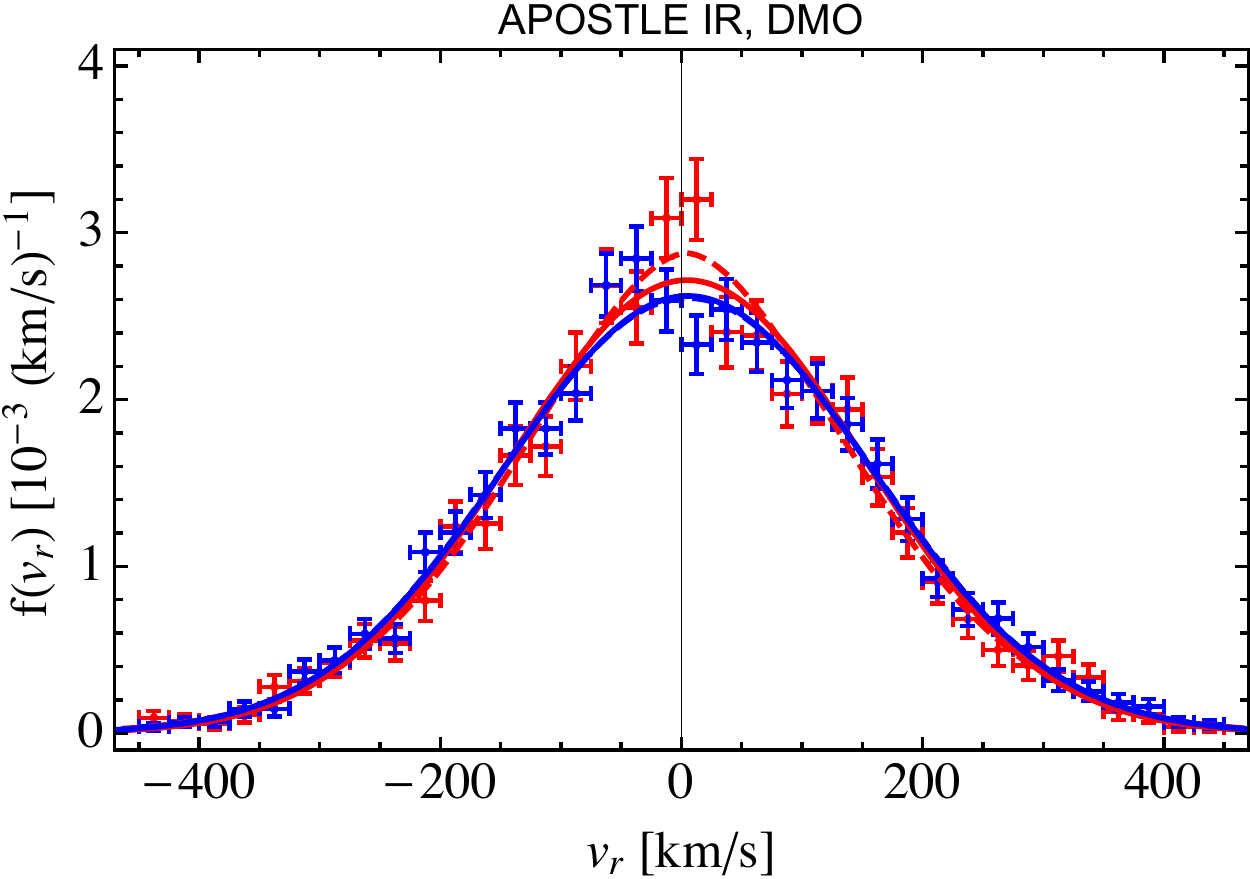}\\
  \vspace{5pt}  \includegraphics[width=0.49\textwidth]{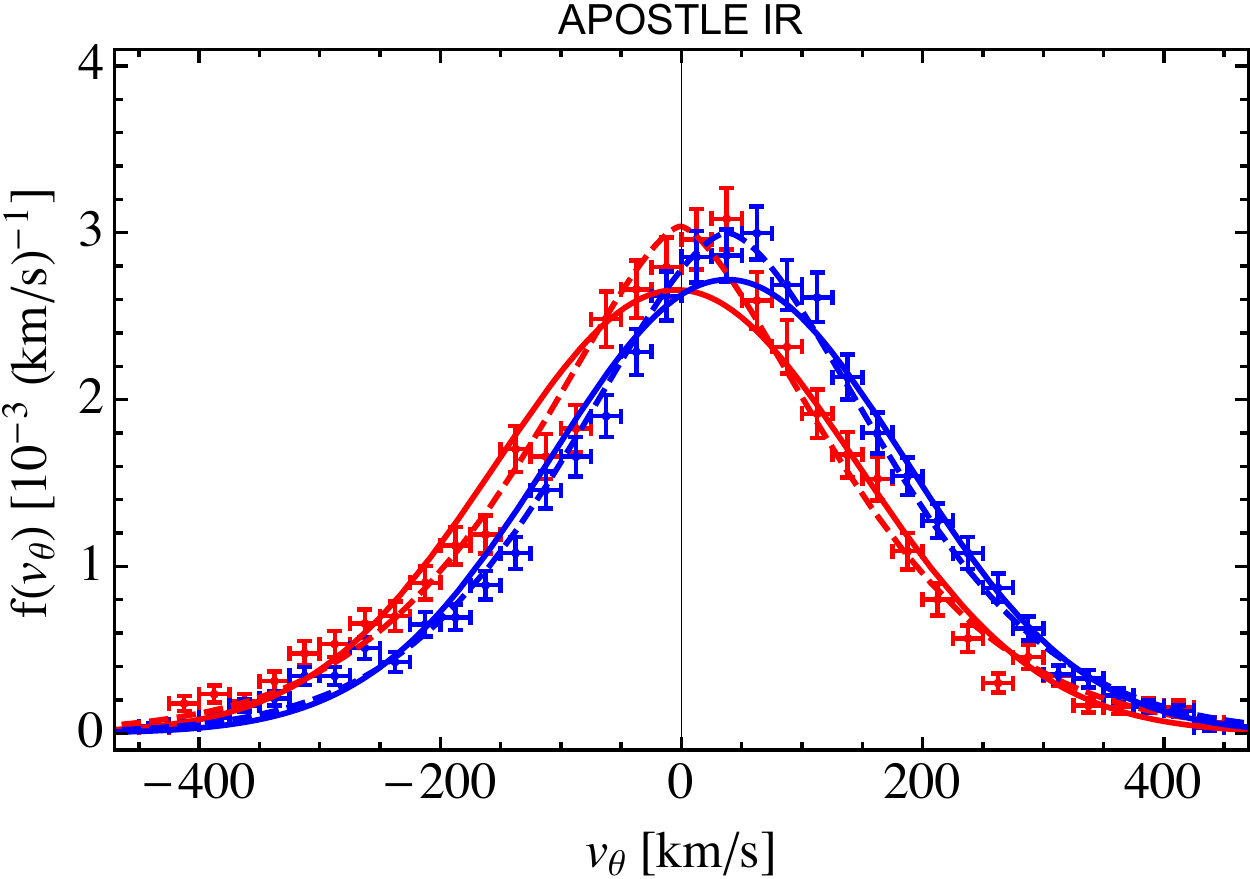}
  \includegraphics[width=0.49\textwidth]{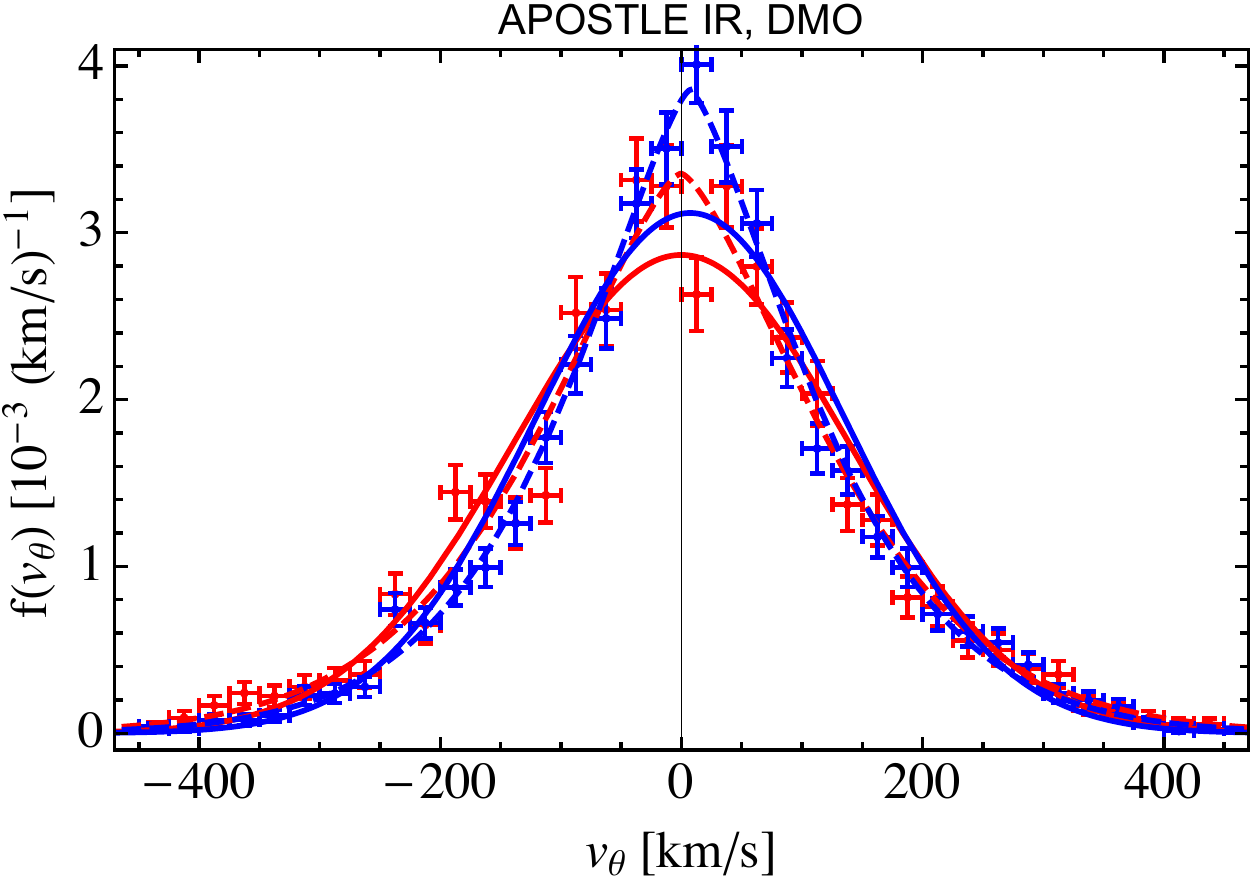}
\caption{\label{fig:fv-comp-APOSTLE-IR} Same as figure~\ref{fig:fv-comp-EAGLE-HR} but
  for haloes A1 and A2 in the \apostle IR simulation.}
\end{center}
\end{figure}

We next show the vertical, radial, and azimuthal components of the DM velocity
distribution in the Galactic rest frame.  Figures~\ref{fig:fv-comp-EAGLE-HR} and
\ref{fig:fv-comp-APOSTLE-IR}  show the DM velocity distribution components for the
same two haloes shown in figure~\ref{fig:fv} in the \eagle HR (haloes E3 and E12) and \apostle IR (haloes A1 and A2)
hydrodynamic (left panels) and DMO (right panels) simulations, respectively.

It is clear from the three panels of figures~\ref{fig:fv-comp-EAGLE-HR} and
\ref{fig:fv-comp-APOSTLE-IR} that the three components of the DM velocity distribution
are not similar, and that there is a clear velocity anisotropy at the Solar
circle. In particular, the radial speed distribution is broader than the vertical and tangential distributions in the hydrodynamic case. Another prominent feature visible in the bottom left panel of
figure~\ref{fig:fv-comp-EAGLE-HR}, is that the azimuthal component of the velocity
distribution for halo E12 (shown in green) is skewed towards positive $v_\theta$,
indicating a significant rotation of the DM particles along the Galactic disc. Note
that such a feature is not visible for the same halo in the DMO simulation (bottom
right panel of figure~\ref{fig:fv-comp-EAGLE-HR}). To quantify the features in the
components of the DM velocity distribution for all the MW analogues, we fit them with
the following fitting functions:

\begin{itemize} 

\item A Gaussian function:
\beq
f(v_i) = \frac{1}{\sqrt{\pi} v_0 } \exp \left[-(v_i - \mu)^2/v_0^2 \right] \, , 
\eeq
with free parameters $v_0$ and $\mu$.

\item A generalized Gaussian function:
\beq
f(v_i) = \frac{1}{2 v_0 \Gamma(1 + 1/(2 \alpha))} \exp \left[ -\left((v_i - \mu)^2/v_0^2 \right)^\alpha \right] \, , 
\eeq
with free parameters $v_0$, $\mu$, and $\alpha$. 

\end{itemize}

The functions are normalised such that the 1D integral in velocity space is equal to
unity (integrated from -$\infty$ to +$\infty$).
 
Additionally, we fit the azimuthal component of the DM velocity distribution with a
double Gaussian function: \beq f(v_\theta) = c_1 f_{\rm 1, Gauss}(v_\theta; v_1,
\mu_1) + c_2 f_{\rm 2, Gauss}(v_\theta; v_2, \mu_2) \, , \eeq with free parameters
$c_1, v_1, \mu_1, v_2$, and $\mu_2$; $c_2$ is instead constrained by requiring
$f(v_\theta)$ to be normalised to 1, $c_1 + c_2 = 1$.

In Appendix~\ref{app:fits}, tables~\ref{tab:fvfit_vr} and~\ref{tab:fvfit_vz} we quote
the best-fit parameters for the Gaussian and generalized Gaussian fits to the radial
and vertical velocity components for the selected MW-like haloes in the \eagle HR and
\apostle IR hydrodynamic simulations.  In tables~\ref{tab:fvfit_vtheta}
and~\ref{tab:fvfit_vtheta_double}, we present the best-fit parameters for the
Gaussian, generalized Gaussian and double Gaussian fits for the azimuthal component
of the DM velocity distribution.  As for the distribution of the radial velocity
component, $v_r$, the fit indicates that the distribution is well described by a
generalized Gaussian with $\alpha \approx 1$ (close to the Gaussian function) and a mean
radial velocity that is always smaller than $\approx 13$ km/s (and for most of the
haloes constrained between -5 km/s and +5 km/s).  Analogous are the conclusions for
the distribution of the vertical velocity component, $v_z$, which is generally peaked
at values around $\mu = 0$ km/s.  In the DMO counterparts of our selected haloes, both the radial and vertical velocity
component distributions peak around zero as well.  The distribution of the azimuthal velocity component,
$v_\theta$, is instead well fitted in most cases by either a Gaussian or a
generalized Gaussian with an $\alpha$ parameter generally smaller than 1 (but
larger than 0.6). The mean azimuthal
speed is typically higher in the case of the hydrodynamic simulation than the mean radial and vertical speeds. In particular, 4 haloes in
the \eagle HR simulation (E4, E7, E10, E12) and 1 in the \apostle IR simulation (A1) have a significant
non-zero mean azimuthal speed ($|\mu|>20$~km/s). From these five haloes, 4 have
positive mean speeds, while one halo is counter-rotating. When
fitting the azimuthal velocity component with a generalized Gaussian, the same five
haloes show evidence of rotation ($|\mu| > 3 \sigma_\mu$). 

To evaluate if baryonic processes are responsible for the net rotation of the DM particles in the 
tangential direction in the torus, 
we compare the azimuthal component of the velocity distribution for haloes in the hydrodynamic
simulations with their counterparts in the DMO simulations. For the four 
haloes with positive mean azimuthal speeds with evidence of rotation in the hydrodynamic
simulations, we find
that their DMO counterparts do not show evidence of rotation at a comparable mean azimuthal speed.
In particular, for these four haloes, the mean of the best fit generalized Gaussian for the 
azimuthal velocity component in the hydrodynamic case is larger than the corresponding mean in the
DMO case, 
$\mu_{\rm hydro} > \mu_{\rm DMO} + 3 \sigma_{\mu_{\rm DMO}}$.

The asymmetry in the azimuthal component of the DM velocity distribution and the
positive mean tangential speed observed for some haloes in the hydrodynamic simulations 
could be pointing to the
existence of a ``dark disc''. The dark disc forms when the stars and gas in the
Galactic disc drag merging satellite galaxies towards the Galactic plane. These
satellites are then disrupted by tidal forces and their accreted material forms thick
stellar and DM discs~\cite{Read:2008fh}. Previous hydrodynamic simulations by Read
{\it et al.}~\cite{Read:2009iv} have found that the density of the dark disc can be
0.25 -- 1.5 times the halo density at $R_0=8$~kpc, and the rotation lag of the
dark disc component is 0 -- 150 km$/$s compared to the stellar disc. Purcell {\it et
  al.}~\cite{Purcell:2009yp} found instead that the contribution of accreted DM material
to the disc at the Solar position is less than $\approx 20\%$ of the host halo density,
and the fraction of DM particles rotating with a lag velocity less than 50 km$/$s
compared to the stellar disc is enhanced by less than $\approx 30\%$  compared to the
SHM. The hydrodynamic simulation performed by Ling {\it et al.}~\cite{Ling:2009eh}
shows instead a dark disc component co-rotating with the Galactic disc which
contributes $\approx 25 \%$ of the total local DM density, and has a mean lag
velocity of $\approx 75$ km$/$s compared to the stellar disc. For the Eris hydrodynamic
simulation~\cite{Kuhlen:2013tra} the dark disc contributes 9.1\% of the
DM density in the disc. Ref.~\cite{Kuhlen:2013tra} quantifies the dark disc by computing the amount of
material in the disc accreted from satellites with positive azimuthal velocities, co-rotating with the stellar disc.

We stress here that a large simulation sample may be necessary to draw robust conclusions regarding the dark disc, since large halo-to-halo scatter in the results is very likely. The conclusions may also be sensitive to stochastic effects and to the 
treatment of baryonic physics and so large samples, ideally drawn from 
simulations with different plausible baryonic physics, are needed to reach 
a definitive conclusion regarding the relevance of dark discs.

The implications of the dark disc for direct detection signals has been explored
before~\cite{Read:2009iv, Bruch:2008rx, Purcell:2009yp, Ling:2009eh,
  Kuhlen:2013tra}. Depending on the density of the dark disc, the event rates in
direct detection experiments could be enhanced, especially in the low recoil energy
range. The annual modulation signal and its phase could also be altered. In
particular, depending on the lag velocity of the DM particles in the disc compared to
the stellar particles, the phase of the annual modulation could be shifted. However, most
previous simulations have shown that the increase in the local DM density due
to the dark disc is relatively small~\cite{Purcell:2009yp, Ling:2009eh,
  Kuhlen:2013tra} and therefore that direct detection signals do not change significantly
due to the presence of a dark disc.

To assess whether a dark disc component might exist for any of our MW analogues, we
extract the azimuthal velocity distribution of the star particles in the torus for
each halo. We then fit the azimuthal components of both the DM and
stellar velocity distributions with a double Gaussian. Our aim is to search for galaxies in which there is a  DM component rotating as fast as the stars. Among the four haloes in our sample which show evidence of rotation with significant positive mean azimuthal speed, only two haloes (E7 and E12) have a
rotating DM component in the disc with mean velocity comparable (within 50 km$/$s) to that of the
stars. The azimuthal component of the DM velocity distribution for the same two haloes in the DMO simulation is symmetric
around zero. Notice that both the E7 and E12 galaxies don't have a prominent stellar disc, as defined in ref.~\cite{Calore:2015oya}.

\begin{figure}
\begin{center}
  \includegraphics[width=0.49\textwidth]{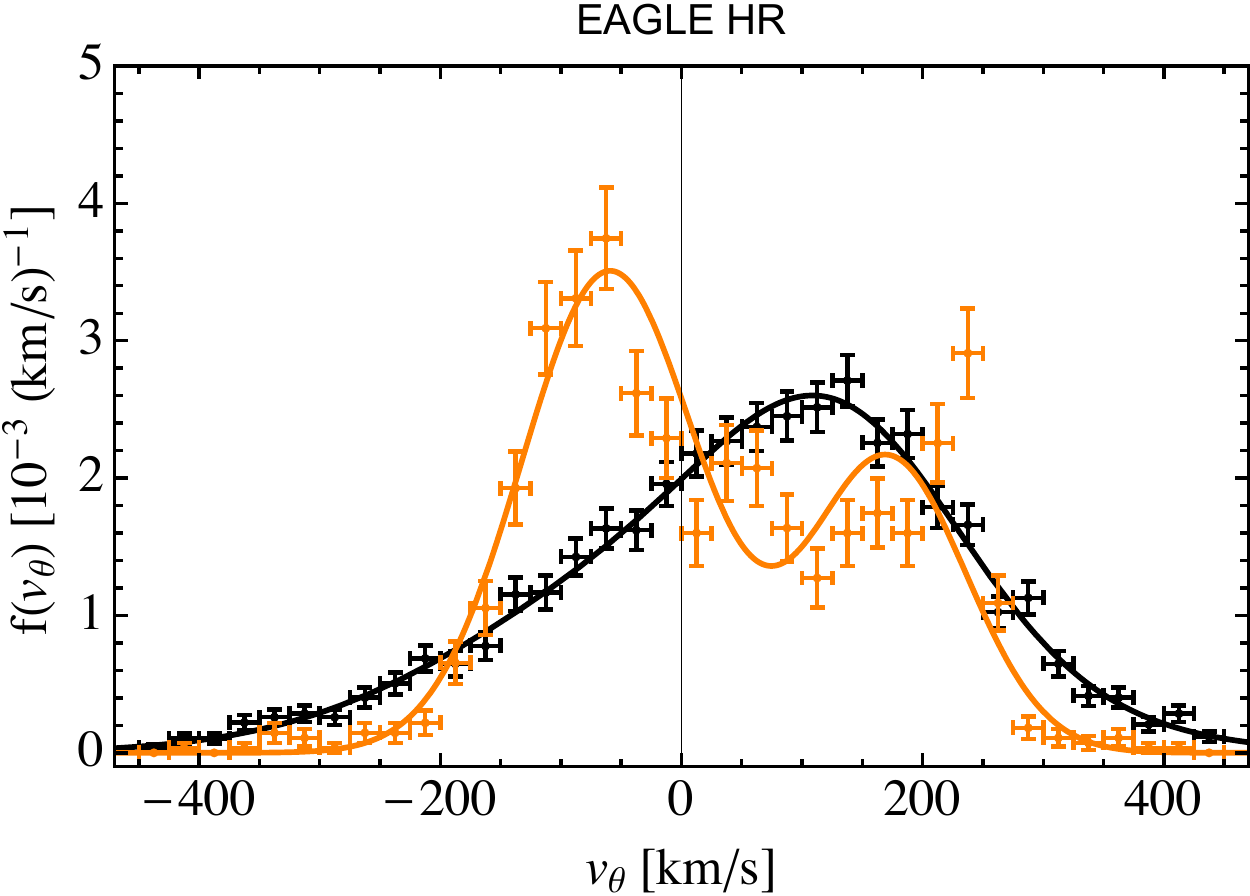}
    \includegraphics[width=0.49\textwidth]{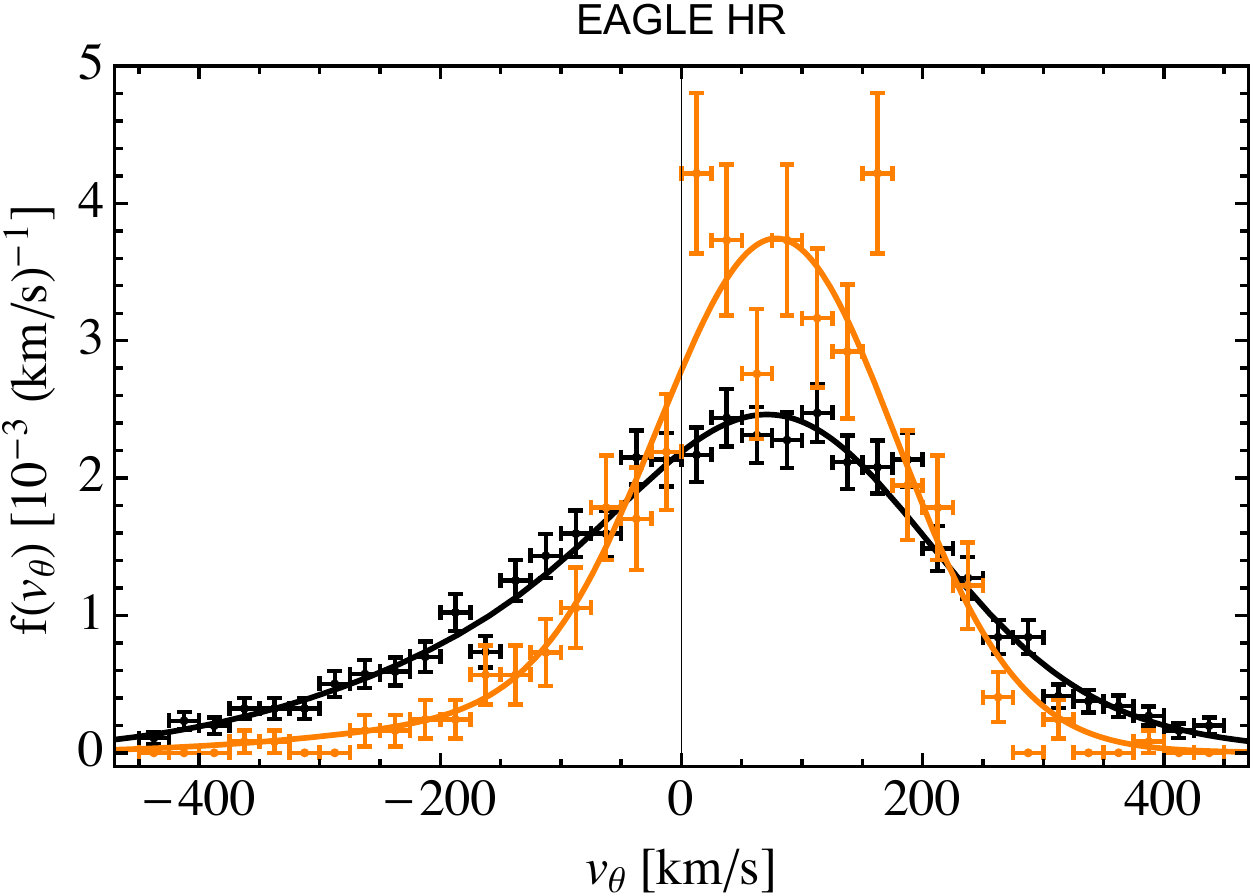}
\caption{\label{fig:fv-comp-EAGLE-HR-DD} Azimuthal components of the local DM (black
  data points with 1$\sigma$ error bars) and stellar (orange data points with
  1$\sigma$ error bars) velocity distribution in the Galactic rest frame for the two
  MW-like haloes in the \eagle HR simulation which have comparable peaks at positive
  azimuthal speeds $v_\theta$. Left panel: halo E12; right panel: halo E7. The best fit double Gaussian distributions to the DM
  and stellar distributions are also plotted as solid black and orange lines,
  respectively.}
\end{center}
\end{figure}

The two panels of figure~\ref{fig:fv-comp-EAGLE-HR-DD} show the azimuthal components
of the local DM (shown as black data points and $1\sigma$ error bars) and stellar
(orange data points and $1\sigma$ error bars) velocity distributions for two MW-like
haloes (E12 and E7) in the \eagle HR simulation with comparable mean azimuthal velocities for the
DM and stellar distributions. The left panel of figure~\ref{fig:fv-comp-EAGLE-HR-DD}
corresponds to the same halo (E12) as shown in the bottom left panel of
figure~\ref{fig:fv-comp-EAGLE-HR}. The black and orange solid lines show the best fit
double Gaussian distribution to the DM and stellar azimuthal velocity distributions,
respectively. We can see from the left panel that there are two kinematically distinct structures in the azimuthal component of the stellar velocity distribution in the torus centered at 8 kpc. The structure with a negative mean azimuthal speed contains most of the mass and is also present in a torus centered at 5 kpc as well as a torus centered at 10 kpc. While the structure with a positive mean azimuthal speed which contains less mass is not present at radii smaller than the Solar position, and may be due to the accretion of a massive satellite into the disc region. Hence, this structure is likely a ``ring" of stars, the relic of a ``ring" of gas that accretes late in the formation of the galaxy. Such features are not infrequent in simulations of galaxy formation (see e.g.~\cite{Algorry:2013xga}).

In the left (right) panel of figure~\ref{fig:fv-comp-EAGLE-HR-DD}, when fitting with a double Gaussian
distribution, the mean azimuthal velocity for the second Gaussian (with higher $\mu$)
is 136 km$/$s (88 km$/$s) for the DM distribution, and 171 km$/$s (81 km$/$s) for the
stellar distribution. Thus, the differences in the second peak of the double Gaussian
between the DM and stellar distributions are 35 km$/$s and 7 km$/$s, in the left and
right panels, respectively. The fraction of the rotating component of the double
Gaussian DM distribution with the larger mean velocity is 32\% and 43\% in the left
and right panels, respectively.

In summary, from a total of 14 MW-like haloes in our sample, evidence for a significant positive mean azimuthal speed, which is comparable
for DM and stars in the torus, exists only for two haloes, and this may hint at the existence of a co-rotating dark disc in those two haloes. Although the mechanism that has been proposed to produce a dark disc may be causing the features discussed for these two haloes, to draw such a conclusion one would need to follow the accretion history of massive satellites into the disc region, and this is beyond the scope of this paper. One instead can conclude that even in the case that the dark disc is responsible for such features, this is a relatively rare occurrence in our halo sample, occurring only in two out of 14 haloes. 
Notice also that the selection of our MW-like halo sample is unbiased with respect to the formation of a dark disc.

\section{Local dark matter density and halo shape}
\label{sec:DMdensity}

The fiducial value of the local DM density in the SHM is 0.3
GeV$/$cm$^3$. By considering a large range of halo profiles and by constraining the mass
model of the MW using global dynamical data sets, local DM densities in the range of
(0.2 -- 0.6) GeV$/$cm$^3$ are obtained~\cite{Catena:2009mf, Weber:2009pt,
  McMillan:2011wd, Iocco:2011jz, Nesti:2013uwa, Sofue:2015xpa, Pato:2015dua}. Local estimates of
the DM density~\cite{Salucci:2010qr, Smith:2011fs, Bovy:2012tw, Garbari:2012ff,
  Zhang:2012rsb, Bovy:2013raa} are consistent with the global estimates, although
they have larger uncertainties.

Here we extract the local DM density using the torus discussed in
section~\ref{sec:veldist}. The average local DM density in the torus is $0.42 -
0.73$~GeV$/$cm$^3$ for the 12 haloes in the \eagle HR simulation, and $0.41 -
0.54$~GeV$/$cm$^3$ for the two haloes in the \apostle IR simulation. When splitting 
the torus into smaller angular regions (8 regions for \eagle HR, and 16 regions for \apostle IR), we find that
the average DM density varies on average by $32\%$ along the torus, a variation
smaller than the halo-to-halo scatter. Notice that the values we find for the local
DM density are in agreement with the global and local estimates discussed above.

To determine if there is an enhancement of the local DM density in the Galactic disc
compared to the halo, we compare the average value of the local DM density in the
torus with the average value in a spherical shell at $7 < R < 9$~kpc. The average DM
density in the shell is in the range $0.41 - 0.66$ GeV/cm$^3$ for the MW-like haloes
in the \eagle HR simulation, and is 0.42 GeV/cm$^3$ for the two \apostle IR MW
analogues. From the 14 MW analogues in both simulations (12 in \eagle HR $+$ 2 in
\apostle IR), the local DM density is higher by 2 -- 27\% for 10 haloes compared to
the shell averaged value, while the enhancement is greater than 10\% for five haloes, 
and greater than 20\% for only two haloes. To evaluate if baryonic physics is responsible for this
enhancement, we compare the average local DM density in an identical torus and  shell 
for the same haloes in the DMO simulations. We find that two of the DMO counterparts (haloes E1 and E11)
show a similar increase of the DM density in the torus compared to the shell, which can be
due to the halo itself not being a perfect sphere. 

The increase in the local DM density in the disc for some haloes could be due to
the DM halo contraction as a result of dissipational baryonic processes in the
hydrodynamic simulations (See \cite{Schaller:2015a} for a measure of dark halo
contraction in the \eagle galaxy formation model). Note that for the two haloes
discussed in section \ref{sec:vel-comp} which have a DM component co-rotating
with the stellar disc with a mean azimuthal speed comparable to that of the
stars, the local DM density in the Galactic disc is 27\% higher for halo E12
(shown in the left panel of figure~\ref{fig:fv-comp-EAGLE-HR-DD}) and 9\% lower
for halo E7 (shown in the right panel of figure~\ref{fig:fv-comp-EAGLE-HR-DD})
compared to the shell averaged value.

To further understand the effect of baryonic physics on the local DM density, we can
study the shape of the inner ($<8$~kpc)  DM halo.
We first calculate the inertia tensor of the DM
particles within two different radii (5~kpc and 8~kpc). The resulting inertia tensor
then describes an ellipsoid with three axes of length $a\ge b\ge c$: these lengths
 provide information about the shape of the DM
distribution. We also repeat this process for the DMO counterparts to our \eagle HR
galaxy sample.

We first calculate the sphericity, $s=c/a$. If the DM
distribution is a perfect sphere, all three axes will have the same length, so $c=a$
and thus sphericity $s=1$; a value of $s<1$ describes an increasing deviation from
sphericity in our sample. We find for our galaxy sample that the sphericities at
5~kpc cluster in the range $s=[0.85,0.95]$ and are typically lower by less than 10\%
at 8~kpc, thus haloes have a larger deviation from a sphere at larger radii. In both cases, however,
due to dissipational baryonic processes the DM sphericity is systematically higher in the hydrodynamic simulations
than is the case for the DMO runs, in which the range in $s$ is $[0.75,0.85]$. This
result is in agreement with earlier detailed simulations
\cite{Dubinski:1994,Bryan:2013} and reproduces what was found in higher resolution
simulations of the \apostle series \cite{Schaller:2015c}.

Where there is a deviation from sphericity, this can be described using
the triaxiality parameter, $T$, which is defined as
\begin{equation}
	T = \frac{a^2-b^2}{a^2-c^2}.
\end{equation}

For very oblate systems, $a\approx b \gg c$ and thus $T \approx 0$, whereas for the opposite,
prolate case $a\gg b\approx c$ and $T \approx 1$. We find that there is little correlation in
our galaxy sample between the values of $T$ at 5~kpc and 8~kpc, with both varying in
the range $[0.02,0.8]$. However, it should be noted that because these inner haloes
are very close to spherical ($s=[0.81,0.96]$), the deviation towards either prolate or oblate
distributions is very small. By contrast, the DMO counterparts almost all have a
preference for inner haloes that are prolate rather than oblate, with $T=[0.4,0.9]$
at both radii for all of our candidates. We therefore conclude that spherical
symmetry is a very good approximation for our galaxy halo centres with little
dependence on radius, and that this marks an important difference from DMO studies.

For halo E12 (shown in the left panel of figure~\ref{fig:fv-comp-EAGLE-HR-DD}), which
has a local DM density in the torus enhanced by 27\% compared to the shell averaged
value, the triaxiality parameter $T=0.02$ at 8 kpc, which implies an
oblate halo, as expected.

\section{Direct detection signals}
\label{sec:signals}

\subsection{Event rates}

We consider a DM particle $\chi$ scattering elastically off a nucleus with atomic mass
number $A$ and atomic number $Z$, and depositing the nuclear recoil energy $E_R$ in
the detector. The differential rate (events per unit energy, per unit detector mass, per unit time) is given by
\beq \label{rate}
\frac{d R}{d E_R} = \frac{\rho_\chi}{m_\chi} \frac{1}{m_A}\int_{v>v_{\rm min}}d^3 v \frac{d\sigma_A}{d{E_R}} v \tilde f_{\rm det}(\vect v, t).
\eeq
Here $\rho_\chi$ is the local DM density, $m_\chi$ and $m_A$ are the DM and nucleus
masses, $\sigma_A$ is the DM--nucleus scattering cross section and $\vect v$ is the
3-vector relative velocity between DM and the nucleus, while $v\equiv
|\vect{v}|$. $\tilde f_{\rm det}(\vect v, t)$ is the DM velocity distribution in the
detector rest frame normalised to unity, such that $\int d^3 v \tilde f_{\rm det}(\vect v, t) =1$. For the case of elastic scattering, the
minimum speed $v_{\rm min}$ required for a DM particle to deposit a recoil energy
$E_R$ in the detector is given by \beq v_{\rm min}=\sqrt{\frac{m_A E_R}{2 {\mu_{\chi
        A}^2}}},
\label{eq:vm}
\eeq
where $\mu_{\chi A}=m_\chi m_A/(m_\chi + m_A)$ is the reduced mass of the DM and nucleus.

The differential cross section is in general a sum of spin-independent and
spin-dependent contributions.  Here we only consider the case of spin-independent
scattering and assume equal couplings of DM to protons and neutrons.
In this case the differential cross section is
\begin{align}
  \frac{d\sigma_A}{dE_R} = \frac{m_A A^2}{2\mu_{\chi p}^2 v^2} {\sigma_{\rm SI}} F^2(E_R) \,, 
  \label{eq:dsigmadE}
\end{align}
where $\sigma_{\rm SI}$ is the spin-independent DM--nucleon scattering cross section,
$\mu_{\chi p}$ is the reduced mass of the DM--nucleon system, and $F(E_R)$ is a
form factor for which we use the expression from Helm~\cite{Helm:1956zz}.

The DM velocity distribution is usually given in the Galactic rest frame, and to
calculate the event rate in Eq.~\eqref{eq:Reta}, one has to transform the velocity
distribution from the Galactic rest frame to the rest frame of the detector, by
\beq
\tilde f_{\rm det} (\vect v, t) = \tilde f_{\rm gal} (\vect v + \vect v_s + \vect v_e (t)),
\eeq
where $\vect{v}_e(t)$ is the velocity of the Earth with respect to the Sun, and
$\vect v_s = \vect v_c + \vect v_{\rm pec}$ is the velocity of the Sun in the Galactic rest frame.
Here $\vect v_c$ is the Sun's circular velocity (or Local Standard of Rest (LSR) velocity), and $\vect v_{\rm pec} \approx (11.10,
12.24, 7.25)$~km/s~\cite{Schoenrich:2009bx} is the peculiar velocity of the Sun with respect to
the LSR. For each simulated halo, we set $\vect v_c = (0, v_\star, 0)$, where
$v_\star$ is the local circular speed for that halo, rather than
using the fiducial value of $(0, 230, 0)$~km/s for the Sun's circular velocity. The
velocities are given in Galactic coordinates where the $x$-axis points towards the
Galactic centre, the $y$-axis points in the direction of the Galactic rotation, and
the $z$-axis points to the North Galactic pole.

The velocity of the Earth with respect to the Sun, $\vect{v}_e(t)$, causes the time
dependence of the differential event rate, and can be written as
\cite{Gelmini:2000dm}
\beq\label{eq:ve}
\vect{v}_e(t) = v_e [\vect{e}_1 \sin\lambda(t) - \vect{e}_2\cos\lambda(t) ] \,,
\eeq
assuming a circular orbit. Here $v_e=29.8$ km/s is the orbital speed of the Earth,
and $\lambda(t)=2\pi(t-0.218)$ with $t$ in units of 1 year and $t=0$ on January
1st. The orthogonal unit vectors $\vect e_1$ and $\vect e_2$ span the plane of the
Earth's orbit, and are given by, $\vect e_1 = (-0.0670,0.4927,-0.8676)$ and $\vect
e_2 =(-0.9931,-0.1170,0.01032)$ in Galactic coordinates.  For simplicity, we neglect
the small eccentricity of the Earth's orbit since Eq.~\eqref{eq:ve} provides an
excellent approximation to describe the annual modulation signal~\cite{Green:2003yh}.

\subsection{Halo integral}
\label{sec:haloint}

The astrophysical uncertainties in the event rate originate from the uncertainties in
the DM velocity distribution and density at the position of the Sun. Let us define
the halo integral as
\beq\label{eq:eta} 
\eta(v_{\rm min}, t) \equiv \int_{v > v_{\rm mim}} d^3 v \frac{\tilde f_{\rm det}(\vect v, t)}{v} \,.
\eeq
Then from Eqs.~\ref{rate} and \ref{eq:dsigmadE}, the differential event rate can be written as
\beq\label{eq:Reta}
\frac{d R}{d E_R} = \frac{\rho_\chi A^2 \sigma_{\rm SI}}{2 m_\chi \mu_{\chi p}^2} \, F^2(E_R) \, \eta(v_{\rm min}, t).
\eeq
The halo integral $\eta (v_{\rm min}, t)$, together with the local DM density
$\rho_\chi$, encapsulates the astrophysics dependence of the event rate.

\begin{figure}
\begin{center}
 \includegraphics[width=0.49\textwidth]{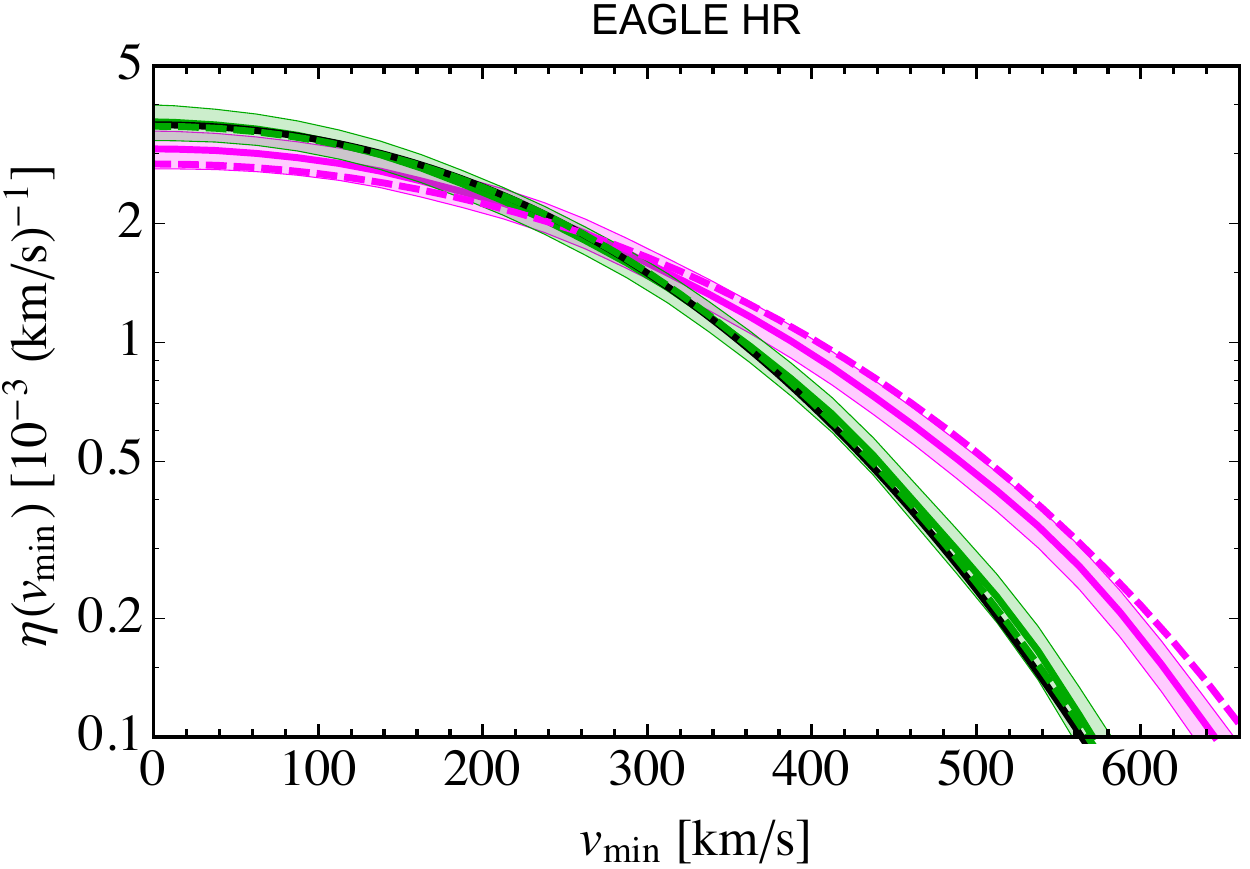}
  \includegraphics[width=0.49\textwidth]{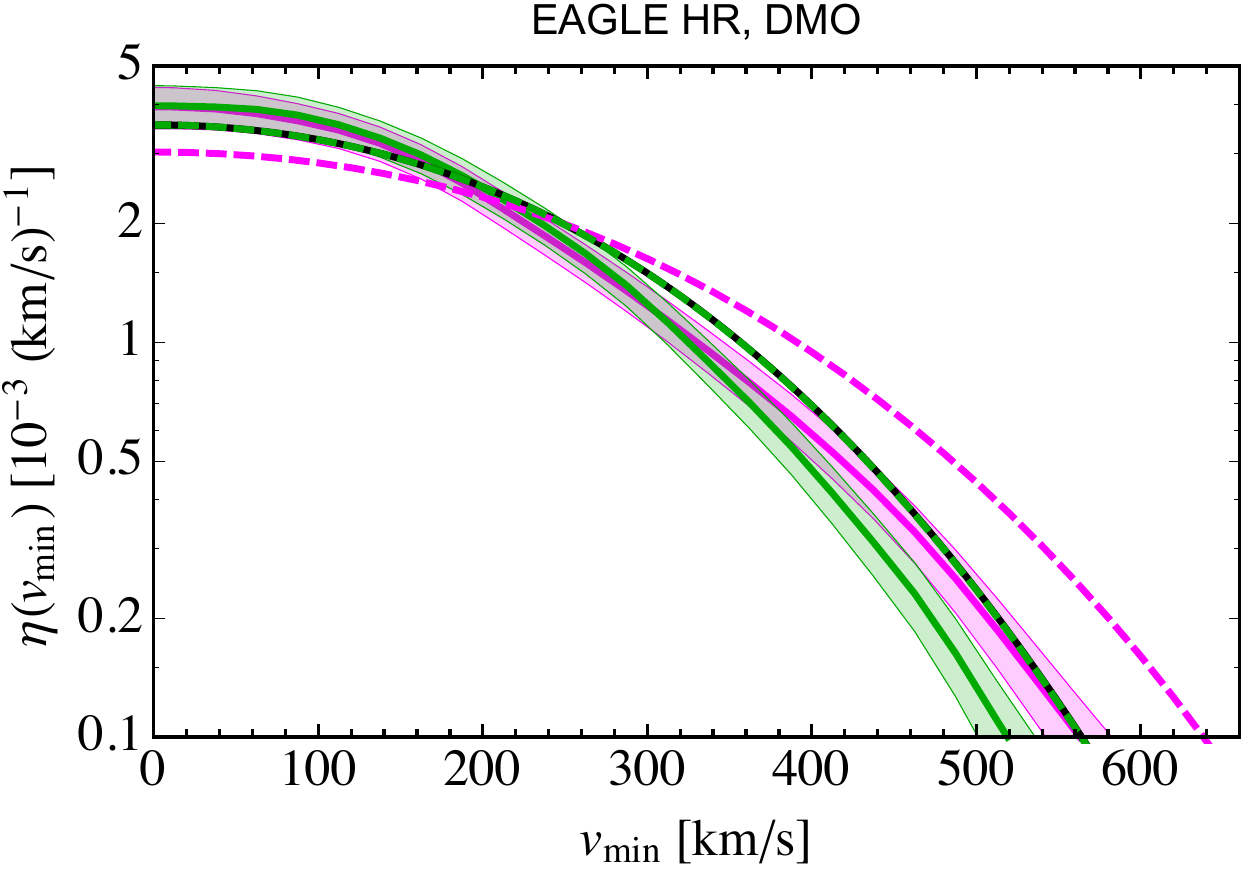}\\
 \vspace{5pt}\includegraphics[width=0.49\textwidth]{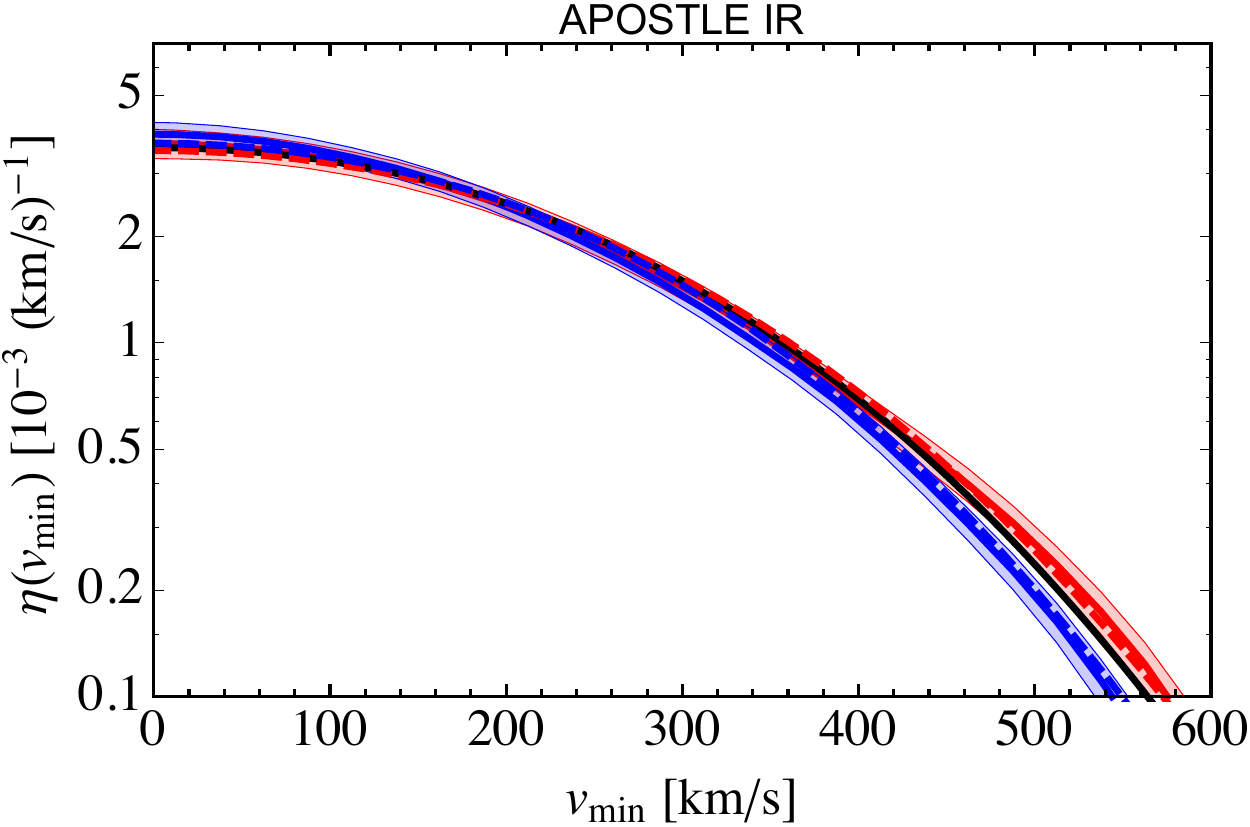}
  \includegraphics[width=0.49\textwidth]{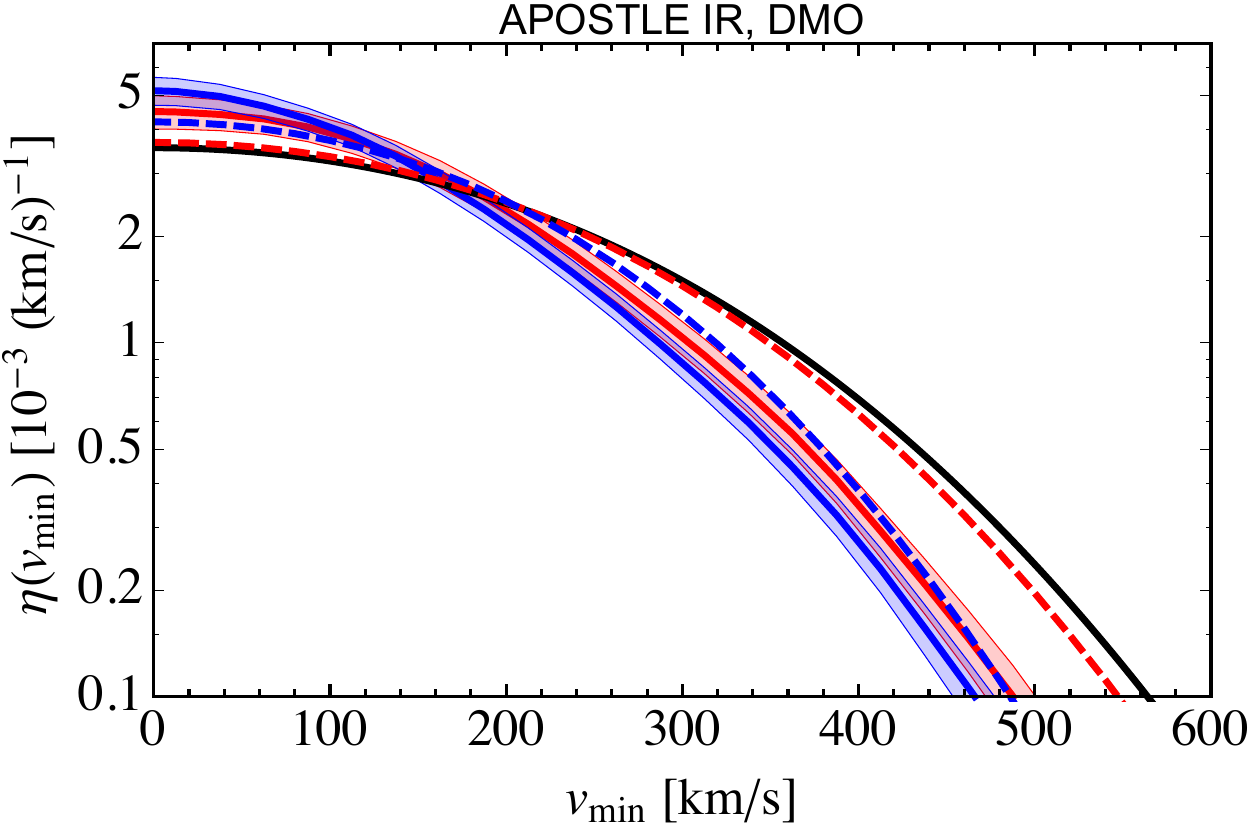}
\caption{\label{fig:eta} Time averaged halo integrals (solid coloured lines) for the
  the two haloes closest to (halo E12 shown in green) and farthest from (halo E3 shown in magenta) the SHM Maxwellian in
  the \eagle HR simulation (top left), and the two haloes (blue and red lines) in the
  \apostle IR simulation (bottom left). The solid coloured lines and the shaded bands
  correspond to the halo integrals obtained from the mean velocity distribution and
  the velocity distribution at $\pm 1\sigma$ from the mean, respectively. The solid black line
  specifies the halo integral for the SHM Maxwellian (with peak speed of 230 km/s), and 
  the dashed coloured lines specify the halo integral obtained from the best fit 
  Maxwellian velocity modulus distribution for each halo (with matching colours). The right
  panels show the halo integrals for the same haloes in the corresponding left panel
  but in a DMO simulation.}
\end{center}
\end{figure}

The top left panel of figure~\ref{fig:eta} shows the time averaged halo integral
as a function of the minimum velocity, $v_{\rm min}$, for the two haloes with speed
distributions that are the closest to (halo E12), and the farthest from (halo E3), the SHM Maxwellian
(specified with a solid black line) in the set of \eagle HR MW-like galaxies. The
bottom left panel of the same figure shows the time averaged halo integral for
the two \apostle IR MW-like haloes. The right panels show the halo integral for the
same haloes in the left panels but in a DMO simulation. The coloured solid lines
specify the halo integral computed from the mean value of the velocity distribution,
while the shaded band is obtained by adding and subtracting one standard deviation to
the mean velocity distribution. The halo integrals obtained from the best fit Maxwellian velocity 
modulus distributions (eq.~\eqref{eq:genMax} with $\alpha=1$) are shown by dashed coloured lines with matching colours for each halo.

From the left panels of figure~\ref{fig:eta}, one can see that the halo-to-halo
scatter in the halo integral is significant. There are MW-like haloes which have halo
integrals similar to the SHM Maxwellian, but there also exists MW
analogues with halo integrals that deviate from the SHM Maxwellian significantly at
higher minimum velocities. The effect is more evident for the \eagle HR sample, for
which we have more haloes. Comparing to the DMO case, where the halo-to-halo
scatter is less pronounced, one can conclude that baryons affect
the velocity distribution strongly at $R_0=8$~kpc, typically resulting in a shift
of the tails of the halo integral to higher velocities with respect to the DMO case.

The difference between the halo integral of the simulated haloes in the hydrodynamic case and the SHM Maxwellian, which is more pronounced at relatively large $v_{\rm min}$, is mainly due to the different 
peak speed of the DM velocity distribution of the simulated haloes compared to the 230 km$/$s assumed for the SHM Maxwellian. Indeed, 
the halo integral for the best fit Maxwellian velocity distribution falls within the $1\sigma$ 
uncertainty band of the halo integrals for all but two MW-like haloes in the hydrodynamic simulation. For those two haloes, there is only
a slight deviation between the halo integral for the best fit Maxwellian and the simulated halo at large $v_{\rm min}$. 
The range of the best fit $v_0$ parameter for the Maxwellian distribution is 223 -- 289 km$/$s in the hydrodynamic case, as can be seen in table 
\ref{tab:fvfit} in Appendix~\ref{app:fits}. 

On the other hand, the shape of the DM velocity modulus distributions for the haloes in the DMO simulations are not well captured by the Maxwellian distribution with a free peak. For most haloes in the DMO case, there are large deficits at the peak, and an excess of particles at low and very high velocities compared to the best fit Maxwellian speed distribution. These large differences result in halo integrals of the simulated haloes which are quite different from the best fit Maxwellian halo integrals, as can be seen from the right panels of figure ~\ref{fig:eta}.

\section{Results}
\label{sec:results}

We discuss below the details of the analysis of direct detection data using directly the local
DM density and velocity distribution of the selected MW analogues. Our aim is to
investigate how the allowed regions and exclusion limits set by different direct
detection experiments in the plane of DM mass and spin-independent cross section
change for our selected haloes compared to the SHM, and if the compatibility between
the hints for a DM signal and null-results improves.

\subsection{Data from direct detection experiments}
\label{sec:data}

In this section we discuss our analysis of the data from multiple direct detection
experiments. In particular, we consider the positive signals from
DAMA/LIBRA~\cite{Bernabei:2013xsa} (DAMA for short) and the silicon detectors in
CDMS-II~\cite{Agnese:2013rvf} (CDMS-Si for short), as well as the the null-results
from the LUX~\cite{Akerib:2013tjd} and SuperCDMS~\cite{Agnese:2014aze} experiments.

\bigskip

{\bf DAMA}: The DAMA experiment observed an annual modulation signal at the
9.3$\sigma$ level over 14 annual cycles. We use the data on the annual modulation
amplitude given in figure~8 of ref.~\cite{Bernabei:2013xsa} for the total cumulative
exposure of 1.33 ton yr of DAMA/LIBRA-phase1 and DAMA/NaI. In our analysis, we
consider 12 bins in the signal region between 2 keVee and 8 keVee, above which the
data is consistent with zero modulation. For the quenching factors of Na and I, we
use $q_{\rm Na}=0.3$ and $q_{\rm I}=0.09$, respectively, as measured by the DAMA
collaboration~\cite{Bernabei:1996vj}.

Our analysis of the DAMA data is similar to the analyses in~\cite{Fairbairn:2008gz,
  Kopp:2009qt, Kopp:2011yr}. In particular, we construct a $\chi^2$ function to fit
the DAMA data,
\be
\chi^2_{\rm DAMA} (m_\chi, \sigma_{\rm SI}) = \sum_{i=1}^{i=12} \left(\frac{A_i^{\rm pred} (m_\chi, \sigma_{\rm SI}) - A_i^{\rm obs}}{\sigma_i} \right)^2,
\label{eq: DAMAchisq}
\ee
where the sum is over the 12 energy bins, $A_i^{\rm pred}$ is the predicted annual
modulation amplitude, while $A_i^{\rm obs}$ and $\sigma_i$ are the experimental data
points and their errors, respectively, from figure~8 of
ref.~\cite{Bernabei:2013xsa}. To find the best fit, we minimize Eq.~(\ref{eq:
  DAMAchisq}) with respect to the DM mass $m_\chi$, and spin-independent DM--nucleon
cross section $\sigma_{\rm SI}$. The allowed regions in the $m_\chi - \sigma_{\rm SI}
$ plane at a given CL are obtained from contours of $\chi^2
(m_\chi, \sigma_{\rm SI}) =\chi^2_{\rm min} + \Delta \chi^2 ({\rm CL})$, where
$\Delta \chi^2 ({\rm CL})$ is evaluated for 2 degrees of freedom (dof), e.g., $\Delta
\chi^2 (99.73 \%) = 11.8$ and $\Delta \chi^2 (90\%) = 4.6$.

\bigskip

{\bf CDMS-Si}: Three DM candidate events with recoil energies of 8.2, 9.5, and 12.3
keV were observed in the 140.2 kg day of data taken with the Si detectors of the
CDMS-II experiment~\cite{Agnese:2013rvf}. The total estimated background was 0.62
events in the recoil energy range of $[7,100]$ keV. Our analysis of the CDMS-Si data
is analogous to the one presented in refs.~\cite{Bozorgnia:2013pua,
  Bozorgnia:2014gsa}. To include the background, we rescale the individual background
spectra from ref.~\cite{McCarthy}, such that 0.41, 0.13, and 0.08 events are expected
from surface events, neutrons, and $^{206}$Pb, respectively. We assume an energy
resolution of 0.3 keV and use the detector acceptance from
ref.~\cite{Agnese:2013rvf}.

\bigskip

{\bf LUX}: The LUX experiment has presented the analysis of 85.3 live-days of
data~\cite{Akerib:2013tjd}, which is consistent with the background-only
hypothesis. Assuming the SHM, the LUX result is in strong tension with positive hints
from DAMA and CDMS-Si. We use the maximum gap method~\cite{Yellin:2002xd} to set
exclusion limits, since with the available information we cannot reproduce the
likelihood analysis performed by the LUX collaboration.  We consider the region below
the mean of the Gaussian fit to the nuclear recoil calibration events (red solid
curve in figure~4 of \cite{Akerib:2013tjd}) as signal region, and assume an
acceptance of 0.5. Figure~4 of \cite{Akerib:2013tjd} shows that one event at 3.1
photoelectrons falls on the red solid curve. Hence, in this analysis we assume zero
events make the cut. To find the relation between S1 and nuclear recoil energy $E_R$,
we use figure~4 of \cite{Akerib:2013tjd} and find the value of S1 at the intersection
of the mean nuclear recoil curve and each recoil energy contour. For the efficiency
as a function of recoil energy, we interpolate the black points in figure~9 of
\cite{Akerib:2013tjd} for events with a corrected S1 between 2 and 30 photoelectrons
and a S2 signal larger than 200 photoelectrons. We multiply the efficiency from
figure~9 of \cite{Akerib:2013tjd} by 0.5 to find the total efficiency, and set it
equal to zero below $E_R=3$ keV. Assuming the SHM with the Maxwellian velocity
distribution and parameters chosen as in \cite{Akerib:2013tjd}, the 90\% CL exclusion
limit we find agrees well with the limit set by the LUX collaboration.

\bigskip

{\bf SuperCDMS}: Eleven events in the recoil energy range of [1.6, 10] keV have been
observed by the SuperCDMS collaboration with an exposure of 577 kg day of data taken
with their Ge detectors~\cite{Agnese:2014aze}. The collaboration sets a 90\% CL on
the spin-independent DM--nucleon cross section of $1.2 \times 10^{-42}$~cm$^2$ at a
DM mass of 8 GeV. We employ the maximum gap method to set an upper limit on the cross
section.  We use the red curve in figure~1 of \cite{Agnese:2014aze} for the detection
efficiency as a function of recoil energy, and assume an energy resolution of 0.2
keV.

\subsection{Dark matter parameter space}

In this section, we present our results for the exclusion limits and allowed regions
in the DM mass and interaction cross section plane set by each direct detection
experiment discussed in section \ref{sec:data} when assuming the DM distribution of
the simulated MW-like galaxies. Notice that we use directly the local DM density and velocity
distributions extracted from the simulations to perform the analysis of direct detection data.

To illustrate the effect of baryons on the interpretation of direct DM detection
results, we consider four benchmark haloes in the \eagle HR hydrodynamic simulation
which satisfy our selection criteria: two haloes with the smallest (halo E6) and largest (halo E4) local
DM densities, and two haloes with speed distributions closest to (halo E12) and farthest from (halo E3) the SHM
Maxwellian. The exclusion limits and allowed regions in the $m_\chi$--$\sigma_{\rm SI}$ plane 
obtained assuming the DM distribution of these benchmark haloes encompass the range of results obtained assuming the DM distribution of all our MW-like analogues. For the \apostle IR simulation, we present the results for the two haloes
which satisfy our selection criteria (haloes A1 and A2).

\begin{figure}
\begin{center}
 \includegraphics[width=0.49\textwidth]{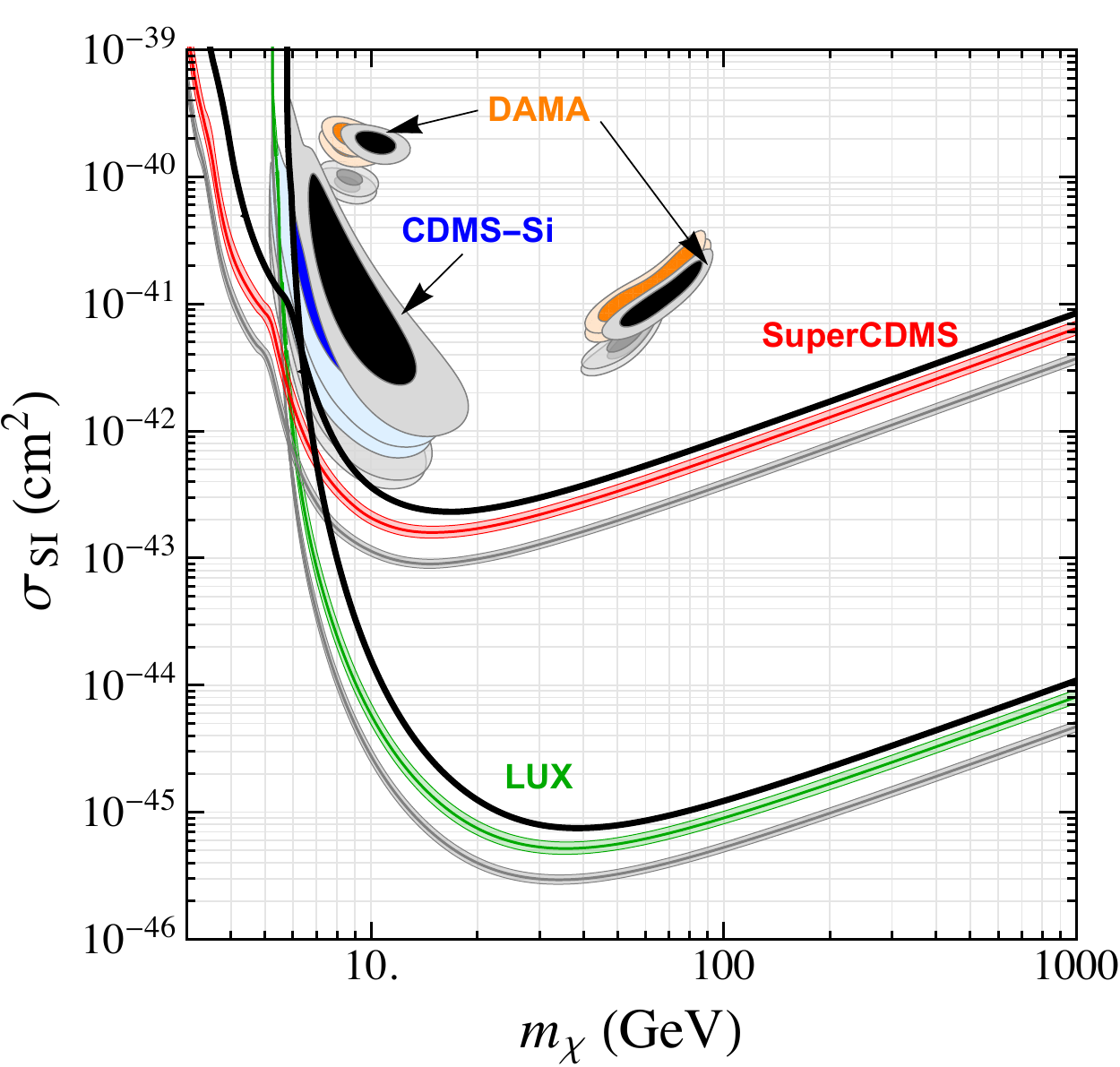}
  \includegraphics[width=0.49\textwidth]{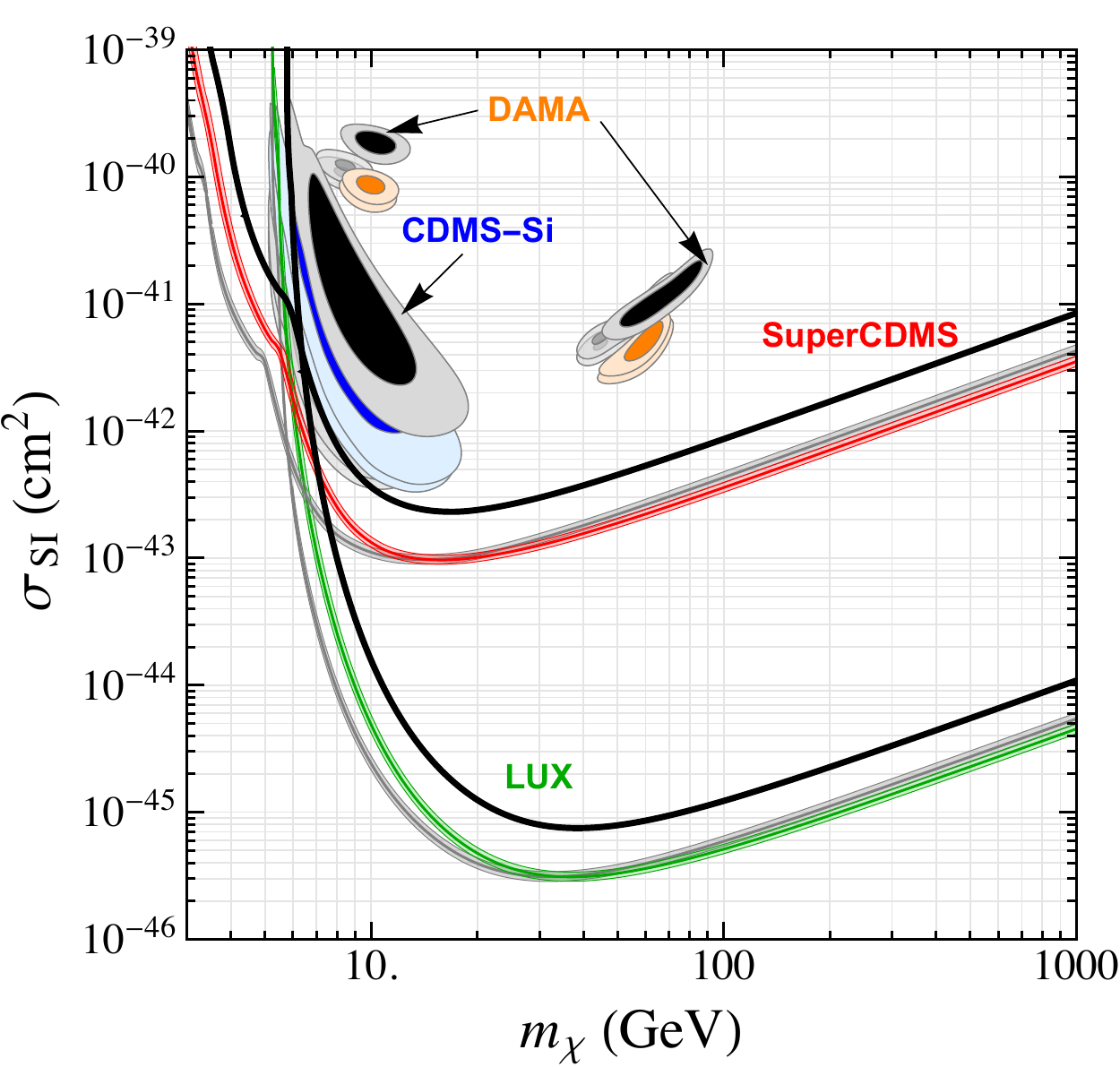}\\
    \includegraphics[width=0.49\textwidth]{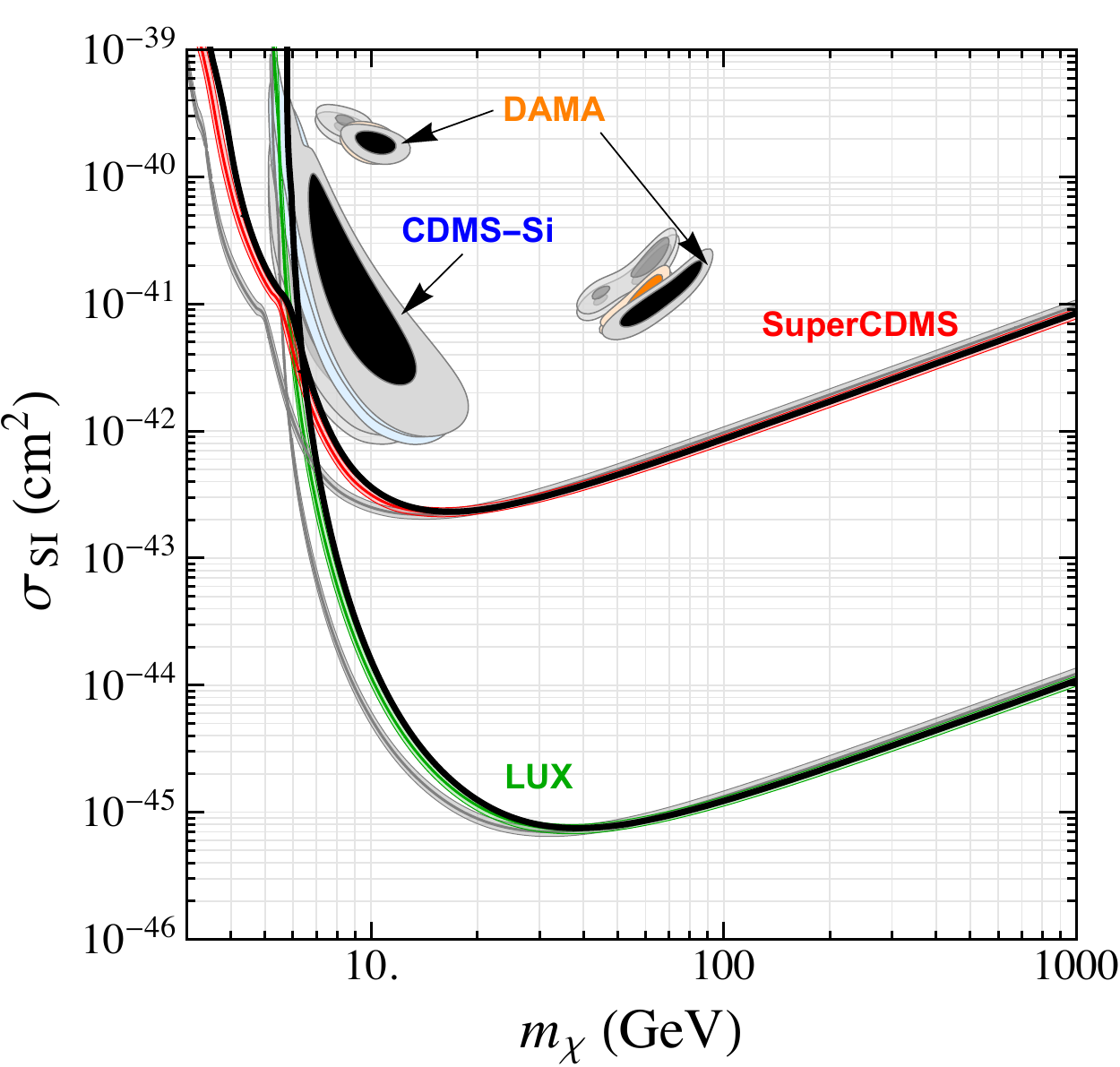}
\caption{\label{fig:ExcLim-EagleHR-h1h2} Exclusion limits from LUX and SuperCDMS (at
  90\% CL) and preferred regions from DAMA (at 90\% CL and 3$\sigma$) and CDMS-Si (at
  68\% and 90\% CL) in the spin-independent DM--nucleon cross section and DM mass
  plane for two haloes in the \eagle HR run with the smallest (halo E6 shown in colour) and
  largest (halo E4 shown in gray) $\rho_{\chi}$ (top left panel), and two haloes with
  velocity distributions closest to (halo E12 shown in colour) and farthest from (halo E3 shown in gray) the
  SHM Maxwellian (top right panel). The bottom panel shows the results for the same
  two haloes in the top right panel (E3 and E12) but assuming $\rho_\chi=0.3$~GeV$/$cm$^3$ for the
  two haloes instead of using the $\rho_\chi$ for each halo from the simulations. The shaded bands in the exclusion limits and the
two adjacent allowed regions of the same colour are obtained from the upper and lower
1$\sigma$ limits of the halo integral for each halo. The
  black exclusion limits and allowed regions correspond to the SHM Maxwellian. The
  local $\rho_{\chi}$ is 0.42 and 0.73 GeV$/$cm$^3$ for the two haloes in the top
  left panel, and it is 0.68 and 0.71 GeV$/$cm$^3$ for the two haloes in the top
  right and bottom panels.
}
\end{center}
\end{figure}

The top panels of figure~\ref{fig:ExcLim-EagleHR-h1h2} show the allowed regions from
the DAMA and CDMS-Si experiments and exclusion limits from the LUX and SuperCDMS
experiments in the plane of DM mass and spin-independent cross section for these four
\eagle HR benchmark haloes. In the left panel, the results are shown for the two
haloes with smallest (E6) and largest (E4) $\rho_\chi$, whereas in the right panel the results
are shown for the two haloes with speed distributions closest to (E12) and farthest from (E3) the
SHM Maxwellian. The bottom panel shows the results for the same two haloes in the top
right panel (E3 and E12), but setting the local DM density to 0.3 GeV$/$cm$^3$ for both
haloes, in order to isolate the effect induced by the deviation of the velocity
distribution from the SHM Maxwellian distribution.

For comparison, the allowed regions and exclusion limits assuming the SHM Maxwellian
(with peak speed of 230 km$/$s) velocity distribution and a local DM density of 0.3
GeV$/$cm$^3$ are shown in black. The shaded band in the exclusion limits and the
two allowed regions of the same colour are obtained from the upper and lower
1$\sigma$ limits of the halo integrals.

One can see from figure~\ref{fig:ExcLim-EagleHR-h1h2} that the general vertical shift of the
preferred regions and exclusion limits with respect to the SHM at all DM masses is due to the
variation in the local DM density, whereas the shift at low DM masses is due to the
high velocity tail of the DM distribution. For the halo with speed distribution farthest from the 
SHM Maxwellian (E3) and assuming a local DM density of 0.3 GeV$/$cm$^3$, the allowed regions 
and exclusion limits at low DM masses ($m_\chi \leq 10$~GeV) shift by a maximum of $\sim 2$ GeV towards smaller DM masses (bottom panel of figure~\ref{fig:ExcLim-EagleHR-h1h2}). When considering the local DM density extracted from simulations for each halo, 
all the regions and exclusion limits shift vertically at most by a factor of $\sim 2 - 3$ in the scattering cross section compared to the SHM (see left panel of figure~\ref{fig:ExcLim-EagleHR-h1h2}). This vertical shift is due to the different local DM densities of the simulated haloes compared to the SHM, which enters as a normalisation factor in the predicted recoil rate in direct detection experiments.

It is also clear from figure~\ref{fig:ExcLim-EagleHR-h1h2} that the halo-to-halo
scatter is larger than the 1$\sigma$ uncertainty for each halo. Notice also that in general the
preferred regions and exclusion limits all shift in the same direction, and thus the
compatibility between the results of different experiments is not improved for any
of the simulated haloes.

\begin{figure}
\begin{center}
 \includegraphics[width=0.49\textwidth]{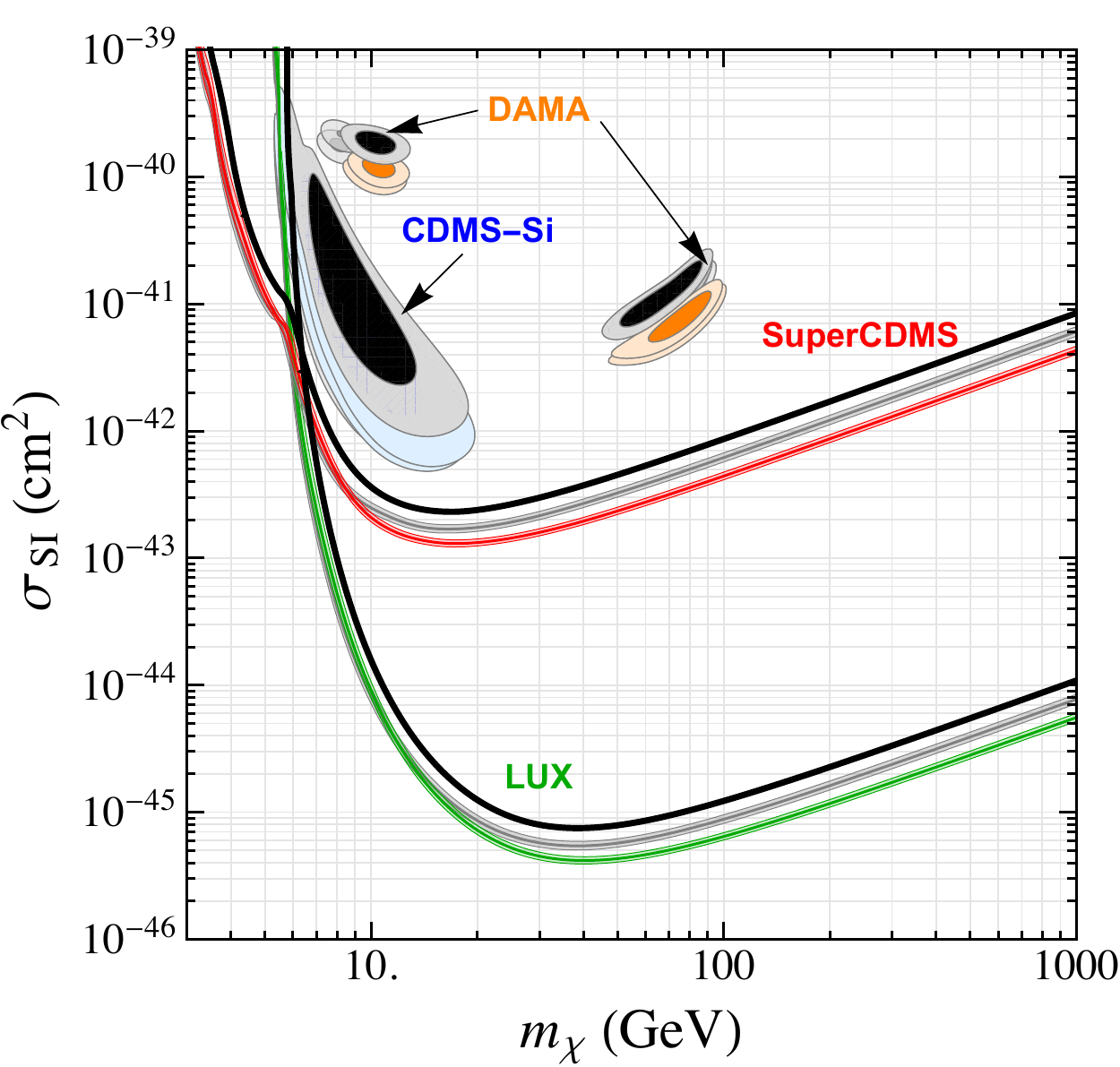}
  \includegraphics[width=0.49\textwidth]{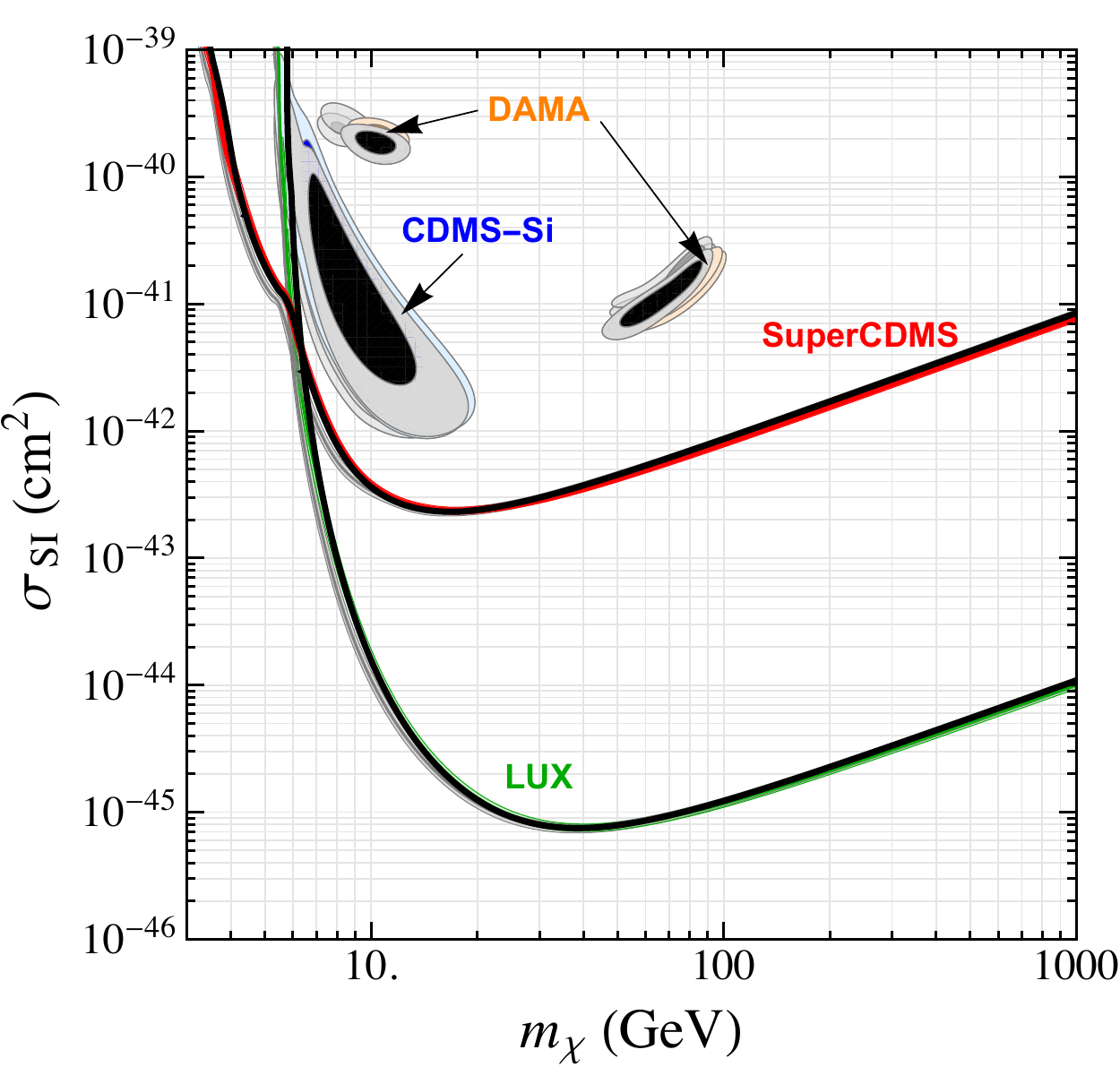}
\caption{\label{fig:ExcLim-LGIR} Same as figure~\ref{fig:ExcLim-EagleHR-h1h2} but for
  the two selected haloes (A1 and A2) in the \apostle IR simulation. Left panel: Using the
  $\rho_\chi$ for each halo (0.41 and 0.54 GeV$/$cm$^3$) from the simulations. Right
  panel: Assuming $\rho_\chi=0.3$~GeV$/$cm$^3$ for both haloes.}
\end{center}
\end{figure}

The left panel of figure~\ref{fig:ExcLim-LGIR} is the same as the top left panel of
figure~\ref{fig:ExcLim-EagleHR-h1h2}, but for the two \apostle IR haloes (A1 and A2). The right
panel shows the results for the same two haloes in the left panel but assuming a
fixed local DM density of 0.3 GeV$/$cm$^3$ for both haloes. The same features
discussed for the \eagle HR haloes are also visible for the \apostle IR haloes in
figure~\ref{fig:ExcLim-LGIR}.

\section{Conclusions}
\label{sec:summary}

In this work we have studied the predictions of the set of cosmological,
hydrodynamic simulations of the \eagle~\cite{Schaye:2015,Crain:2015} and
\apostle~\cite{Sawala:2015,Fattahi:2015} projects for DM direct
detection. Considering first all simulated haloes in the mass range $\mathcal{O}(10^{12} - 10^{13}) \, \Msun$, we applied two selection criteria for
choosing galaxies that are MW-like: (i) the rotation curve of the simulated
galaxy provides a good fit to the recent compilation of observed MW rotation
curves from ref.~\cite{Iocco:2015xga}, and (ii) the stellar mass of the simulated
galaxy is within the $3\sigma$ observed stellar mass range of the MW, $4.5
\times10^{10}<M_{*}/\Msun<8.3 \times10^{10}$~\cite{McMillan:2011wd}. We have
shown that these two selection criteria are especially relevant for accurate
predictions of the local DM density and velocity distribution, which are important
quantities for investigating the implications for DM direct detection searches.

For our selected MW analogues (12 \eagle HR and 2 \apostle IR), we extracted the
DM distributions and used them directly to perform an analysis of current dark
matter direct detection experiments data. In particular, we studied the
implications of the simulated MW-like haloes for direct detection considering
the allowed regions from DAMA~\cite{Bernabei:2013xsa} and
CDMS-Si~\cite{Agnese:2013rvf} and exclusion limits from
LUX~\cite{Akerib:2013tjd} and SuperCDMS~\cite{Agnese:2014aze} assuming
spin-independent elastic scattering. We summarize our main findings below:

\begin{itemize}

\item Including baryons in the hydrodynamic simulations results in a shift of the
  local DM velocity distributions to larger velocities compared to the DMO
  simulations. This results in a shift in the tails of the halo integral
  (eq.~\eqref{eq:eta}) to higher velocities in the hydrodynamic compared to the
  DMO simulations.
  
\item The halo integrals obtained from the best fit Maxwellian velocity distribution for each halo fall
within the $1\sigma$ uncertainty band of the halo integrals obtained directly from the simulated velocity distributions for all but two
  MW-like haloes in the hydrodynamic simulations. The range of the best fit peak speed for the Maxwellian velocity distribution is 223 -- 289 km$/$s.
  The deviation in the best fit peak speed values from the SHM Maxwellian results in a shift of the halo integral tail at large velocities.

\item Anisotropy in the DM velocity distribution is present at the Solar
  circle. In particular, four haloes in our sample show evidence of rotation
  with significant positive mean azimuthal speed ($\mu > 20$~km$/$s), but only two of them have comparable mean
  azimuthal speeds for DM and star particles. This may be hinting towards the existence of a dark disc for two out of 14 MW analogues.
 
\item The average local DM density in a torus aligned with the stellar disc and
  at $7~{\rm kpc}<R<9$~kpc and $|z|<1$~kpc is 0.41 -- 0.73 GeV$/$cm$^3$ for our
  halo sample. Most haloes in our sample show an enhancement of the local DM
  density in the torus compared to the spherical shell at the Solar circle
  (with $7~{\rm kpc}<R<9$~kpc). In particular, the local DM density in the torus
  is enhanced by more than 10\% for five haloes, and by more than 20\% for
  two haloes compared to the shell-averaged value.  Haloes in the hydrodynamic
  simulation are in general more spherical compared to the DMO case. However, they are not
  perfect spheres, and  are typically slightly more oblate compared to their DMO counterparts.

\item The variation in the local DM density from halo-to-halo causes the largest
  shift in the exclusion limits and allowed regions set by different direct
  detection experiments in the plane of DM mass and scattering cross section
  compared to the SHM. Future data from space observatories such as Gaia can in
  principle set stringent constraints on the value of the local DM density, and
  such constraints could be used in future numerical simulations.
  
\item The shift in the tail of the halo integral to higher minimum velocities compared to
  the SHM Maxwellian for some haloes results in a shift of a few GeV at low DM
  masses in the allowed regions and exclusion limits.

\item The overall shift in all regions and exclusion limits due to the different
  local DM distribution of the MW-like haloes compared to the SHM, occurs in the
  same direction and compatibility between different experiments is not  improved.

\end{itemize}

We have therefore shown that, when baryons are included in the \eagle and \apostle high resolution hydrodynamic simulations in such a way that the main galaxy population properties are broadly reproduced, a Maxwellian velocity distribution with a most probable speed larger than the standard value assumed in the SHM describes well the local DM velocity distribution of simulated MW analogues. The local DM density obtained from the simulations is in agreement with global and local estimates.

Shortly after the submission of this work, the preprints by Kelso {\it et al.}~\cite{Kelso:2016qqj} and Sloane {\it et al.}~\cite{Sloane:2016kyi} appeared, studying the local DM distribution of MW-like galaxies in hydrodynamic simulations and the implications for direct detection. In general we agree with the conclusions of ref.~\cite{Kelso:2016qqj}, which  studies two MW-like galaxies from the MaGICC cosmological simulations. They find that the SHM with velocity distribution inferred from the mass distribution of each simulated halo provides a reasonable fit in both of the MW-like galaxies they considered. Analogous to our findings, they also find that when baryons are included in the simulation, the best fit Maxwellian velocity distribution provides a good fit to the velocity distribution of each simulated halo. Ref.~\cite{Sloane:2016kyi} considers four MW-like galaxies with different merger histories in high resolution simulations. Their results agree with ours and those of ref.~\cite{Kelso:2016qqj} in finding that including baryons in the simulations leads to a shift of the DM speed distributions to larger speeds. However, our results differ in that ref.~\cite{Sloane:2016kyi} finds a deficit of high speed DM particles in their simulations with baryons compared to the SHM (with a local circular speed of 220 km$/$s). The halo integrals obtained from their best fit Maxwellian velocity distributions (with peak speeds smaller than 220 km$/$s), however, show only small discrepancies  at high speeds compared to those obtained from the simulations. The difference between the best fit peak speeds of the Maxwellian distribution in our works is likely due to the different stellar masses of our MW-like galaxies, given the strong correlation between the stellar mass and local circular velocity as shown in figure~\ref{fig:vc}.  

We would like to stress again that the results of this work as well as refs.~\cite{Kelso:2016qqj, Sloane:2016kyi} show that the halo integrals and hence direct detection event rates obtained from a Maxwellian velocity distribution with a free peak speed are similar to those obtained directly from the local DM velocity distributions of the simulated MW analogues. Thus a Maxwellian velocity distribution with a peak speed constrained by hydrodynamic simulations as well as observations could be used by the community when analysing the results of direct detection experiments.

\subsection*{Acknowledgements}

We thank Christopher McCabe and Christoph Weniger for useful discussions on the results of this work. We especially thank Fabio Iocco and Miguel Pato for providing the extensive compilation of rotation curve 
measurements used in this paper. G.B. (P.I.) and N.B.~acknowledge support from the European Research Council
through the ERC starting grant WIMPs Kairos. The research of F.C. is part of the VIDI research programme ``Probing the Genesis of Dark Matter", which is financed by the Netherlands Organisation for Scientific Research (NWO). This work is part of the
D-ITP consortium, a program of the NWO that is funded by the Dutch Ministry of Education, Culture and Science (OCW). This work was supported by the Science and Technology Facilities Council (grant
number ST/F001166/1); European Research Council (grant numbers GA 267291 ``Cosmiway''
and GA 278594 ``GasAroundGalaxies'') and by the Interuniversity Attraction Poles
Programme initiated by the Belgian Science Policy Office (AP P7/08 CHARM).  R.A.C.~is
a Royal Society Research Fellow.\\
This work used the DiRAC Data Centric system at Durham University, operated by
the Institute for Computational Cosmology on behalf of the STFC DiRAC HPC
Facility (\url{www.dirac.ac.uk}). This equipment was funded by BIS National
E-infrastructure capital grant ST/K00042X/1, STFC capital grant ST/H008519/1,
and STFC DiRAC Operations grant ST/K003267/1 and Durham University. DiRAC is
part of the National E-Infrastructure.  We acknowledge PRACE for awarding us
access to the Curie machine based in France at TGCC, CEA,
Bruy\`eres-le-Ch\^atel. 
This work makes use of PyMinuit\footnote{\url{http://code.google.com/p/pyminuit}}.

\clearpage
%%%%%%%%%%%%%%%%%%%%%%%%%%%%%%%%%%%%%%%%%%%%%%%%%%%%%%%%%%
\appendix

\section{Alternative selection of Milky Way analogues}
\label{app:criteria}

In this appendix we present the velocity distributions and halo integrals for the two
haloes in the \eagle HR simulation that satisfy all three selection criteria
discussed in section~\ref{sec:selection}. These two haloes contain a substantial disc
component, in addition to satisfying criteria (i) and (ii).

\begin{figure}
\begin{center}
 \includegraphics[width=0.49\textwidth]{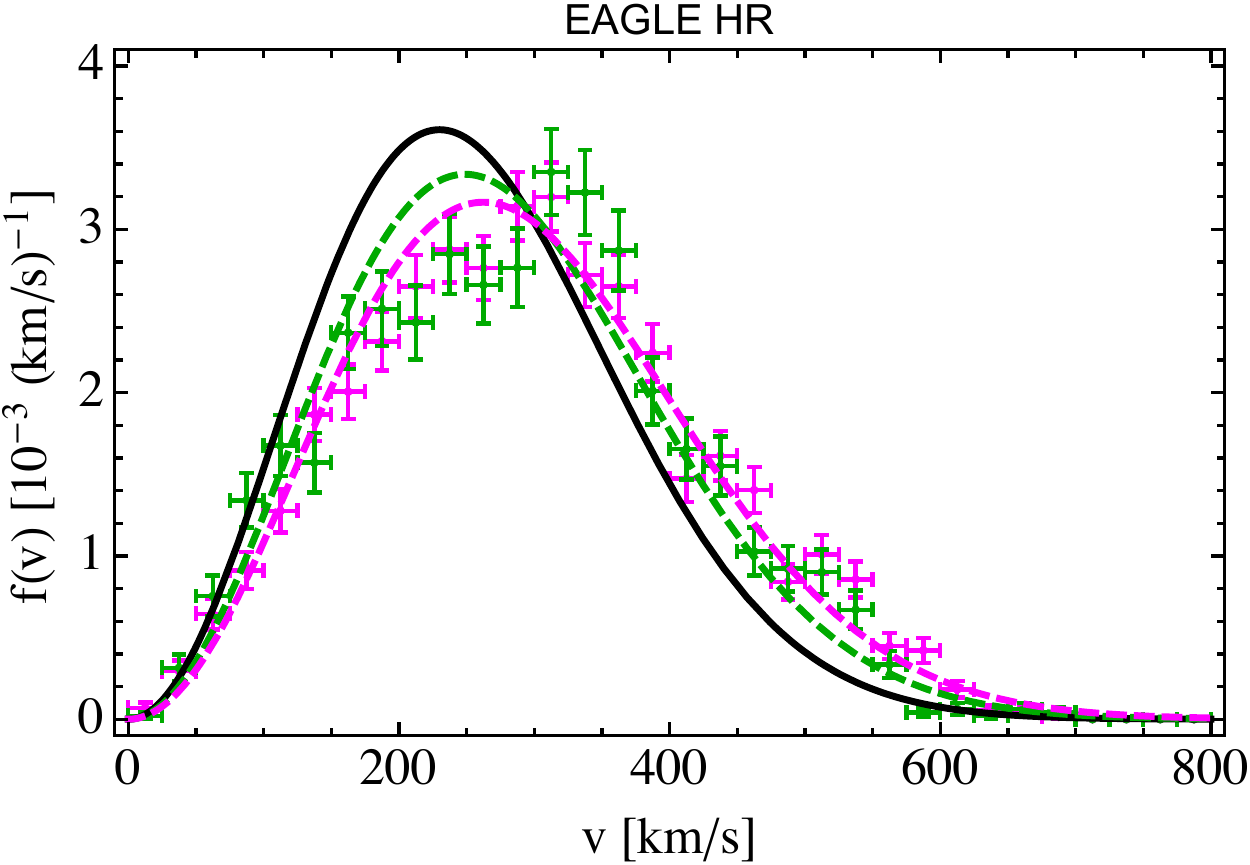}
  \includegraphics[width=0.49\textwidth]{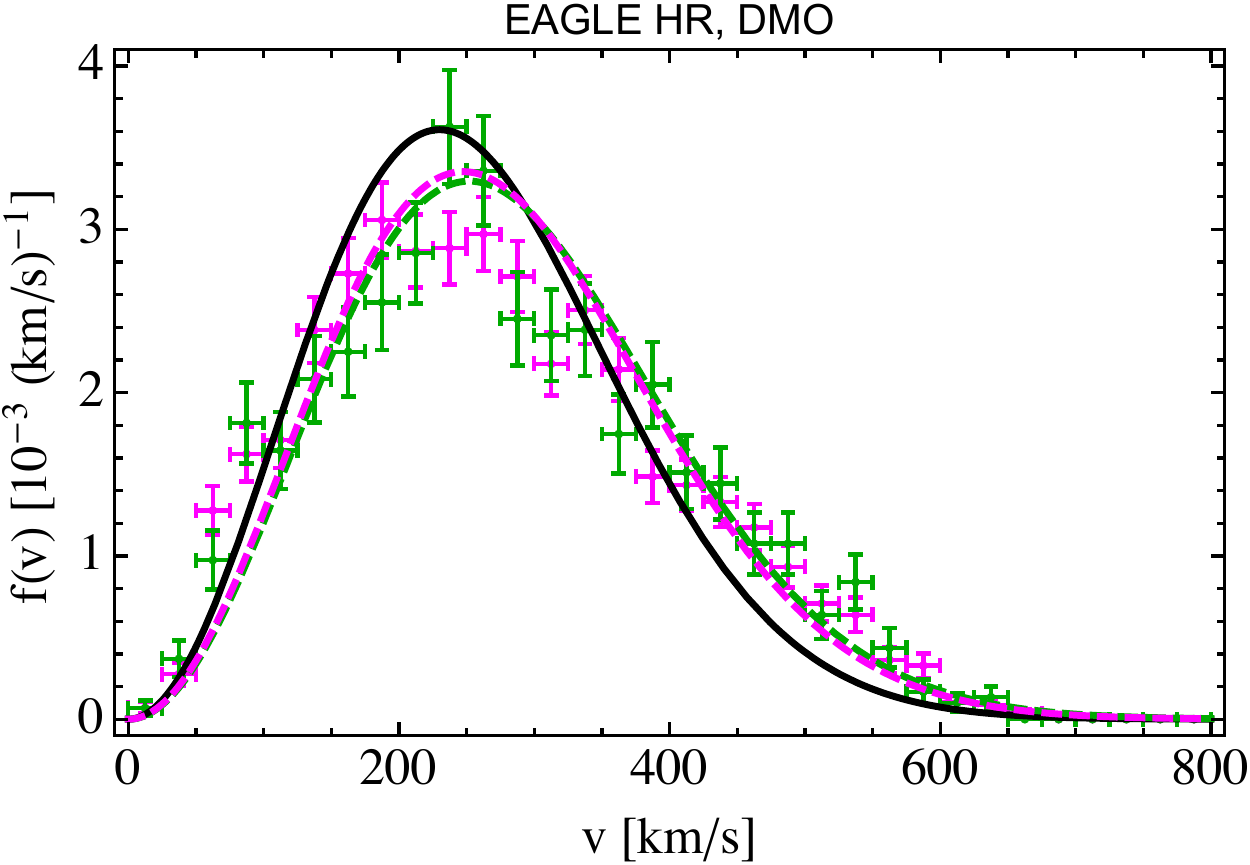}\\
\caption{\label{fig:fv-EAGLE-HR-all} Same as figure~\ref{fig:fv}, but for the two
  haloes in the \eagle HR simulation (E9 shown in green and E11 shown in magenta) which satisfy all three selection criteria
  discussed in section~\ref{sec:selection}.}
\end{center}
\end{figure}

In figure~\ref{fig:fv-EAGLE-HR-all}, we show the local DM velocity modulus
distribution in the Galactic rest frame for the two haloes (E9 and E11) satisfying criteria (i),
(ii), and (iii) in the \eagle HR hydrodynamic simulation (left panel) and DMO
simulation (right panel). The SHM Maxwellian with peak speed of 230 km$/$s (solid black line), as well as the best
fit Maxwellian distributions (dashed coloured lines) are also shown for
comparison. The velocity distributions are not well fit by the SHM Maxwellian and
are instead best fitted by the empirical fitting function of Mao {\it et
  al.}~\cite{Mao:2012hf} (eq.~\eqref{eq:Mao}). The best fit parameters in
eq.~\eqref{eq:Mao} are $v_0=250$~km$/$s and $p=3.1$ with a reduced $\chi^2$ of 1.6
for the best fit halo (E11), and $v_0=394$~km$/$s, and $p=4.8$ with a reduced $\chi^2$ of
2.6 for the other halo (E9). By comparing the two panels of 
figure~\ref{fig:fv-EAGLE-HR-all}, one observes that the peak of the
distributions in the DMO simulation is shifted to smaller velocities compared to the
hydrodynamic case.

\begin{figure}
\begin{center}
  \includegraphics[width=0.49\textwidth]{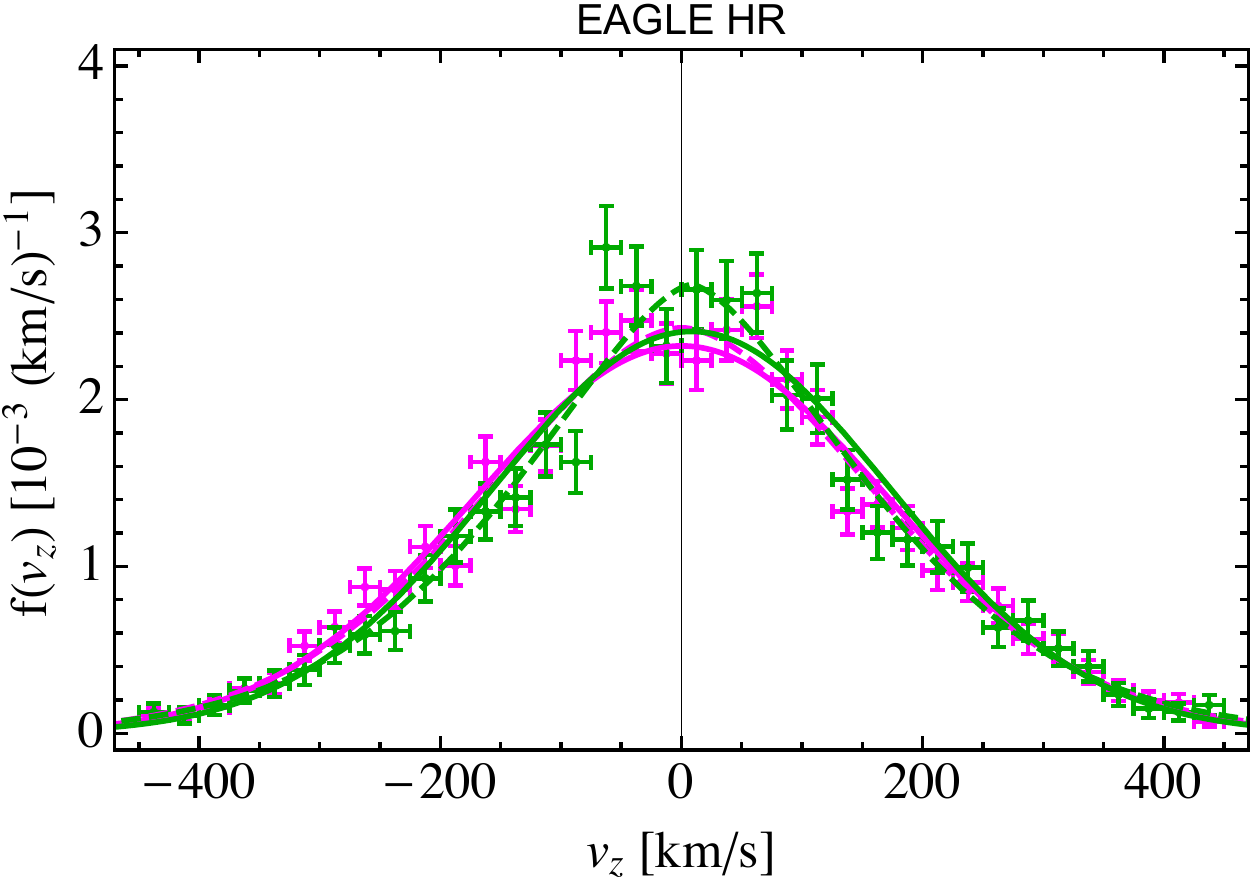}
  \includegraphics[width=0.49\textwidth]{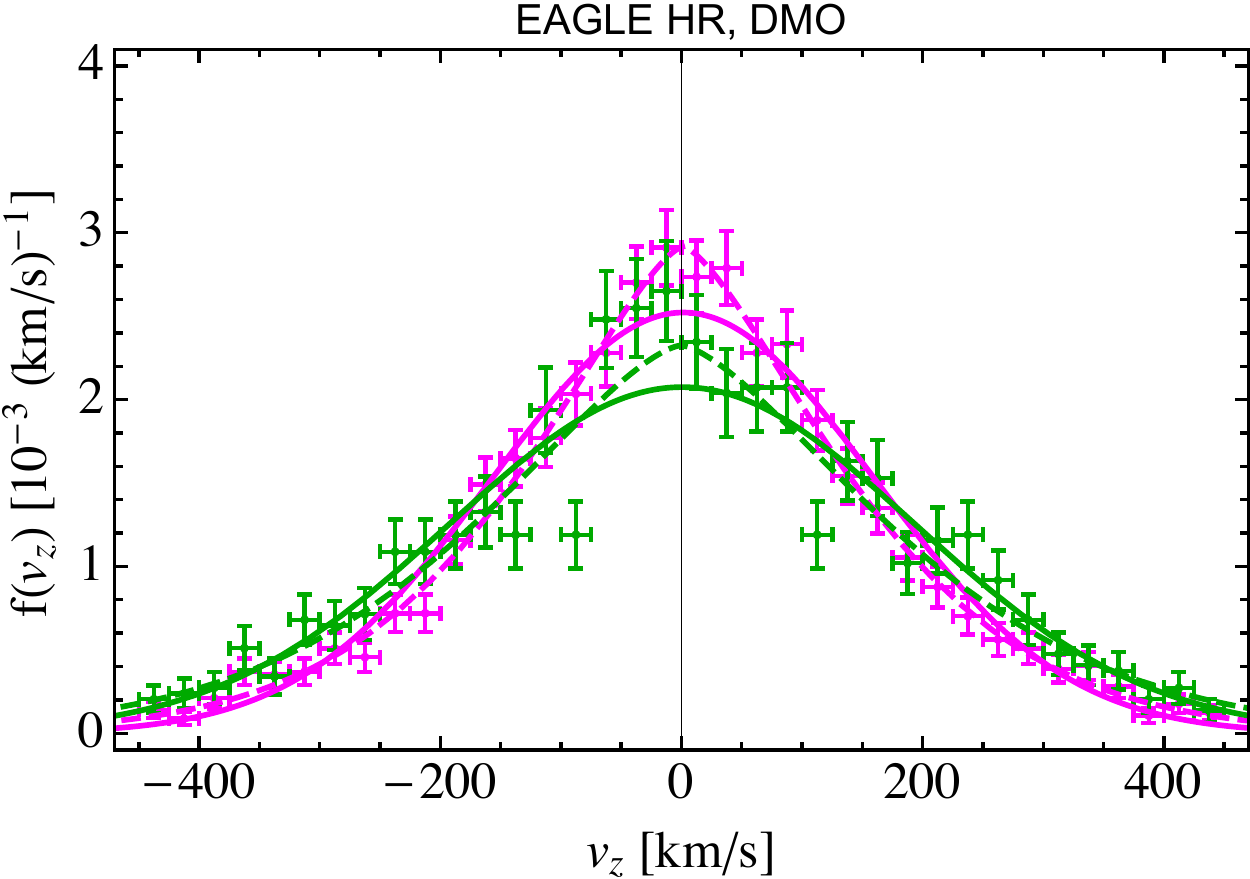}\\
    \vspace{5pt} \includegraphics[width=0.49\textwidth]{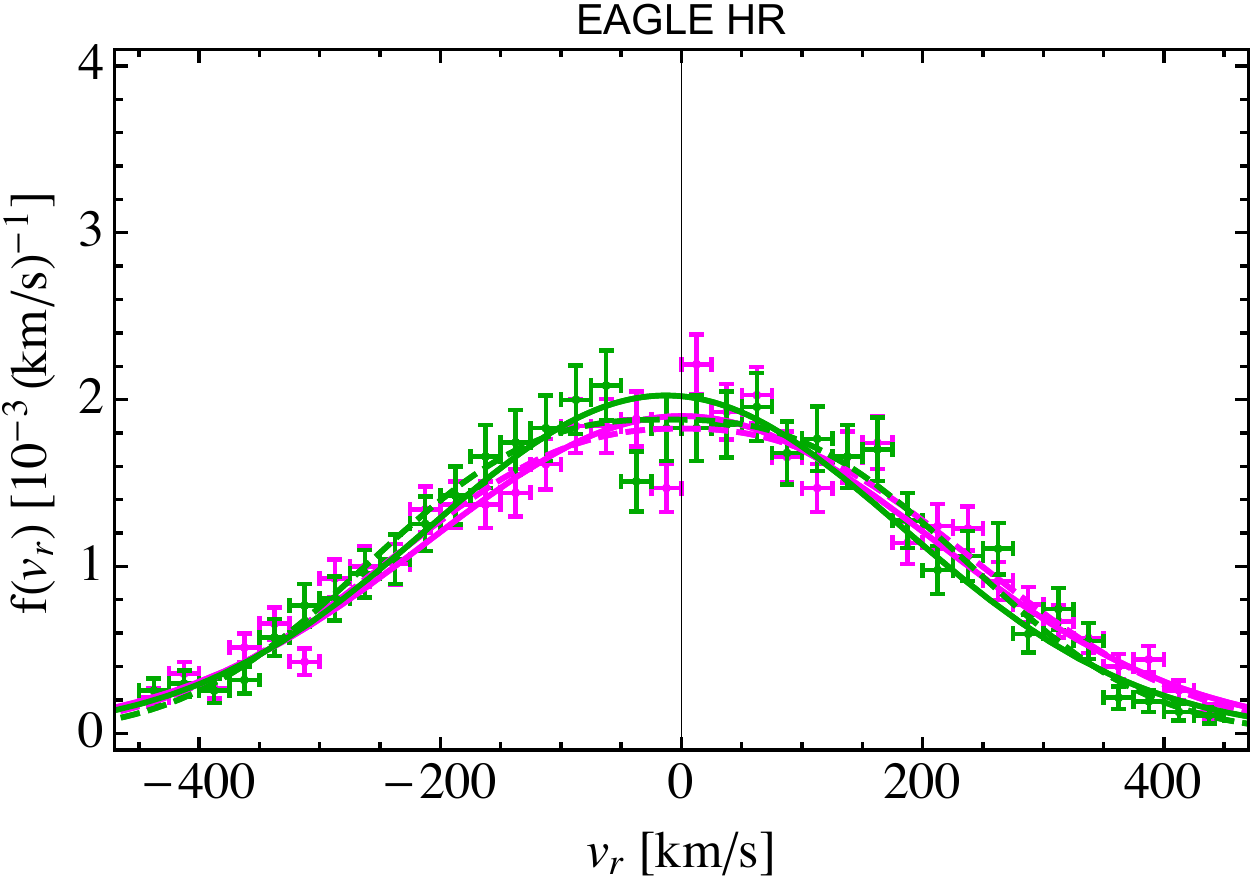}
  \includegraphics[width=0.49\textwidth]{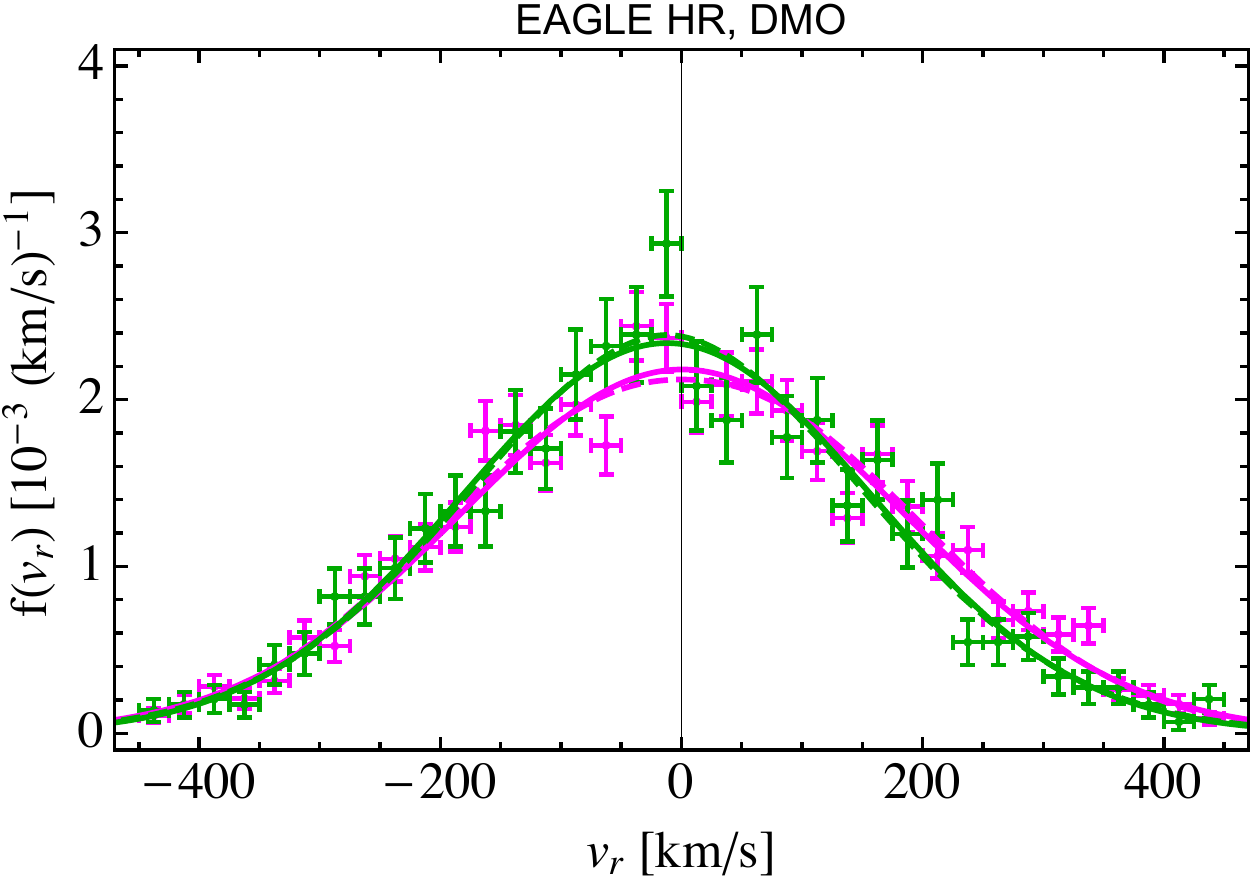}\\
  \vspace{5pt}  \includegraphics[width=0.49\textwidth]{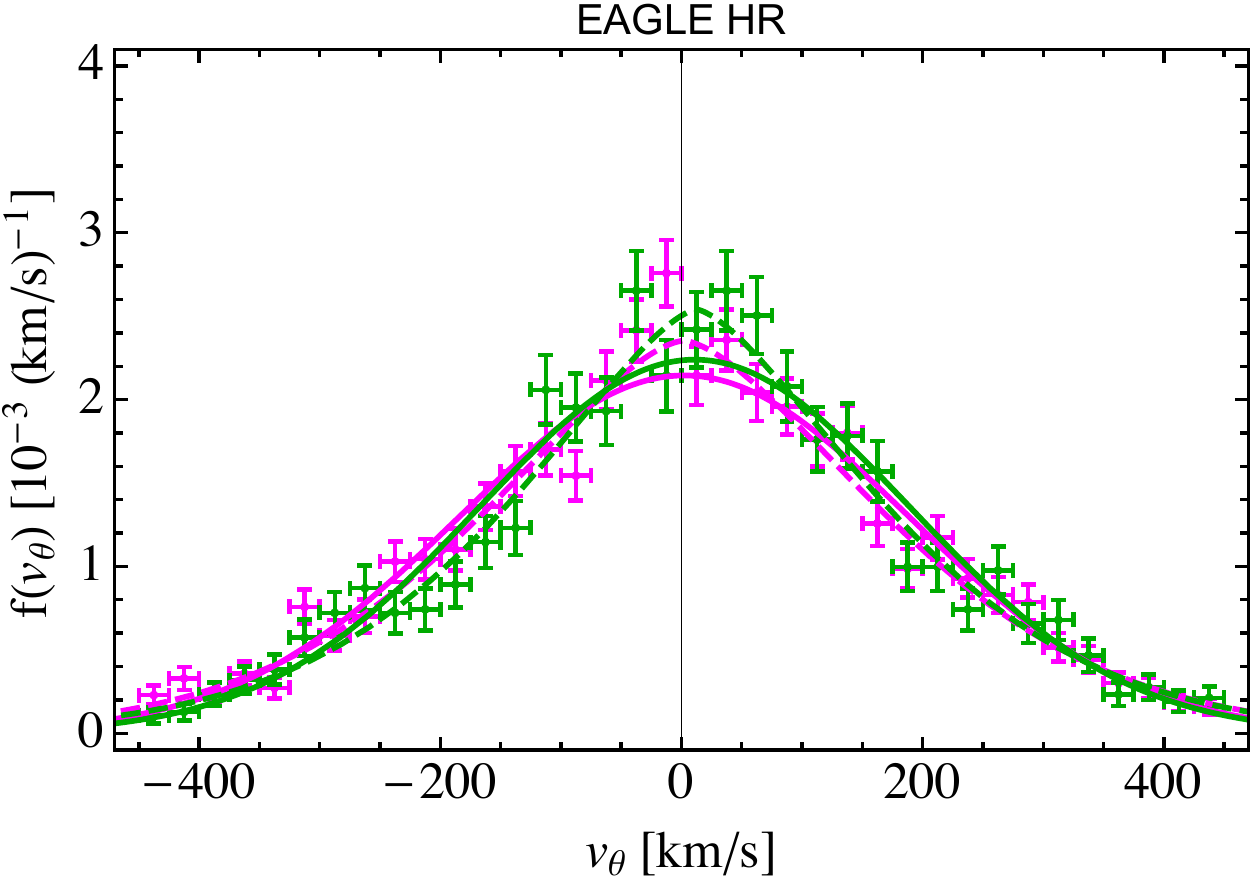}
  \includegraphics[width=0.49\textwidth]{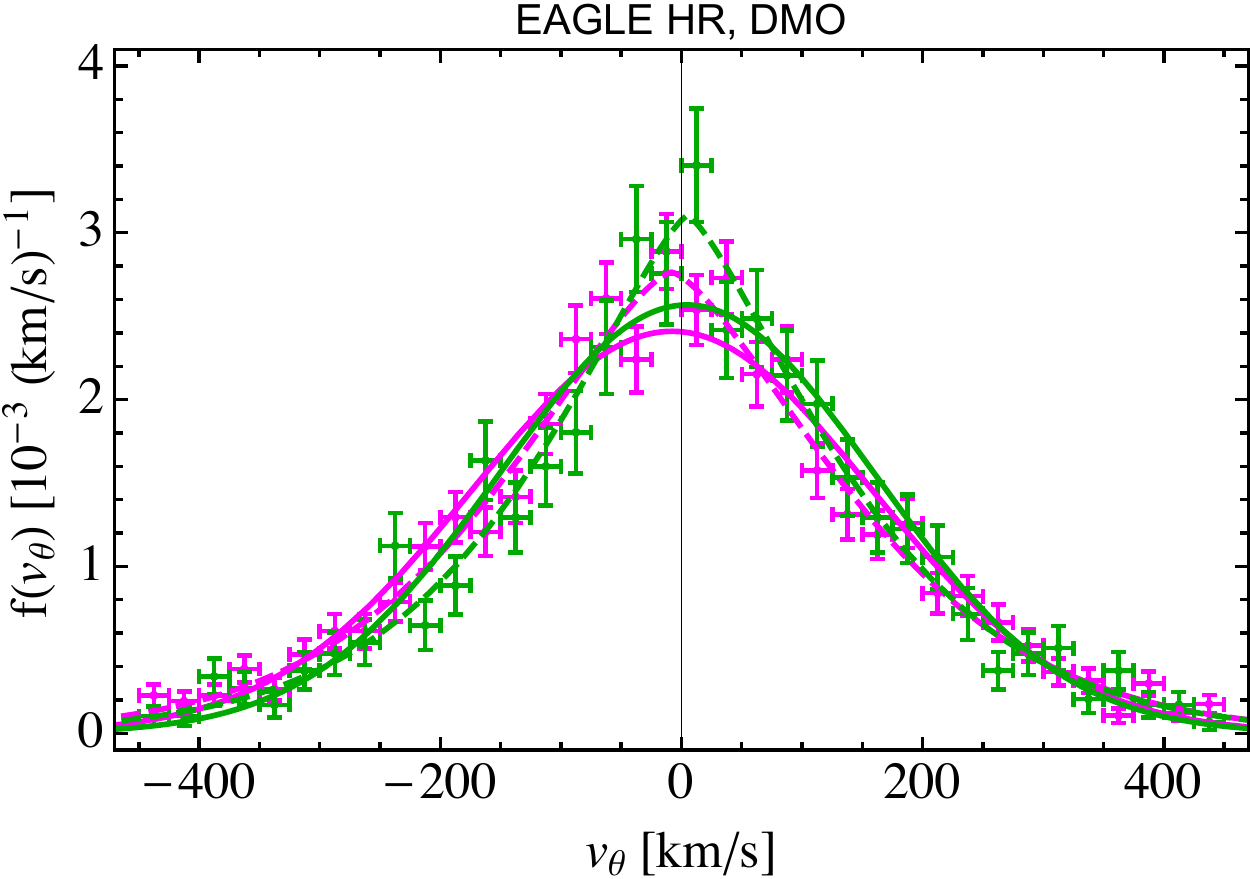}
\caption{\label{fig:fv-comp-EAGLE-HR-all} Same as figure~\ref{fig:fv-comp-EAGLE-HR},
  but for the two haloes in the \eagle HR simulation which satisfy all three
  selection criteria discussed in section~\ref{sec:selection}.}
\end{center}
\end{figure}

Figure~\ref{fig:fv-comp-EAGLE-HR-all} shows the components of the DM velocity
distribution for the same two haloes in the hydrodynamic (left panels) and DMO (right
panels) simulations. The best fit Gaussian and generalized Gaussian distributions are
also shown as solid and dashed coloured lines, respectively. The generalized Gaussian
distribution gives the best fit to the radial and vertical components of the DM
velocity distribution, while the double Gaussian gives a slightly better fit to the
azimuthal velocity distribution (for a few haloes there is evidence for overfitting
meaning that there is too much freedom in the fitting function). The two haloes do
not have a significant non-zero mean azimuthal speed and hence there is no evidence
for a dark disc component.

The average DM density in the torus located at the Solar position is 0.44 -- 0.65
GeV$/$cm$^3$ for the hydrodynamic case and 0.27 -- 0.53 GeV$/$cm$^3$ for the DMO
case.  While the average DM density in a shell centred at 8 kpc is 0.45 -- 0.63
GeV$/$cm$^3$ for the hydrodynamic case, comparable to the DM density in the torus.

\begin{figure}
\begin{center}
 \includegraphics[width=0.49\textwidth]{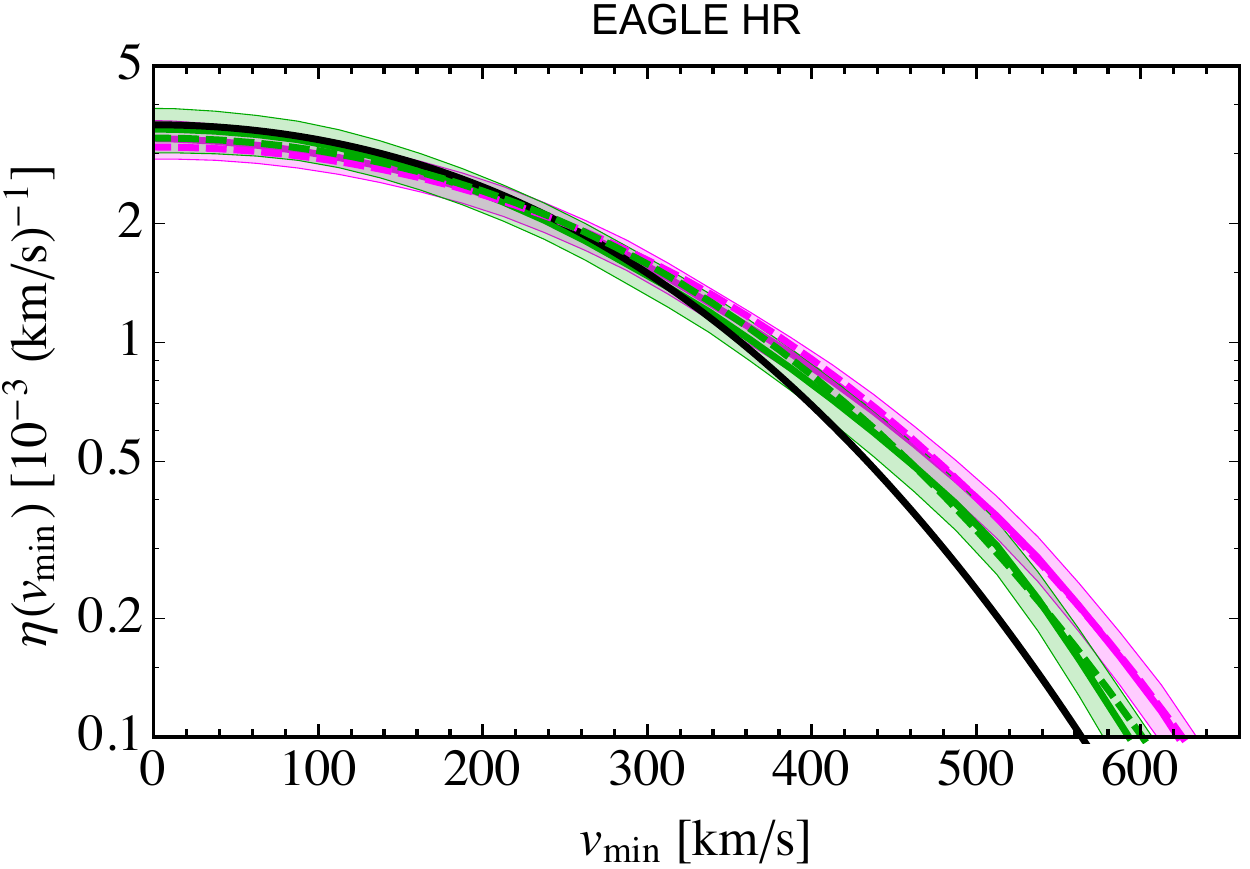}
  \includegraphics[width=0.49\textwidth]{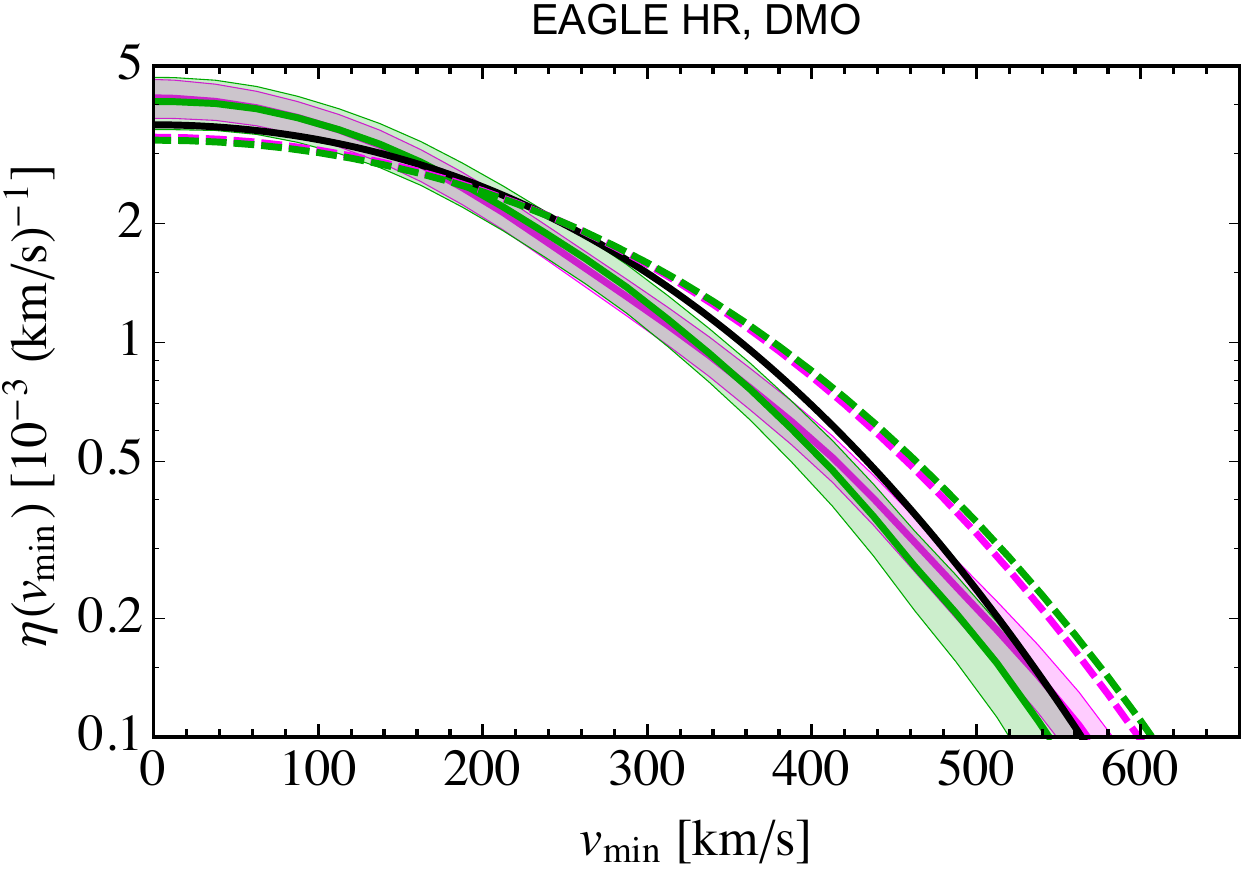}\\
\caption{\label{fig:eta-EAGLE-HR-all} Same as figure~\ref{fig:eta}, but for the two
  haloes in the \eagle HR simulation which satisfy all three selection criteria
  discussed in section~\ref{sec:selection}.}
\end{center}
\end{figure}

In figure~\ref{fig:eta-EAGLE-HR-all} we show the time-averaged halo integrals
(Eq.~\ref{eq:eta}) as a function $v_{\rm min}$ for the two haloes in the hydrodynamic
(left panel) and DMO (right panel) simulations. The halo integrals computed from the
mean value of the velocity distribution are shown by solid coloured lines, while the
shaded bands are obtained by adding and subtracting $1\sigma$ to the mean velocity
distribution. The halo integral for the SHM Maxwellian and the 
best fit Maxwellian  are shown by solid black, and dashed 
coloured lines for each halo (with matching colours), respectively. 
One can see from the left panel that the dashed lines fall within the $1 \sigma$ uncertainty band of the
halo integrals for both haloes in the hydrodynamic simulation. In the DMO simulation however, 
the halo integrals deviate from the best fit Maxwellian at higher minimum velocities.

\begin{figure}
\begin{center}
 \includegraphics[width=0.49\textwidth]{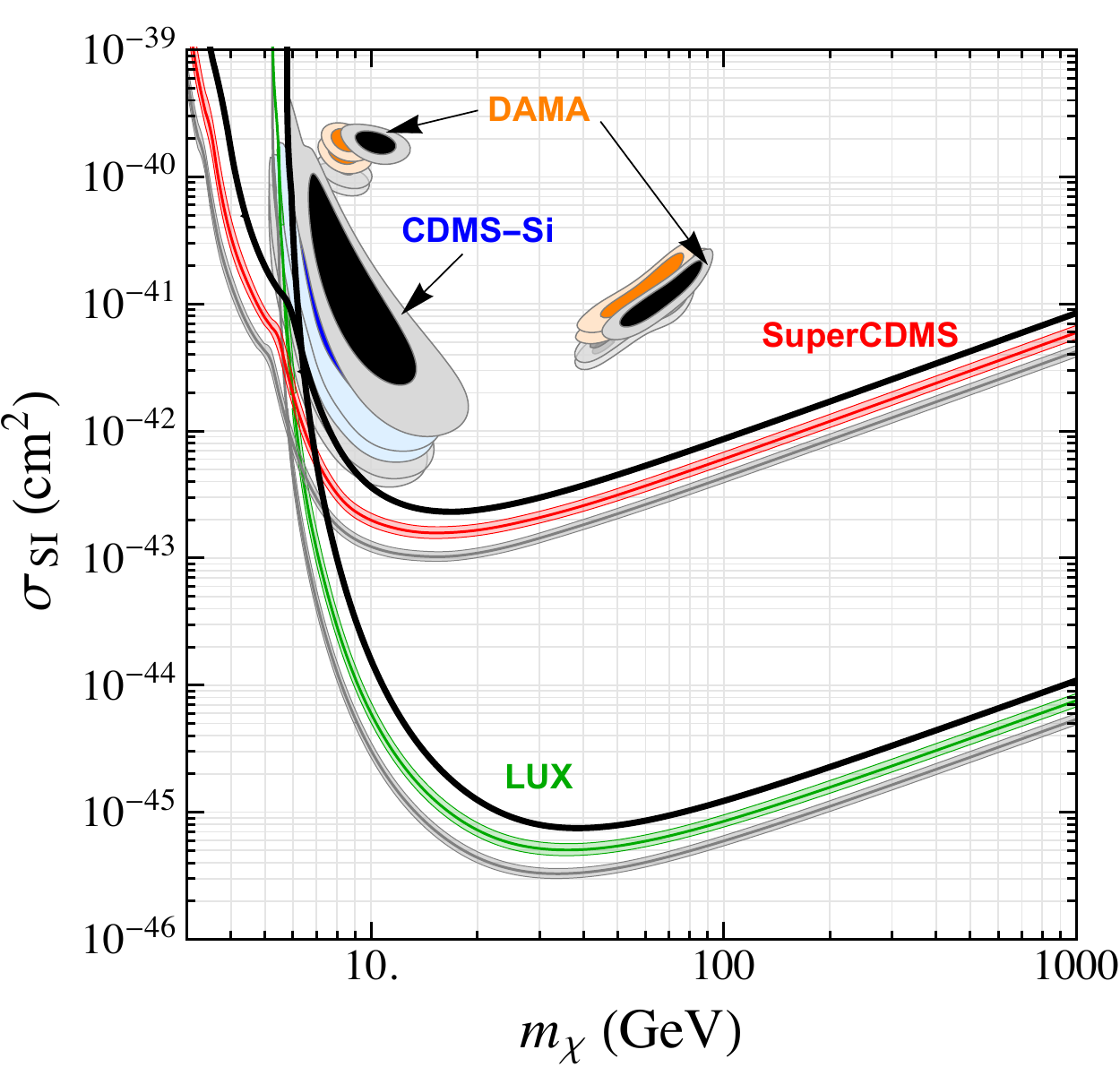}
  \includegraphics[width=0.49\textwidth]{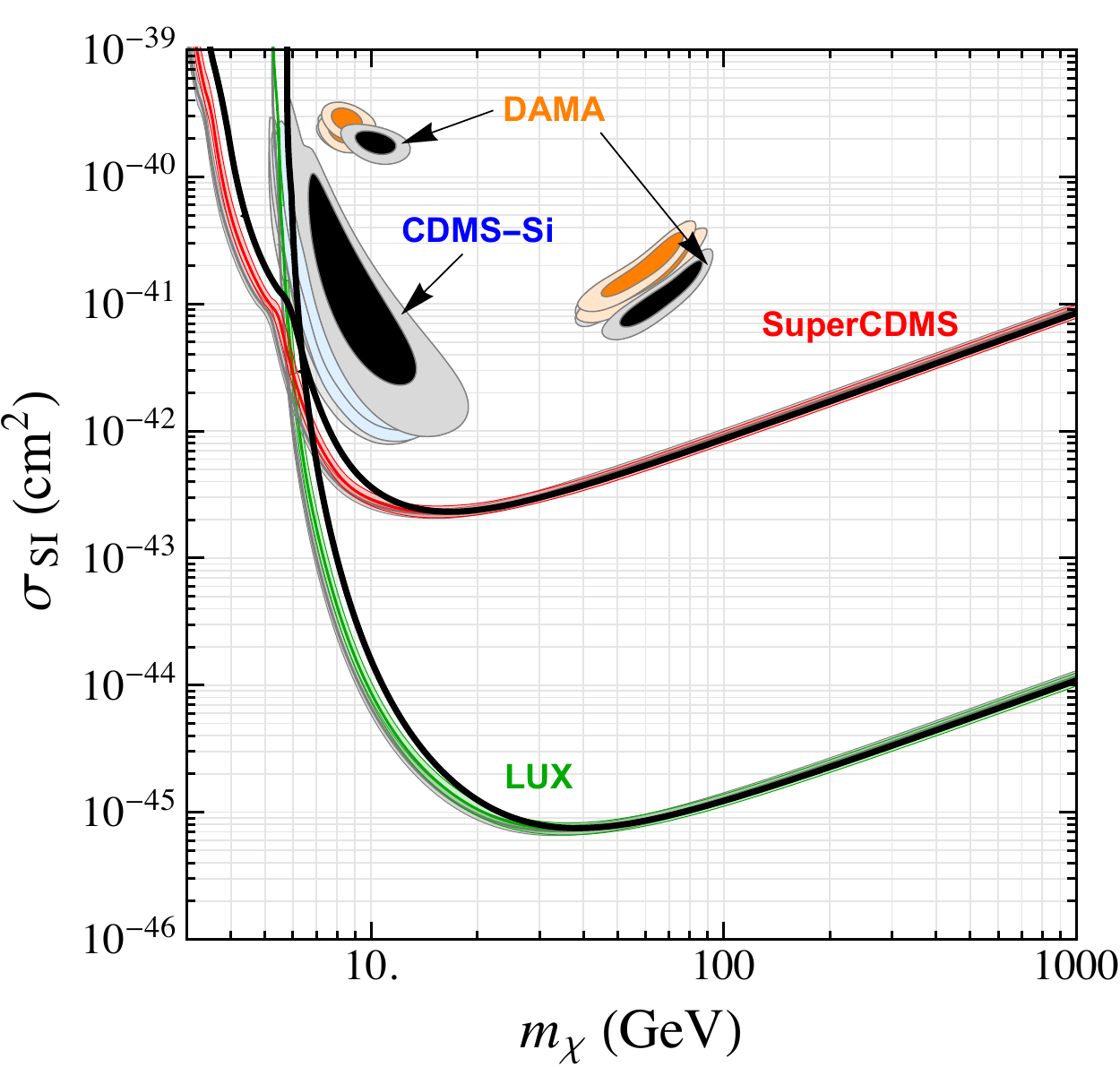}
\caption{\label{fig:ExcLim-EagleHR-all} Same as figure~\ref{fig:ExcLim-LGIR} but for
  the two haloes in the \eagle HR simulation which satisfy all three selection criteria.}
\end{center}
\end{figure}

Finally, in figure~\ref{fig:ExcLim-EagleHR-all}, we use the DM distribution for the
two haloes to analyse the data from direct detection experiments. The left panel of
figure~\ref{fig:ExcLim-EagleHR-all} shows the allowed regions from DAMA and CDMS-Si
and exclusion limits from LUX and SuperCDMS in the DM mass and spin-independent cross
section plane for the two haloes. In the right panel, we show the results for the
same two haloes in the left panel but assuming the value of 0.3 GeV$/$cm$^3$ for the
local DM density of both haloes. The conclusions remain the same as before. Namely,
when fixing the local DM density to the value assumed in the SHM, the shift in the
exclusion limits at large DM masses disappears. The high velocity tail of the DM
distribution is responsible for the shift in the preferred regions and exclusion
limits at low DM masses. While the general shift with respect to the SHM at all
masses is due to the variation in the local DM density.


\section{Dark matter velocity distribution}
\label{app:f(v)}

\begin{figure}
\begin{center}
  \includegraphics[width=0.49\textwidth]{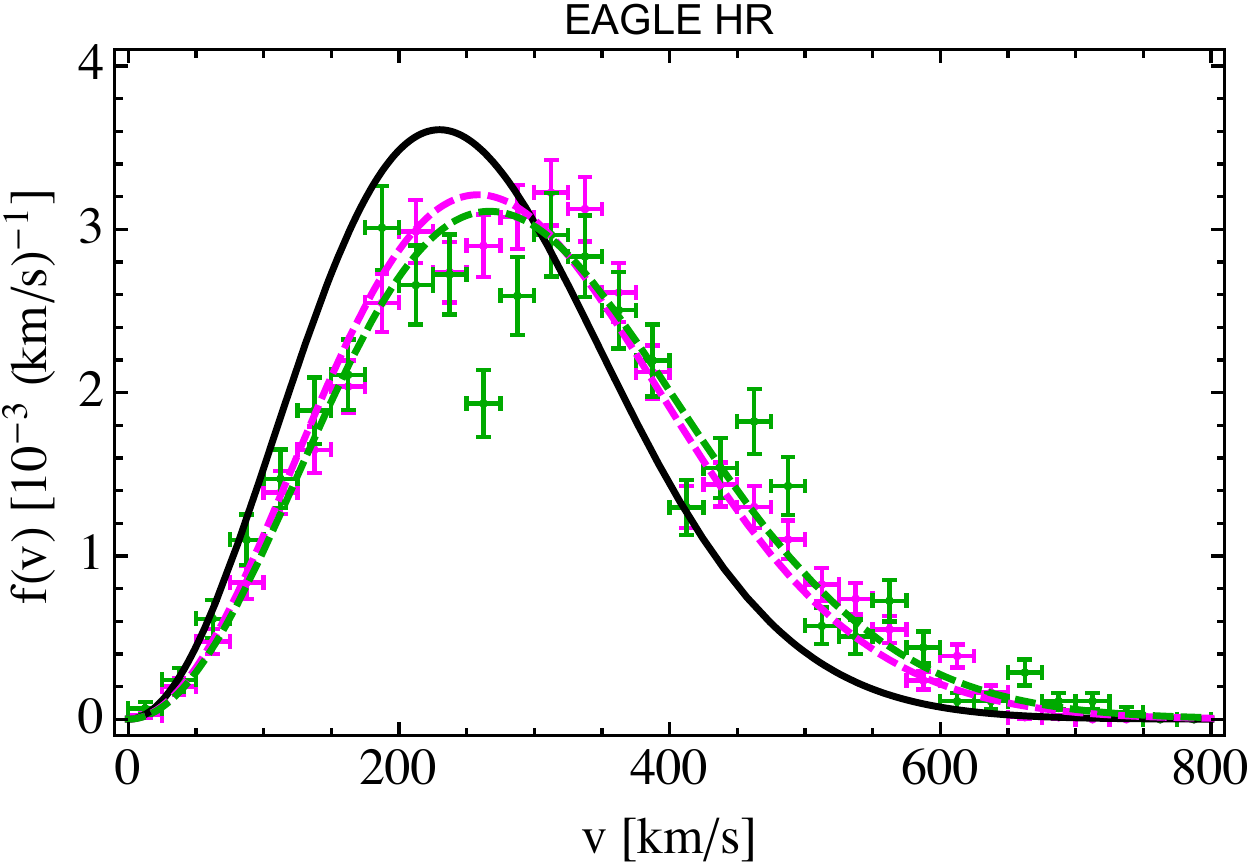}
  \includegraphics[width=0.49\textwidth]{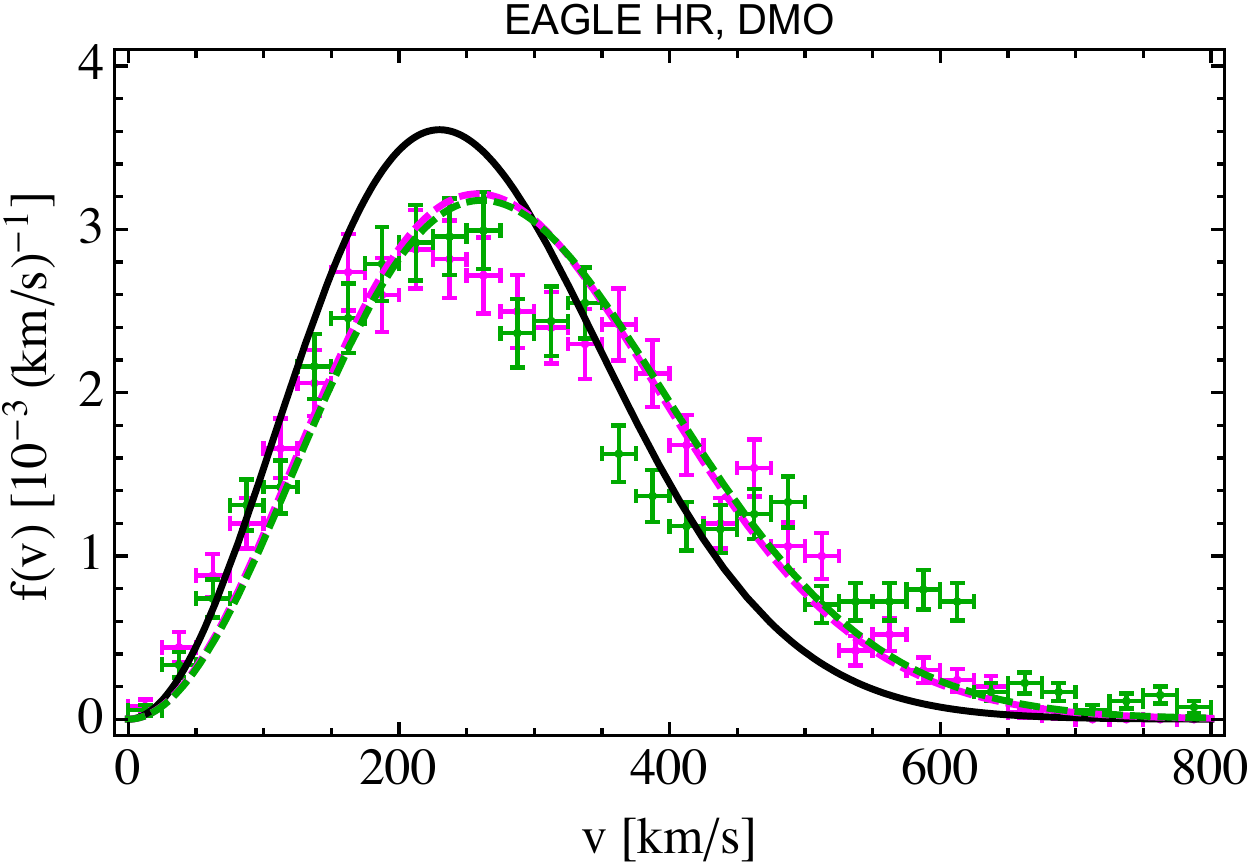}\\
\caption{\label{fig:fvrho} Same as figure~\ref{fig:fv}, but for the two haloes in the
  \eagle HR simulation with smallest (halo E6) and largest (halo E4) local DM density in the hydrodynamic
  case (left panel), and their DMO counterparts (right panel).}
\end{center}
\end{figure}

In figure~\ref{fig:fvrho} we show the DM velocity modulus distributions in the
Galactic rest frame for the haloes in the \eagle HR hydrodynamic simulation which
satisfy our selection criteria (i) and (ii) and have the smallest (halo E6) and largest (halo E4) local
DM density in our halo sample, 0.42 and 0.73 GeV$/$cm$^3$, respectively. The local DM
densities in the torus for the same haloes in the DMO simulation are 0.49 and 0.46 GeV$/$cm$^3$,
respectively. The left panel of figure~\ref{fig:fvrho} corresponds to the
hydrodynamic case, while the right panel shows the results for the DMO case. The SHM
Maxwellian (solid black line) and the best fit Maxwellians (dashed coloured lines)
are also plotted for comparison.

\section{Best fit parameters for velocity distributions}
\label{app:fits}

In tables~\ref{tab:fvfit} and~\ref{tab:fvfitDMO} we quote the best fit parameters for
three fitting functions, eq.~\eqref{eq:genMax} (with free parameter $\alpha$, and
$\alpha=1$), eq.~\eqref{eq:Mao}, and eq.~\eqref{eq:Lisanti}, presented in
section~\ref{sec:vel-Mod} to fit the DM velocity modulus distribution of the
hydrodynamic and DMO haloes, respectively.

\begin{sidewaystable}
%\begin{table}[t!]
    \centering
       \begin{tabular}{|c|c|c|c|c|c|c|c|c|c|c|c|}
        \hline
      & \multicolumn{2}{|c|}{Maxwellian} & \multicolumn{3}{|c|}{Generalized Maxw.} & \multicolumn{3}{|c|}{Mao {\it et al.}} & \multicolumn{3}{|c|}{Lisanti {\it et al.}} \\
      \hline
        Halo Name &  $v_0 \,  [\rm km/s]$ &  $\tilde{\chi}^2$ & $v_0 \,  [\rm km/s]$ & $\alpha$ &  $\tilde{\chi}^2$ & $v_0 \,  [\rm km/s]$ & $p$  & $\tilde{\chi}^2$& $v_0 \,  [\rm km/s]$ & $k$  &  $\tilde{\chi}^2$ \\
       \hline
      E1 & 287.73 & 3.77 & 189.56 & 0.68 & 1.05 & 154.48 & 3.18 & 1.03 & 287.73 & 0.03 & 3.90 \\
E2 & 266.01 & 2.68 & 227.12 & 0.85 & 2.14 & 206.14 & 4.98 & 2.00 & 266.01 & 0.02 & 2.77 \\
E3 & 288.64 & 1.47 & 294.02 & 1.02 & 1.51 & 267.35 & 3.76 & 0.82 & 296.49 & 1.86 & 1.26 \\
E4 & 258.53 & 3.81 & 284.30 & 1.12 & 3.35 & 317.74 & 4.21 & 2.51 & 282.08 & 2.65 & 2.70 \\
E5 & 249.99 & 4.39 & 277.30 & 1.13 & 4.10 & 408.10 & 5.79 & 2.81 & 321.47 & 4.80 & 3.39 \\
E6 & 266.97 & 3.52 & 227.81 & 0.85 & 3.24 & 198.93 & 3.54 & 2.98 & 266.97 & 0.02 & 3.65 \\
E7 & 270.84 & 3.67 & 254.67 & 0.93 & 3.68 & 240.96 & 3.69 & 3.87 & 270.96 & 0.77 & 3.81 \\
E8 & 258.23 & 2.86 & 217.95 & 0.84 & 2.10 & 166.91 & 2.69 & 1.61 & 258.30 & 0.63 & 2.97 \\
E9 & 248.81 & 4.19 & 288.70 & 1.20 & 3.73 & 393.63 & 4.82 & 2.58 & 342.49 & 4.36 & 3.05 \\
E10 & 261.78 & 2.49 & 274.11 & 1.06 & 2.51 & 281.33 & 3.41 & 1.17 & 282.77 & 2.21 & 1.92 \\
E11 & 262.27 & 2.77 & 279.94 & 1.08 & 2.67 & 250.06 & 3.14 & 1.57 & 276.42 & 1.87 & 2.10 \\
E12 & 231.90 & 1.43 & 252.25 & 1.11 & 0.99 & 467.84 & 5.69 & 1.53 & 245.72 & 2.67 & 1.02 \\
       \hline
       A1 & 222.93 & 3.74 & 212.25 & 0.95 & 3.73 & 186.32 & 3.29 & 2.02 & 224.89 & 0.54 & 3.40 \\
A2 & 234.35 & 10.35 & 213.58 & 0.91 & 10.50 & 136.41 & 1.07 & 3.52 & 243.23 & 0.64 & 8.25 \\
        \hline
    \end{tabular}
    \caption{Best-fit parameters for the velocity modulus distribution of the selected galaxies in 
      the \eagle HR (E1 to E12) and \apostle IR (A1 and A2) simulations, adopting the three different
      fitting functions explained in the main text (eqs.~\eqref{eq:genMax}, \eqref{eq:Mao}, \eqref{eq:Lisanti}) and, additionally, a standard
      Maxwellian distribution ($\alpha$ fixed to 1 in eq.~\eqref{eq:genMax}). Besides the best-fit parameters,
      we quote the reduced $\chi^2$, $\tilde{\chi}^2 \equiv \chi^2/(N - dof)$, where
      $dof = 2$ for the three fitting functions and $dof = 1$ for the standard
      Maxwellian. }
    \label{tab:fvfit}
 % \end{table}
  \end{sidewaystable}

 \begin{sidewaystable}
% \begin{table}[t!]
    \centering
    \begin{tabular}{|c|c|c|c|c|c|c|c|c|c|c|c|}
     \hline
      & \multicolumn{2}{|c|}{Maxwellian} & \multicolumn{3}{|c|}{Generalized Maxw.} & \multicolumn{3}{|c|}{Mao {\it et al.}} & \multicolumn{3}{|c|}{Lisanti {\it et al.}} \\
      \hline
        Halo Name &  $v_0 \,  [\rm km/s]$ &  $\tilde{\chi}^2$ & $v_0 \,  [\rm km/s]$ & $\alpha$ &  $\tilde{\chi}^2$ & $v_0 \,  [\rm km/s]$ & $p$  & $\tilde{\chi}^2$& $v_0 \,  [\rm km/s]$ & $k$  &  $\tilde{\chi}^2$ \\
       \hline
      E1 DMO & 256.70 & 10.01 & 111.77 & 0.53 & 1.67 & 102.54 & 0.72 & 1.71 & 256.70 & 0.10 & 10.34 \\
E2 DMO & 225.25 & 4.51 & 130.74 & 0.63 & 1.39 & 108.62 & 3.27 & 1.23 & 225.25 & 0.82 & 4.66 \\
E3 DMO & 269.90 & 4.70 & 136.30 & 0.58 & 1.03 & 118.62 & 1.09 & 0.89 & 269.89 & 0.07 & 4.87 \\
E4 DMO & 258.08 & 2.95 & 214.39 & 0.83 & 2.37 & 163.45 & 2.58 & 1.13 & 259.07 & 1.06 & 3.04 \\
E5 DMO & 255.68 & 2.64 & 257.69 & 1.01 & 2.75 & 317.89 & 5.21 & 2.08 & 286.05 & 3.68 & 2.59 \\
E6 DMO & 261.30 & 6.02 & 161.41 & 0.65 & 2.31 & 123.45 & 1.20 & 2.47 & 261.27 & 0.02 & 6.22 \\
E7 DMO & 242.97 & 8.95 & 137.28 & 0.62 & 3.26 & 110.09 & 1.43 & 2.50 & 242.96 & 0.02 & 9.27 \\
E8 DMO & 252.84 & 6.54 & 139.17 & 0.61 & 1.52 & 112.50 & 0.91 & 1.45 & 252.81 & 0.02 & 6.77 \\
E9 DMO & 252.01 & 2.21 & 192.51 & 0.77 & 1.57 & 152.16 & 2.35 & 1.16 & 251.99 & 0.02 & 2.31 \\
E10 DMO & 258.78 & 5.06 & 182.65 & 0.74 & 3.99 & 147.44 & 2.00 & 2.83 & 258.72 & 0.01 & 5.26 \\
E11 DMO & 247.62 & 3.98 & 183.95 & 0.76 & 2.20 & 139.50 & 2.08 & 1.23 & 247.60 & 0.02 & 4.15 \\
E12 DMO & 229.90 & 1.61 & 200.62 & 0.87 & 1.33 & 163.08 & 3.35 & 0.65 & 230.09 & 0.90 & 1.68 \\
       \hline
      A1 DMO & 193.01 & 7.12 & 135.37 & 0.72 & 3.12 & 98.90 & 1.23 & 1.42 & 192.94 & 0.51 & 7.18 \\
A2 DMO & 221.86 & 3.50 & 168.10 & 0.77 & 2.16 & 116.88 & 1.12 & 0.94 & 221.78 & 0.66 & 3.57 \\
      \hline
    \end{tabular}
    \caption{Same as table~\ref{tab:fvfit} for the corresponding DMO simulations.}
    \label{tab:fvfitDMO}
%  \end{table}
  \end{sidewaystable}

In tables~\ref{tab:fvfit_vr} and~\ref{tab:fvfit_vz} we present the best-fit
parameters for the fit with Gaussian and generalized Gaussian functions to the radial
and vertical velocity components for the selected MW-like haloes in the hydrodynamic
simulations.  In tables~\ref{tab:fvfit_vtheta} and~\ref{tab:fvfit_vtheta_double}, we
present the results of the fits with Gaussian, generalized Gaussian and double
Gaussian functions for the azimuthal component of the DM velocity distribution. The
results are discussed in section~\ref{sec:vel-comp}.

\begin{sidewaystable}
    \centering
    \begin{tabular}{|c|c|c|c|c|c|c|c|c|c|c|c|}
     \hline
      & \multicolumn{4}{|c|}{Gaussian} & \multicolumn{5}{|c|}{Generalized Gaussian} \\
      \hline
      Halo Name & $v_0 \,  [\rm km/s]$ &  $\mu \,  [\rm km/s]$ & $\sigma_\mu \,  [\rm km/s]$ &$\chi^2$ &  $v_0 \,  [\rm km/s]$ &  $\mu \,  [\rm km/s]$   & $\sigma_\mu \,  [\rm km/s]$  & $\alpha$ & $\chi^2$  \\
       \hline
     E1  & 291.24 & 1.60 & 4.66 & 1.41 & 280.09 & 2.02 & 4.81 & 0.88 & 1.38\\
E2 & 282.12 & -5.32 & 4.47 & 1.15 & 288.89 & -5.39 & 4.39 & 1.08 & 1.15\\
E3 & 319.49 & 1.89 & 5.02 & 1.58 & 340.66 & 2.13 & 4.57 & 1.35 & 1.31\\
E4 & 304.65 & 5.39 & 4.44 & 1.91 & 328.51 & 4.51 & 3.93 & 1.44 & 1.39\\
E5 & 295.21 & -6.11 & 4.57 & 0.83 & 307.12 & -5.96 & 4.36 & 1.17 & 0.73\\
E6 & 282.13 & -7.24 & 5.21 & 0.96 & 289.14 & -7.11 & 5.10 & 1.09 & 0.97\\
E7 & 276.18 & 6.20 & 4.51 & 1.14 & 287.58 & 6.27 & 4.35 & 1.14 & 1.09\\
E8 & 274.78 & 1.23 & 4.35 & 1.44 & 293.53 & -1.08 & 4.14 & 1.26 & 1.15\\
E9 & 278.69 & -13.06 & 4.99 & 1.15 & 299.24 & -10.96 & 4.59 & 1.32 & 0.83\\
E10 & 298.25 & 12.26 & 4.89 & 0.87 & 312.75 & 12.11 & 4.56 & 1.23 & 0.71\\
E11 & 296.73 & 0.24 & 4.48 & 1.50 & 308.95 & 0.32 & 4.26 & 1.18 & 1.41\\
E12 & 235.80 & 1.76 & 3.09 & 1.44 & 257.97 & 1.25 & 2.98 & 1.26 & 0.83\\
       \hline
      A1   & 235.69 & -1.35 & 2.49 & 1.29 & 243.91 & -0.88 & 2.48 & 1.08 & 1.22\\
A2 & 273.70 & -1.21 & 3.52 & 1.07 & 286.41 & -1.08 & 3.41 & 1.15 & 0.95\\
      \hline
    \end{tabular}
    \caption{Best-fit parameters for the distribution of the radial
      velocity component, $v_r$, of the selected galaxies in the \eagle HR (haloes E1 to E12) and
      \apostle IR (haloes A1 and A2) hydrodynamic simulations, adopting a Gaussian and a generalized
      Gaussian function. Besides the best-fit parameters, we quote the reduced
      $\chi^2$, $\tilde{\chi}^2 \equiv \chi^2/(N - dof)$, where $dof = 2$ for the
      Gaussian function and $dof = 3$ for the generalized Gaussian.}
    \label{tab:fvfit_vr}
\end{sidewaystable}
  
\begin{sidewaystable}
    \centering
    \begin{tabular}{|c|c|c|c|c|c|c|c|c|c|c|c|}
         \hline
      & \multicolumn{4}{|c|}{Gaussian} & \multicolumn{5}{|c|}{Generalized Gaussian} \\
      \hline
        Halo Name & $v_0 \,  [\rm km/s]$ &  $\mu \,  [\rm km/s]$  & $\sigma_\mu \,  [\rm km/s]$  &$\chi^2$ &  $v_0 \,  [\rm km/s]$ &  $\mu \,  [\rm km/s]$  & $\sigma_\mu \,  [\rm km/s]$  & $\alpha$ & $\chi^2$  \\
       \hline
     E1  & 262.08 & -1.52 & 3.98 & 1.03 & 243.43 & -1.02 & 4.11 & 0.83 & 0.84\\
E2 & 237.27 & -2.98 & 3.50 & 1.41 & 198.65 & -2.74 & 3.60 & 0.70 & 0.54\\
E3 & 261.94 & -4.63 & 3.66 & 1.96 & 220.15 & -5.47 & 3.81 & 0.69 & 0.96\\
E4 & 219.83 & 1.13 & 2.81 & 0.92 & 205.46 & 1.82 & 2.86 & 0.87 & 0.69\\
E5 & 236.11 & 0.57 & 3.31 & 1.66 & 204.89 & -0.38 & 3.39 & 0.75 & 1.00\\
E6 & 259.93 & 5.05 & 4.63 & 0.90 & 243.11 & 4.08 & 4.78 & 0.85 & 0.80\\
E7 & 283.42 & -11.74 & 4.69 & 1.62 & 245.79 & -11.82 & 4.97 & 0.71 & 1.07\\
E8 & 239.72 & 0.39 & 3.57 & 0.71 & 229.54 & 0.59 & 3.64 & 0.90 & 0.64\\
E9 & 234.25 & 8.29 & 3.89 & 1.23 & 206.72 & 8.08 & 3.98 & 0.77 & 0.85\\
E10 & 237.75 & 0.05 & 3.53 & 0.78 & 228.90 & -0.06 & 3.57 & 0.92 & 0.75\\
E11 & 243.00 & 0.13 & 3.35 & 1.12 & 231.29 & 0.22 & 3.40 & 0.89 & 1.04\\
E12 & 197.79 & 2.68 & 2.52 & 1.91 & 173.20 & 2.94 & 2.54 & 0.77 & 1.14\\
       \hline
      A1   & 210.32 & -3.68 & 2.17 & 1.88 & 186.41 & -3.31 & 2.21 & 0.79 & 0.81 \\
A2 & 222.02 & 4.87 & 2.64 & 2.68 & 183.71 & 6.04 & 2.64 & 0.72 & 1.21 \\
      \hline
    \end{tabular}
    \caption{Same as table~\ref{tab:fvfit_vr} for the distribution of the vertical velocity component, $v_z$.}
    \label{tab:fvfit_vz}
\end{sidewaystable}
  
\begin{sidewaystable}
    \centering
    \begin{tabular}{|c|c|c|c|c|c|c|c|c|c|c|c|}
           \hline
      & \multicolumn{4}{|c|}{Gaussian} & \multicolumn{5}{|c|}{Generalized Gaussian} \\
      \hline
        Halo Name & $v_0 \,  [\rm km/s]$ &  $\mu \,  [\rm km/s]$   & $\sigma_\mu \,  [\rm km/s]$ &$\chi^2$ &  $v_0 \,  [\rm km/s]$ &  $\mu \,  [\rm km/s]$   & $\sigma_\mu \,  [\rm km/s]$ & $\alpha$ & $\chi^2$  \\
       \hline
     E1  & 277.18 & 2.82 & 4.31 & 1.30 & 250.03 & 1.31 & 4.52 & 0.77 & 0.96\\
E2 & 257.62 & -4.63 & 3.94 & 1.39 & 244.29 & -3.11 & 4.12 & 0.88 & 1.33\\
E3 & 276.68 & 7.02 & 3.98 & 1.18 & 259.37 & 6.55 & 4.10 & 0.84 & 1.05\\
E4 & 265.46 & -20.43 & 3.59 & 1.21 & 266.29 & -20.44 & 3.58 & 1.01 & 1.25\\
E5 & 247.12 & 8.42 & 3.52 & 1.46 & 219.77 & 6.90 & 3.62 & 0.78 & 1.08\\
E6 & 247.07 & 3.50 & 4.30 & 1.22 & 215.85 & 3.74 & 4.36 & 0.76 & 0.88\\
E7 & 241.74 & 39.93 & 3.86 & 2.09 & 222.41 & 41.61 & 4.03 & 0.81 & 1.84\\
E8 & 235.53 & 13.86 & 3.47 & 2.55 & 180.26 & 12.03 & 3.43 & 0.61 & 0.88\\
E9 & 251.95 & 10.92 & 4.27 & 1.70 & 217.46 & 12.36 & 4.59 & 0.73 & 1.19\\
E10 & 266.72 & 23.90 & 4.13 & 1.39 & 248.16 & 24.18 & 4.31 & 0.82 & 1.20\\
E11 & 263.07 & 2.29 & 3.74 & 1.67 & 236.30 & 1.72 & 3.87 & 0.78 & 1.31\\
E12 & 230.57 & 64.55 & 3.13 & 2.63 & 218.91 & 66.16 & 3.28 & 0.87 & 2.54\\
       \hline
      A1  & 207.47 & 38.15 & 2.19 & 2.67 & 186.13 & 37.84 & 2.22 & 0.81 & 1.82\\
A2 & 212.30 & -3.95 & 2.55 & 3.09 & 182.27 & -0.88 & 2.61 & 0.75 & 2.02 \\
      \hline
    \end{tabular}
    \caption{Same as table~\ref{tab:fvfit_vr} for the distribution of the azimuthal velocity component, $v_\theta$.}
    \label{tab:fvfit_vtheta}
\end{sidewaystable}
  
\begin{sidewaystable}
    \centering
    \begin{tabular}{|c|c|c|c|c|c|c|c|c|c|c|c|}
      \hline
        Halo Name & $c_1$ & $v_1 \,  [\rm km/s]$ &  $\mu_1 \,  [\rm km/s]$   & $\sigma_{\mu_1} \,  [\rm km/s]$ & $c_2$ &  $v_2 \,  [\rm km/s]$ &  $\mu_2 \,  [\rm km/s]$  & $\sigma_{\mu_2} \,  [\rm km/s]$  & $\chi^2_{\rm double \, Gauss}$  \\
       \hline
      E1 & 0.13 & 111.28 & -21.95 & 14.40 & 0.87 & 303.24 & 6.99 & 5.99 & 0.87 \\
E2 & 0.88 & 275.08 & -14.91 & 6.77 & 0.12 & 112.76 & 54.91 & 15.24 & 1.01 \\
E3 & 0.95 & 287.09 & 7.63 & 4.41 & 0.05 & 62.73 & -6.72 & 13.79 & 0.90 \\
E4 & 0.99 & 262.54 & -21.85 & 3.58 & 0.01 & 13.20 & 420.94 & 2.95 & 1.08 \\
E5 & 0.09 & 71.29 & -10.48 & 9.70 & 0.91 & 261.96 & 11.39 & 4.25 & 0.85 \\
E6 & 0.09 & 70.39 & -4.38 & 11.62 & 0.91 & 262.29 & 5.23 & 5.11 & 0.81 \\
E7 & 0.57 & 295.43 & -10.19 & 22.52 & 0.43 & 165.06 & 88.48 & 12.74 & 0.87 \\ 
E8 & 0.63 & 311.89 & 8.13 & 8.25 & 0.37 & 132.95 & 17.15 & 6.56 & 0.97 \\
E9 & 0.81 & 284.78 & 9.11 & 6.46 & 0.19 & 116.02 & 17.27 & 11.53 & 1.14 \\
E10 & 0.30 & 177.56 & 12.35 & 17.25 & 0.70 & 316.24 & 30.77 & 10.29 & 1.24 \\
E11 & 0.92 & 277.68 & 1.79 & 4.68 & 0.08 & 76.26 & 2.45 & 20.46 & 1.32 \\
E12 & 0.68 & 254.78 & 26.62 & 15.07 & 0.32 & 139.89 & 135.57 & 11.56 & 0.88 \\
       \hline
       A1   & 0.53 & 252.62 & 7.61 & 10.75 & 0.47 & 153.16 & 63.98 & 5.84 & 1.00 \\
A2 & 0.74 & 242.65 & -16.23 & 4.62 & 0.26 & 112.55 & 28.11 & 6.90 & 1.44\\
      \hline
    \end{tabular}
    \caption{Best-fit parameters for the distribution of the azimuthal
      velocity component, $v_\theta$, of the selected galaxies in the \eagle HR (E1 to E12) and
      \apostle IR (A1 and A2) hydrodynamic simulations, adopting a double Gaussian, see text for
      more details.}
    \label{tab:fvfit_vtheta_double}
\end{sidewaystable}

\bibliographystyle{JHEP}
\bibliography{./refs}

\end{document}